\documentclass{aa}  

\usepackage{graphicx}
\usepackage{txfonts}
\usepackage{hyperref}
\hypersetup{colorlinks=true,citecolor=blue}
\usepackage{color}
\usepackage[dvipsnames]{xcolor}

\usepackage{multirow}
\usepackage{amsmath}
\usepackage{lscape}
\usepackage{supertabular}
\usepackage{longtable}

\makeatletter
\DeclareRobustCommand{\HII}{%
  \mbox{H\check@mathfonts\fontsize\sf@size\z@\selectfont\ II\ }%
}
\makeatother

\begin{document}

   \title{The early evolution of young massive clusters:}

   \subtitle{The kinematic history of NGC6611 / M16}

   \author{M. Stoop\inst{1}
          \and
          L. Kaper\inst{1}
          \and
          A. de Koter\inst{1,2}
          \and
          D. Guo\inst{1}
          \and
          H. J. G. L. M. Lamers\inst{1}
          \and
          S. Rieder\inst{3,4}
          }

   \institute{Anton Pannekoek Institute for Astronomy, University of Amsterdam, Science Park 904, 1098 XH Amsterdam, the Netherlands\\
              \email{m.p.stoop@uva.nl}
         \and
             Institute of Astronomy, KU Leuven, Celestijnenlaan 200 D, 3001 Leuven, Belgium
         \and
             Geneva Observatory, University of Geneva, Chemin Pegasi 51, 1290 Sauverny, Switzerland
         \and
              School of Physics and Astronomy, University of Exeter, Stocker Road, Exeter EX4 4QL, United Kingdom
             }

   \date{\today}

 
  \abstract
{Young massive clusters provide the opportunity to study the outcome of the star formation process and the early evolution of star clusters. In the first few million years the (massive) stars dynamically interact, producing runaways and affecting the initial (binary) population.}
{Observing and interpreting the dynamics of young massive clusters is key to our understanding of the star formation process and predicting the outcome of stellar evolution, for example the number of gravitational wave sources.}
{We have studied NGC6611 in the Eagle Nebula (M16), a young massive cluster hosting $\sim$ 19 O stars. We used \textit{Gaia} EDR3 data to determine the membership, age, cluster dynamics, and the kinematics of the massive stars including runaways.} 
{The membership analysis yields 137 members located at a mean distance of 1706 $\pm$ 7~pc. The colour - absolute magnitude diagram reveals a blue and a red population of pre-main-sequence stars, consistent with two distinct populations of stars. In line with earlier studies, the youngest (reddest) population has a mean extinction of $A_{\rm{V}}$ = 3.6 $\pm$ 0.1 mag and an age of 1.3 $\pm$ 0.2 Myr, while the older population of stars has a mean extinction of $A_{\rm{V}}$ = 2.0 $\pm$ 0.1 mag and an age of 7.5 $\pm$ 0.4 Myr. The latter population is more spatially extended than the younger generation of stars. We argue that most of the OB stars belong to the younger population. We identify eight runaways originating from the centre of NGC6611, consistent with the dynamical ejection scenario.}
{We have studied the kinematics of the O stars in detail and show that $\gtrsim$ 50\% of the O stars have velocities comparable to or greater than the escape velocity. These O stars can be traced back to the centre of NGC6611 with kinematic ages ranging from 0 to 2 Myr. These results suggest that dynamical interactions played an important role in the early evolution of NGC6611, which is surprising considering the relatively low current stellar density (0.1-1 $\times$ 10$^{3}$ M$_{\odot}$ pc$^{-3}$). Comparing our results to simulations of young massive clusters, the initial radius of 0.1-0.5 pc (needed to produce the observed O star runaway fraction) is not consistent with that of NGC6611. We propose a scenario where the O stars initially form in wide binaries or higher order systems and possibly harden through dynamical interactions.}


   \maketitle
%

\section{Introduction}
Stars that end their lives in core-collapse supernovae are referred to as massive stars. They have a mass $\gtrsim$ 8 M$_{\odot}$ at the end of the formation process and begin their lives as O or early B-type stars. Most of the O stars are in young massive clusters and OB associations. It has become increasingly clear that most, if not all, O stars in the field originate from these young massive clusters or OB associations, but have been ejected as runaway stars \citep{deWit2005,Gvaramadze2012}.

The two most important mechanisms for these ejections are the dynamical ejection and binary supernova scenarios, or a combination of both \citep{Zwicky1957,Blaauw1961,Hoogerwerf2001}. In the first scenario, two massive stars are initially in a binary orbit. After the primary star explodes as a supernova, the secondary star receives a kick velocity comparable to its original orbital velocity. If the binary remains bound, such a system can for example be observed as a high mass X-ray binary \citep[e.g.][]{vanderMeij21}. If the binary is disrupted, the secondary can be observed as a single runaway and the compact object possibly as a fast moving radio pulsar. In several Galactic supernova remnants, runaway stars are identified which may have been ejected after the supernova \citep{Dincel2015,Boubert2017}. 

In the dynamical ejection scenario, the gravitational interaction between single-binary or binary-binary stars can produce a runaway star \citep{Leonard1988,Fujii2011}. Due to their nature, these interactions are most efficient at high stellar density. Young massive star clusters are ideal environments, where a high stellar density is present before possible expansion sets in. For example, runaway O-type stars have been identified coming from Westerlund 2 and NGC 3603 \citep{Drew2018,Drew2019}. Both clusters are estimated to be younger than $\sim$ 2.5 Myr \citep{Harayama2008,Pfalzner2009}, making these clusters too young to have produced runaways through the binary supernova scenario. The overall importance of dynamical interactions has remained elusive and for the intra-cluster dynamics, we often rely on simulations \citep[see e.g.][]{Fujii2011,Fujii2012,Fujii2013,Fujii2014,Oh2015,Oh2016}.

Typically, we assume that stars in young massive clusters originate in a burst of star formation. Even accounting for an age spread of approximately several 100 kyr, it has become difficult to reconcile this notion for several clusters. For example, the Orion nebula cluster is consistent with having three discrete episodes of star formation each $\sim$ 0.5 to 1.0 Myr apart \citep{Beccari2017,Jerabkova2019}. Similarly, Westerlund 1 and its stellar content are inconsistent with having a single age, but require an age spread of several million years \citep{Baesor2021}. The R136 cluster in the 30 Doradus region in the Large Magellanic Cloud also shows an age spread among the massive stars of several million years \citep{Schneider2018}.

The study of young massive and open clusters has seen great improvement with the advent of \textit{Gaia} \citep{GaiaCollaboration2016}. With the precise \textit{Gaia} astrometry, the member and field stars can now be separated in greater detail than ever before without making strong assumptions on the photometric content of the cluster \citep[see e.g.][]{CantatGaudin2018}. Runaway stars can not only be directly identified based on their relative motion, but can also be traced back to their cluster of origin to accurately constrain their ejection time and kinematic age.

The Eagle Nebula (M16) is a well-known and well-studied \HII region spanning over approximately 30$^\prime$ by 30$^\prime$ on the southern sky. Near the centre of M16 lies the young massive cluster NGC6611, about $\sim$ 5$^{\prime}$ in apparent diameter. Currently, NGC6611 hosts 15 O stars and about 50 early B-type stars. These massive stars have shaped, through their ionising power, the natal molecular cloud, carving out so-called `elephant trunks' \citep[see e.g.][]{Hester1996}. A spectroscopic study of the O star population has revealed a minimal binary fraction of 0.44 \citep{Sana2009,Sana2012}. O stars are not only found inside the \HII region, \citet{Gvaramadze2008} find three nearby O stars which could have been ejected from NGC6611 in the past.

Various photometric and/or spectroscopic studies have been performed throughout the years \citep{Walker1961,Hillenbrand1993,Belikov1999,Evans2005,Guarcello2007b,Martayan2008,Sana2009}. The mass of NGC6611 is estimated to be about $\sim$ 2 $\times$ 10$^{4}$ M$_{\odot}$ \citep{Pfalzner2009}. The age of the cluster varies throughout the literature depending on the method used and ranges from 0.5 to 5 Myr, with most estimates between 1 and 3 Myr \citep{Walker1961,Hillenbrand1993,Belikov2000,Oliveira2008}. Indications for an older population of low-mass pre-main sequence stars has been uncovered, with a rough age estimate somewhere between $\sim$ 8 to 32 Myr \citep{DeMarchi2013}. This older generation of stars is spatially more extended, has a lower X-ray luminosity, and occupies a  different regime in the colour-magnitude diagram compared to the `younger' population \citep{Guarcello2012,Bonito2013}.

With \textit{Gaia} Early Data Release 3 \citep[EDR3;][]{GaiaCollaboration2021}, we have re-analysed the stellar content of NGC6611 and M16. We present a kinematic history of the O star content of the open cluster. In Section~\ref{sec:gaiadata}, we describe the corrections and filters applied to the \textit{Gaia} data and the clustering algorithm to obtain the members of NGC6611. Sections~\ref{sec:ngc6611_members} to \ref{sec:res_Ostarkin} present our main results regarding its astrometric, photometric and kinematic properties. We discuss our results in Section~\ref{sec:discussion} and provide a summary in Section~\ref{sec:conclusion}.

\section{\textit{Gaia} EDR3 data selection}
\label{sec:gaiadata}
We have searched for candidate members of NGC6611 in the \textit{Gaia} Early Data Release 3 \citep[EDR3;][]{GaiaCollaboration2016,GaiaCollaboration2021} catalogue by performing a cone-search centred on ($\alpha$, $\delta$) = (274.7 deg, --13.8 deg) or ($l$, $b$) $\simeq$ (16.96 deg, +0.80 deg) with a radius of 0.25 deg, resulting in 5583 sources. We note that the astrometry and photometry for these sources in the full DR3 is exactly the same, except for the additional radial velocities. To ensure the quality of the \textit{Gaia} data, we have excluded sources with two-parameter astrometric solutions (\texttt{astrometric\_params\_solved} = 3) and sources with missing photometry entries in either the G, Bp, or Rp-band, leaving 4886 sources. These filters are necessary as the proper motion, parallax and photometry play a key role in the membership selection and subsequent analysis. This dataset was next subjected to a set of corrections before applying any filters. First, we have corrected the G-band flux and magnitude for sources with 2 and 6-parameter astrometric solutions \citep{GaiaCollaboration2021,Riello2021}. Second, we have applied the \citet{Lindegren2021b} correction to the parallax to account for the zero-point bias. Third, the \citet{CantatGaudin2021} correction is applied, adjusting the proper motion of stars with G < 13 mag. Last, we have applied the \citet{ElBadry2021} correction to the parallax error for stars with G-magnitudes in the range of 7 $<$ G $<$ 21 mag.

Next, a set of filters was applied to the data. We briefly describe these and list the remaining number of sources after each step:
\begin{itemize}
    \item We excluded sources with a renormalised unit weight error (\texttt{ruwe}) larger than 1.4, likely indicating poor quality astrometric solutions for these sources, possibly as a result of binarity. This leaves 4626 sources.
    \item We removed sources for which, in 10\% of the cases, more than one peak was identified in the windows used by \textit{Gaia} (\texttt{ipd\_frac\_multi\_peak}). This could imply that the source may be a visually resolved double star. This leaves 4548 sources.
    \item We excluded sources for which the visibility periods (\texttt{visibility\_periods\_used}) used were 9 or less, possibly indicating that the astrometric parameters (such as the parallax) could be subject to larger systematic errors \citep[see e.g.][]{Lindegren2018}. This leaves 4522 sources.
    \item We removed sources which had more than one source identifier during the data processing (\texttt{duplicated\_source}) potentially indicating astrometric or photometric issues. This leaves 4520 sources.
    \item We excluded sources for which the statistic measuring the amplitude of the image parameter determination goodness of fit (\texttt{ipd\_gof\_harmonic\_amplitude}) is larger than 0.15, implying that the source could be a double star. This leaves 4310 sources.
    \item We excluded sources for which the fractional parallax uncertainty (\texttt{parallax} / \texttt{parallax\_error}) is smaller than 5, in order to properly distinguish stars belonging to NGC6611 from fore or background stars. This leaves 2059 sources.
\end{itemize}
With these filters, we may create biases by excluding binaries and fainter sources. We have not filtered on the \texttt{astrometric\_excess\_noise} ($\epsilon_{i}$) and \texttt{astrometric\_excess\_noise\_sig} ($D$), which measure the disagreement between the observations and best-fitting standard astrometric model \citep[see e.g.][]{Lindegren2012}. This was not done as this excluded nearly every OB star present in NGC6611. The \textit{Gaia} Documentation\footnote{\url{https://gea.esac.esa.int/archive/documentation/GEDR3/Gaia\_archive/chap\_datamodel/sec\_dm\_main\_tables/ssec_dm_gaia\_source.html}} also notes: "In the early data releases $\epsilon_{i}$ will however include instrument and attitude modelling errors that are statistically significant and could result in large values of 
$\epsilon_{i}$ and $D$". We have also not filtered on the corrected G$_{\rm{Bp}}$ and G$_{\rm{Rp}}$ flux excess factor $C^{*}$ \citep{Riello2021}, which is expected to be close to 0 \citep{Riello2021}, as this would exclude a major fraction of the OB stars present in NGC6611.

\subsection{Separating cluster members from field stars}
\label{sec:upmask}
With the 2059 high quality sources, we can separate the cluster members and field stars. Similar to \citet{CantatGaudin2018}, we have applied the unsupervised membership algorithm UPMASK to the candidate sources \citep{KroneMartins2014}. UPMASK was applied to the 5-dimensional space ($\alpha$, $\delta$, $\mu_{\alpha^{*}}$, $\mu_{\delta}$, $\varpi$), the observables, in the following manner:
\begin{enumerate}
    \item For each star, the observables are randomly drawn from a multivariate normal distribution with means equal to the values of the observables and covariance matrix based on the uncertainties and correlations between the observables.
    \item A grouping algorithm, $k$-means clustering, is applied to the randomly drawn observables in the 3D astrometric space ($\mu_{\alpha^{*}}$, $\mu_{\delta}$, $\varpi$), with $k$ equal to the number of groups.
    \item The groups in (2.) are tested whether they are closely distributed in sky-coordinates ($\alpha$, $\delta$). This is done by evaluating the total branch length $l_{\rm{obs}}$ of the minimum spanning tree (MST) of each group and comparing it to the total branch length $l$ of the MST of randomly drawn uniform distributions. This gives for each group and thus each star a binary membership assignment (true or false).
    \item Step (1.) to (3.) are repeated $n$ times, resulting in each star having a membership `probability' $p$; how many times a star was assigned member relative to the total number of iterations $n$.
\end{enumerate}
In practice, instead of self-assigning the number of groups $k$ in the $k$-means clustering algorithm, we can set the average number of stars per group since we know the total amount of candidate stars we start out with. Here, we set the average number of stars per group to 15 as recommended by \citet{KroneMartins2014}. The total branch length $l$ and standard deviation $\sigma_{l}$ of randomly uniformly drawn distributions are pre-computed containing 3 to 100 stars in circular distributions with radii equal to the previously specified cone-search radius. The stars in each group with $l_{\rm{obs}}$ are assigned member for that iteration if $l_{\rm{obs}}$ deviates more than $\sigma_{l}$ from the expected $l$.

We adopted a total of 10,000 iterations for $n$ and a cut-off membership probability $p$ of 0.9, balancing the sensitivity to the number of member stars found and the field stars excluded. After applying the UPMASK algorithm, we obtain the probable cluster members. A last step is to exclude outliers by removing stars that deviate more than three median absolute deviations from the median value for $\mu_{\alpha^{*}}$, $\mu_{\delta}$ and $\varpi$, leaving 137 stars brighter than G $\simeq$ 17.5 mag that we identify as members of NGC6611.

\subsection{O stars in NGC6611}
We have gathered all known O stars in and around NGC6611 in Table~\ref{tab:Ostars} and list their identifiers and spectral type from the literature. This gives 19 O stars in a total of 16 systems, where we have included the three distant runaway O stars found in Section~\ref{sec:res_runaway} outside of our cone-search region. Only 7 of the 16 O star systems were found to be members of NGC6611. This is a combination of the position of O stars being outside the cone-search region (due to their runaway nature as we subsequently show), \texttt{ruwe} $>$ 1.4 due to their binary nature, or proper motion directions not intersecting with the cluster proper motion. We have added the missing 9 apparent non-cluster O star systems to the sample of 137 members. This gives 130 B or later-type and 16 O star systems for a total of 146 members (Table~\ref{tab:members_all}). Since several of these O stars are significantly separated from NGC6611, and/or have different proper motions, only the O stars which were initially identified as a member (`T' in Table~\ref{tab:Ostars}) are used to calculate the cluster centre, radius, motion and distance in the following section.

\begin{table*}[ht]
\centering
\caption{O stars originating from NGC6611.}
\label{tab:Ostars}
\begin{tabular}{l l l l l l l}
\hline
\hline
\multirow{2}{*}{Identifier} & \multirow{2}{*}{Spectral type} & Projected distance & \multirow{2}{*}{Distance} & \multirow{2}{*}{ruwe} & \multirow{2}{*}{Member$^{b}$} & \multirow{2}{*}{Reference}\\
& & from centre & & & & \\
- & - & arcminute (pc) & pc & - & - & - \\
\hline
\multicolumn{7}{c}{O stars}\\
\hline
HD 168076 & O3.5 V((f+)) + O7.5 V & 1.7 (0.8) & - & 8.382 & F & 1\\ \vspace{0.5mm}
BD--13$^{\circ}$ 4923 & O4 V((f+)) + O7.5 V & 2.6 (1.3) & - & 2.544 & F & 1\\ \vspace{0.5mm}
HD 168075 & O6.5 V((f)) + B0-1 V & 1.4 (0.7) & 1589$^{+86}_{-47}$ & 1.110 & T & 1\\ \vspace{0.5mm}
BD--13$^{\circ}$ 4927 & O7 II(f) & 1.5 (0.7) & 1654$^{+77}_{-60}$ & 1.056 & F & 1\\ \vspace{0.5mm}
BD--13$^{\circ}$ 4929 & O7 V + (B0.5 V + B0.5 V) & 1.3 (0.6) & 1609$^{+133}_{-89}$ & 1.746 & F & 1\\ \vspace{0.5mm}
HD 168137 & O7 V + O8 V & 4.2 (2.1) & 1767$^{+100}_{-78}$ & 0.933 & F & 1\\ \vspace{0.5mm}
W222 & O7 V((f)) & 3.3 (1.6) & 1653$^{+81}_{-62}$ & 1.203 & T & 2\\ \vspace{0.5mm}
W161 & O8.5 V & 4.4 (2.2) & 1708$^{+82}_{-65}$ & 1.265 & T & 2\\ \vspace{0.5mm}
LS IV --13 14 & O9 V & 2.9 (1.5) & 1716$^{+60}_{-51}$ & 0.913 & T & 2\\ \vspace{0.5mm}
W584 & O9 V & 11.2 (5.5) & 1768$^{+76}_{-61}$ & 1.002 & T & 2\\ \vspace{0.5mm}
HD 168183 & O9.5 III + B3-5 V/III & 13.4 (6.6) & 1935$^{+198}_{-122}$ & 1.415 & F & 1\\ \vspace{0.5mm}
BD--13$^{\circ}$ 4928 & O9.5 V & 0.5 (0.2) & 1604$^{+51}_{-42}$ & 0.903 & T & 1\\ \vspace{0.5mm}
BD--13$^{\circ}$ 4930 & O9.5 Vp & 4.2 (2.1) & 1580$^{+57}_{-40}$ & 0.843 & T & 1\\
\hline
\multicolumn{7}{c}{Distant runaway O stars}\\
\hline
UCAC2 27149134$^{a}$ & O3-4 V & 56.2 (27.8) & 1750$^{+74}_{-60}$ & 1.023 & F & 3\\ \vspace{0.5mm}
BD--14 5040 & O5.5 V(n)((f)) & 119.7 (59.2) & 1615$^{+45}_{-40}$ & 0.855 & F & 4\\ \vspace{0.5mm}
HD 168504 & O7.5 V(n)z & 30.2 (14.9) & 1623$^{+134}_{-90}$ & 1.670 & F & 4\\
\hline
\end{tabular}\\
\footnotesize{$^{(a)}$}{Spectral type estimated from photometry;}
\footnotesize{$^{(b)}$}{Membership based on \textit{Gaia} analysis}
\tablebib{(1) \citet{Sana2009,Sana2012}; (2) \citet{Evans2005}; (3) \citet{Gvaramadze2008}; (4) \citet{MaizApellaniz2016};}
\end{table*}

\begin{figure*}[ht]
\centering
\includegraphics[width=0.95\linewidth]{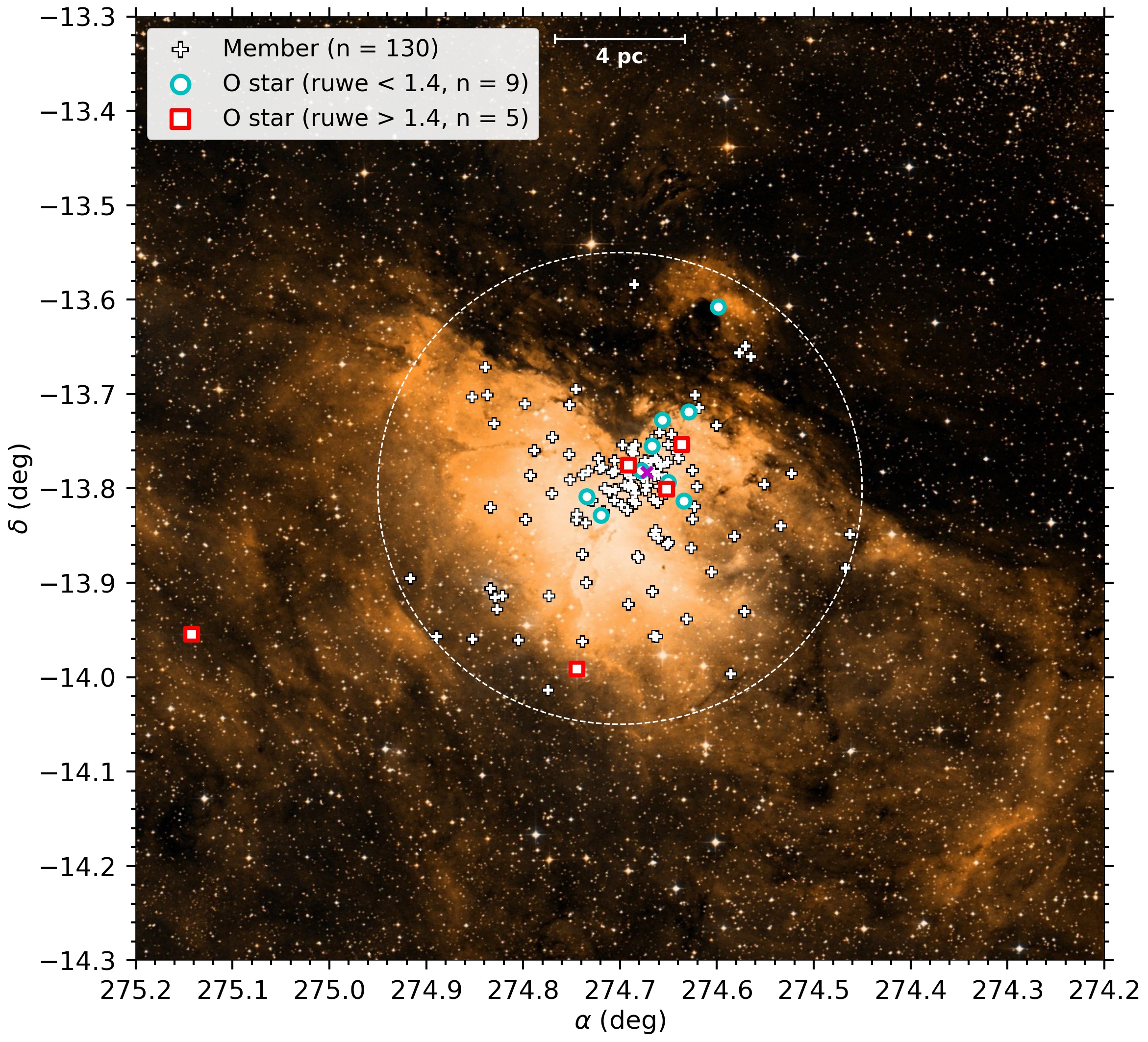}
\caption{DSS2 B, R and I colour image of NGC6611 and the surrounding nebula (M16). The member stars are shown as white plus symbols. North is up and east is to the left. We highlight the O stars in this image and show these with poor quality astrometry with the red squares, while the ones with good quality astrometry are shown with the cyan circles. Two runaway O stars are not visible in this image. The centre of NGC6611, the position with the highest source density, is indicated with the magenta cross. At the top of the figure the angular distance equivalent to 4 pc is given for a distance of 1.7 kpc. We show the cone-search region with the dashed circle, with a radius equivalent to 0.25 deg or 7.4 pc. The open cluster Trumpler 32 \citep[age $\sim$ 0.3 Gyr;][]{Kharchenko1995} can be seen in the top-right of the image.}
\label{fig:RaDecImage}
\end{figure*}

\begin{figure*}
\centering
\begin{minipage}{.49\textwidth}
\centering
\includegraphics[width=1.0\columnwidth]{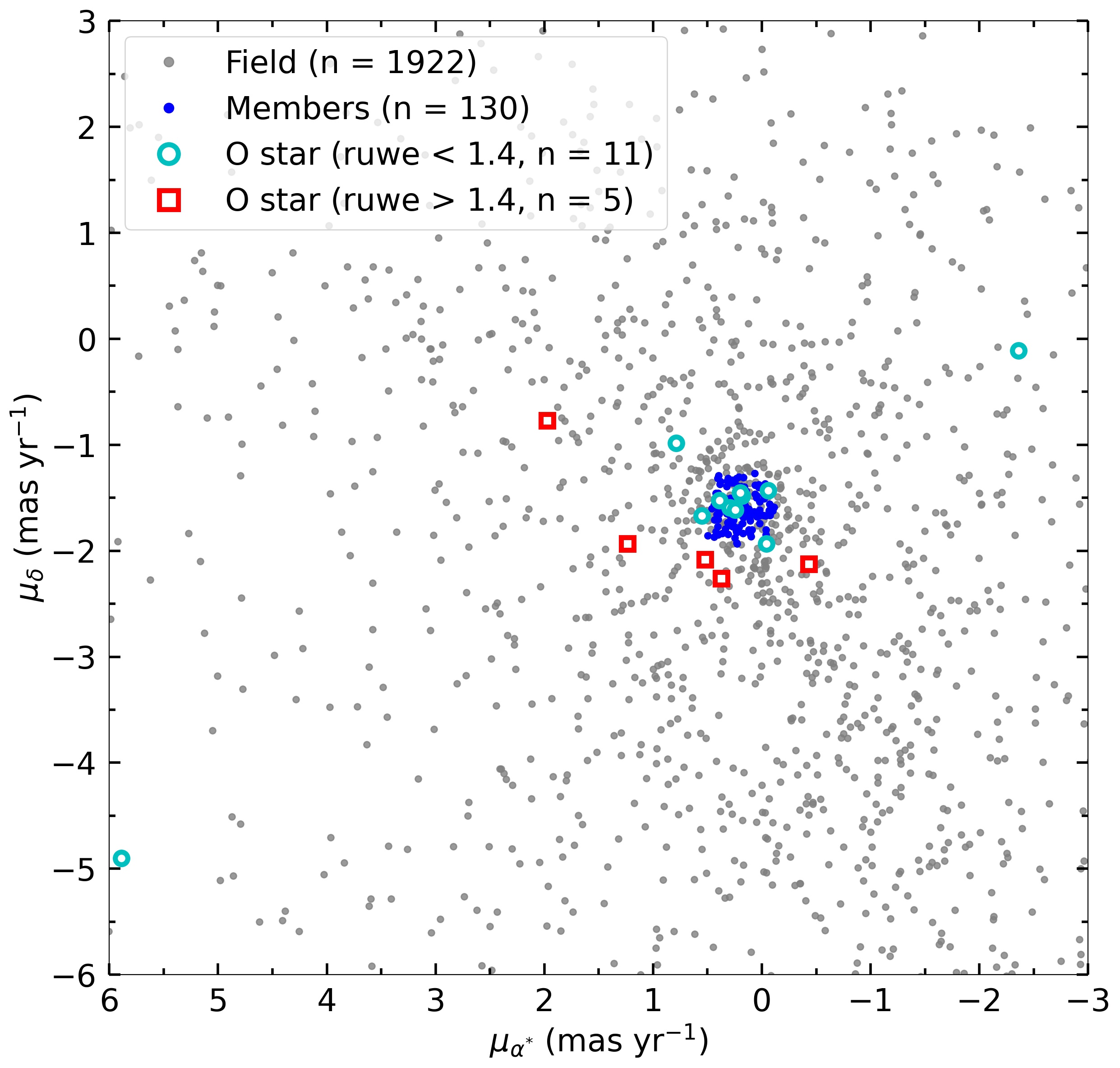}
\end{minipage}
\begin{minipage}{.49\textwidth}
\centering
\includegraphics[width=1.0\columnwidth]{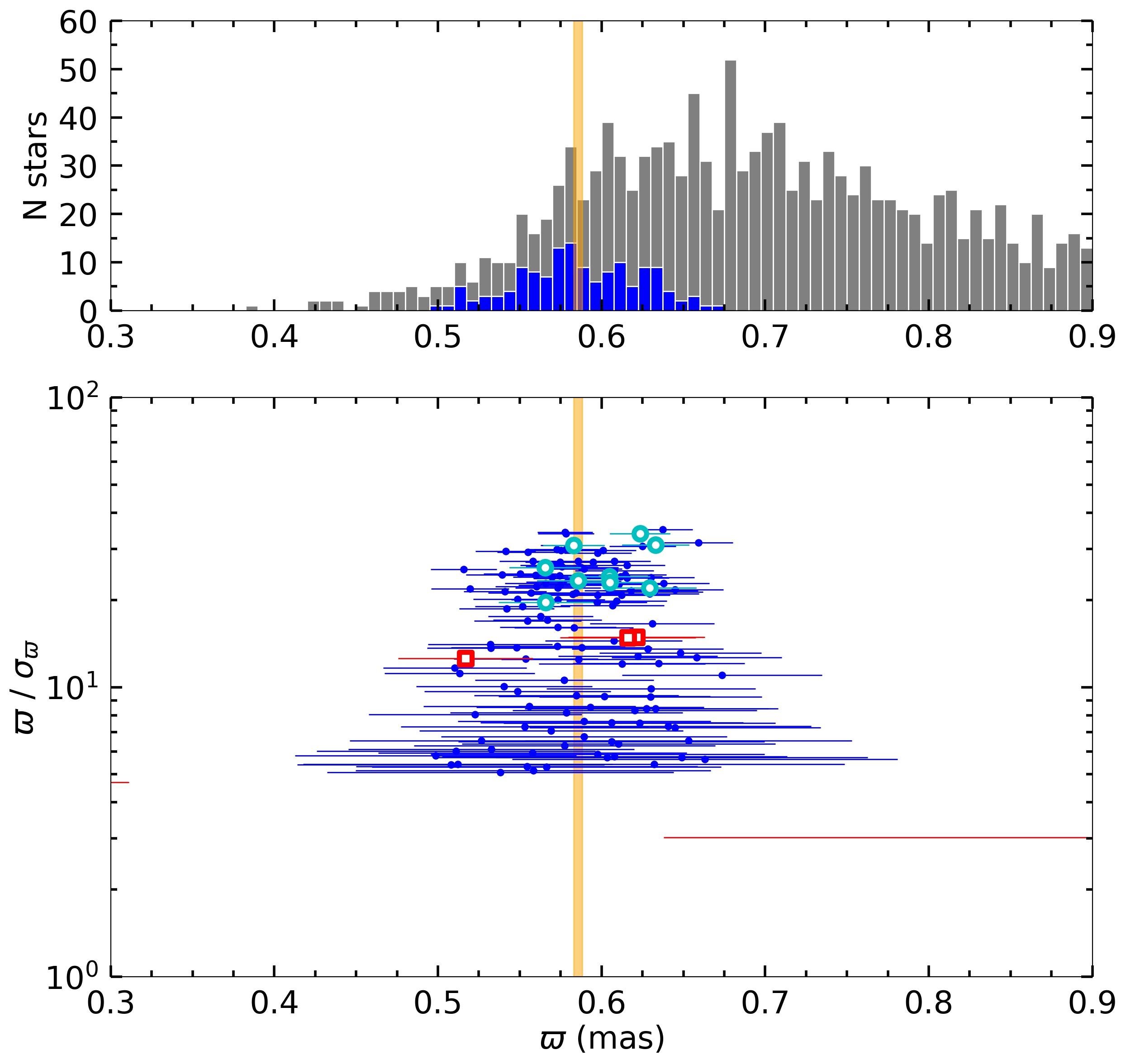}
\end{minipage}
\begin{minipage}{.49\textwidth}
\centering
\includegraphics[width=1.0\columnwidth]{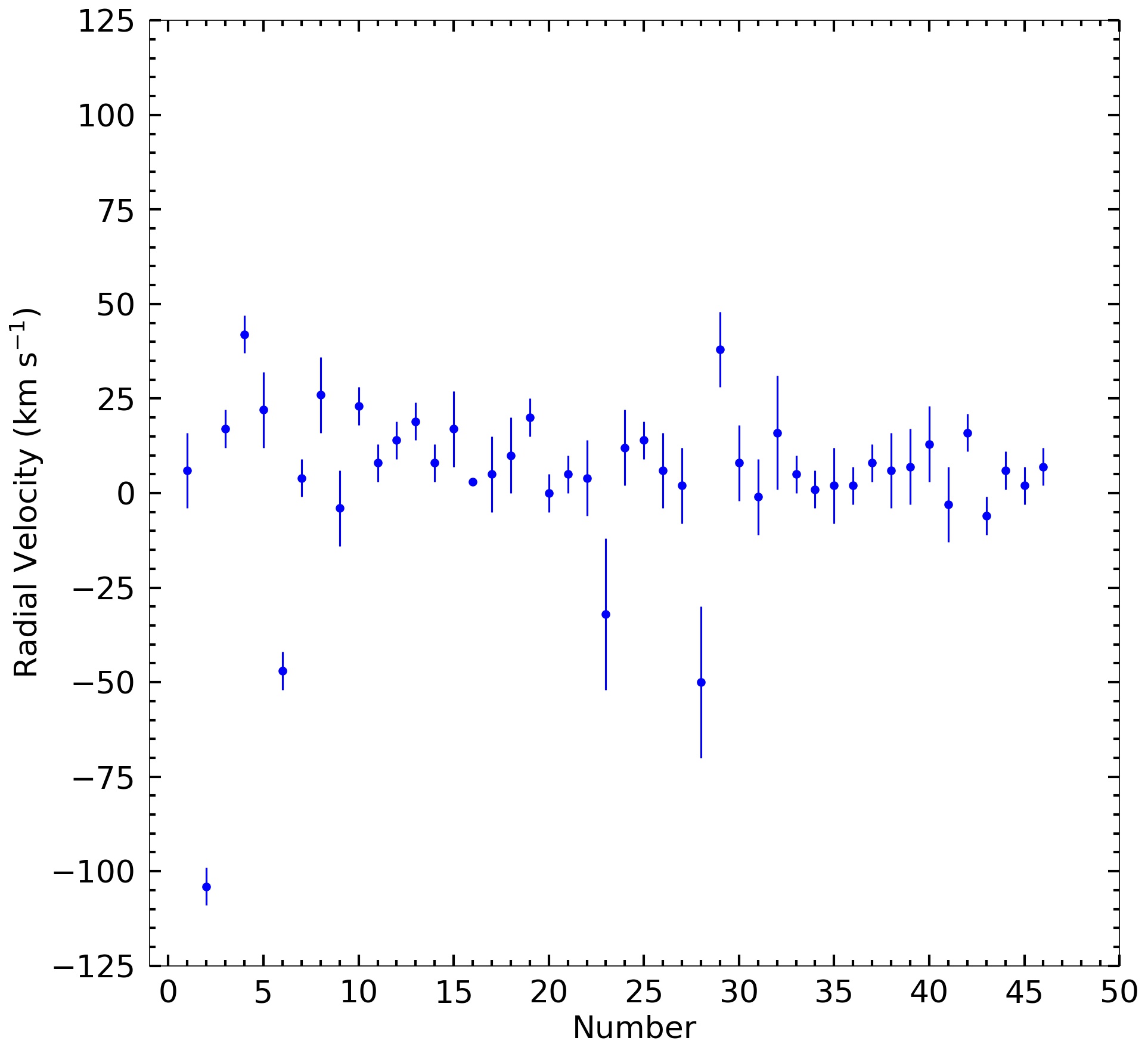}
\end{minipage}
\begin{minipage}{.49\textwidth}
\centering
\includegraphics[width=1.0\columnwidth]{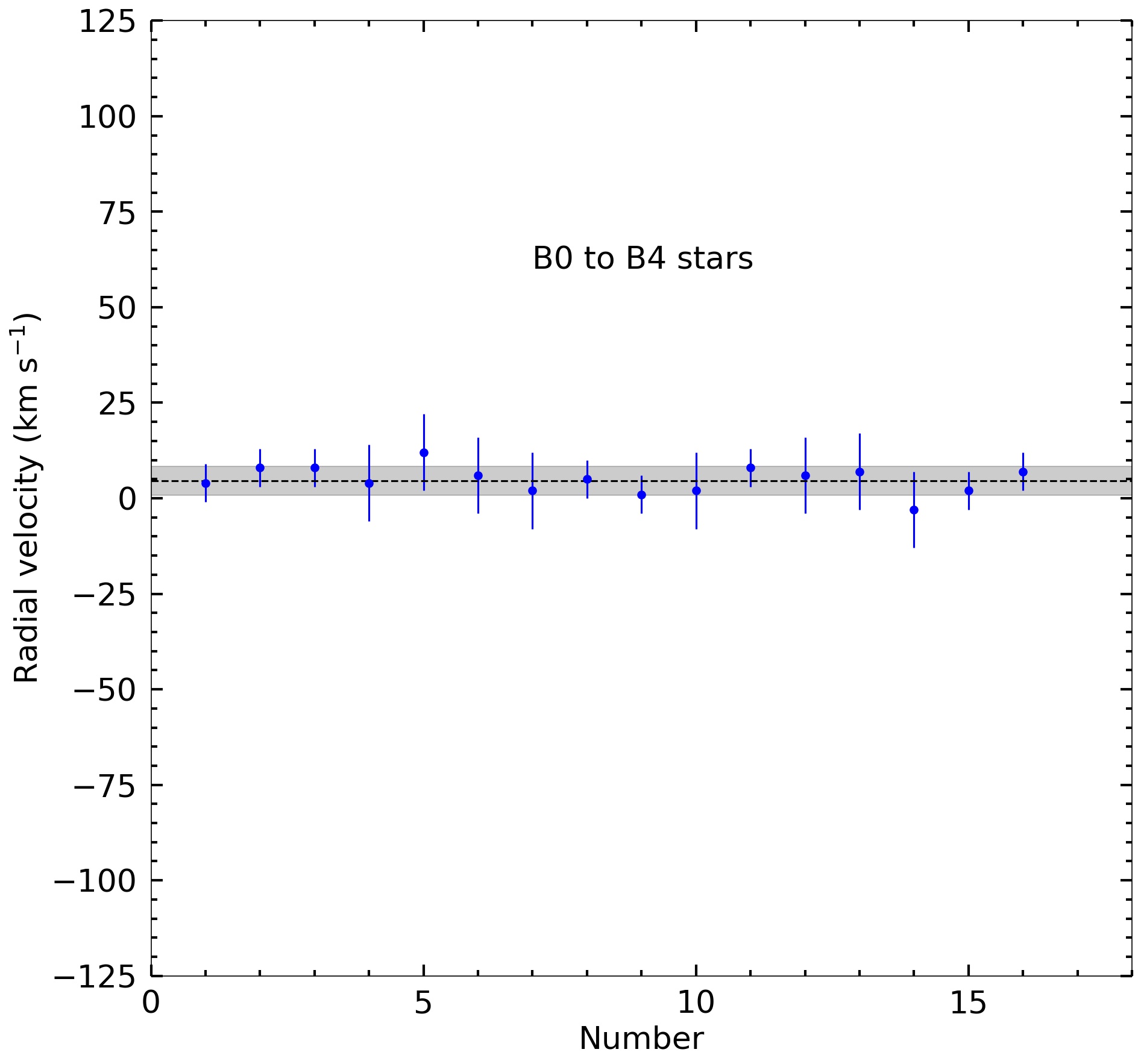}
\end{minipage}
\caption{Astrometric parameters of NGC6611. \textit{Top-left}: The proper motion distribution of the field stars (grey) and members of NGC6611 (blue). We show the O stars present in NGC6611 with poor and good quality astrometry with the red squares and cyan circles, respectively. \textit{Top-right}: The parallax distribution of the field stars and members, coloured similarly as in the top-left panel. We also show the fractional parallax uncertainty ($\varpi$/$\sigma_{\varpi}$) of the members as a function of their parallax. Two O star systems have $\textsc{ruwe}$ = 2.544 and 8.382, and have parallaxes outside this panel, with only part of the error-bar visible. The 1$\sigma$ error on the parallax of NGC6611 is shown with the orange bar. \textit{Bottom-left}: Radial velocity of the members in the heliocentric frame collected from the literature. \textit{Bottom-right}: Radial velocity in the heliocentric frame of the B0 to B4 stars after removal of the outliers in the bottom-left panel. Outliers have been excluded as described in Section~\ref{sec:ngc6611_members}. The mean and standard deviation of the radial velocity of these stars are shown with the dashed black line and shaded region, respectively.}
\label{fig:members_astrometry}
\end{figure*}

\section{Members of NGC6611}
\label{sec:ngc6611_members}
In order to validate the results of the UPMASK clustering algorithm, we will first evaluate the 5 astrometry parameters used ($\alpha$, $\delta$, $\mu_{\alpha^{*}}$, $\mu_{\delta}$, $\varpi$).
We show a 0.8 by 0.8 deg Digital Sky Survey 2 B, R and I colour image of NGC6611 and the surrounding nebula in Figure~\ref{fig:RaDecImage}. The dashed circle indicates the cone-search region, in which the members and O stars are shown. Most of the members are not concentrated towards the centre of the H II region, but several arcminutes north-west of this position. We define the centre of NGC6611 as the position with the highest source density. The centre is determined to be ($\alpha_{\rm{NGC6611}}$, $\delta_{\rm{NGC6611}}$) = (274.67 $\pm$ 0.01$^{\circ}$, --13.78 $\pm$ 0.01$^{\circ}$), indicated with the purple cross. We highlight the spectroscopically confirmed O stars visible in this image and make a distinction between O stars with poor and good quality astrometry (in this case \textsc{ruwe} $>$ 1.4 and $<$ 1.4 respectively). We will motivate that these originate from NGC6611 and we include them for visualisation.

We show the proper motion distribution in the top-left panel of Figure~\ref{fig:members_astrometry}. The members of NGC6611 (blue) are clearly concentrated in proper motion space, relative to the field stars (grey). As in Figure~\ref{fig:RaDecImage}, we highlight the proper motion of all O stars. To determine the average cluster proper motion, we follow step 1 in Section~\ref{sec:upmask} by drawing the proper motion randomly from a multivariate normal distribution with means equal to the observed values and a covariance matrix based on the uncertainties and correlations. In this step we ignore correlations between the stars and add in quadrature the systemic uncertainties of 0.0112 and 0.0107 mas yr$^{-1}$ from \citet{Lindegren2021a} to \texttt{pmra\_error} and \texttt{pmdec\_error}, respectively. Each iteration gives a mean proper motion, with a negligible difference between the mean and median proper motion. The average cluster proper motion is determined with the 50$^{\rm{th}}$ percentile over the distribution of all mean proper motions (1000 iterations), while the 1$\sigma$ uncertainty is determined with the 16$^{\rm{th}}$ and 84$^{\rm{th}}$ percentiles. This results in ($\mu_{\alpha^{*},\rm{NGC6611}}$, $\mu_{\delta,\rm{NGC6611}}$) = (0.21 $\pm$ 0.01 mas yr$^{-1}$, $-$1.59 $\pm$ 0.01 mas yr$^{-1}$). This derived cluster proper motion is consistent with other estimates based on \textit{Gaia} \citep[see e.g.][]{Flynn2022,MaizApellaniz2022}.

The next astrometry parameter is the parallax. We show the parallax distribution of the field stars and members (including the O stars) in the top-right panel of Figure~\ref{fig:members_astrometry}, adopting equal colouring and marking as before. Of the 146 members, 72 sources have $\varpi$/$\sigma_{\varpi}$ $\leq$ 20 and 74 sources have $\varpi$/$\sigma_{\varpi}$ $>$ 20, providing (extremely) accurate parallaxes for nearly half of the members. Assuming that the members are at the same distance, we expect that the parallaxes of these stars are normally distributed around the true parallax of NGC6611. Note that the parallaxes of several of the O stars should not be trusted due to \texttt{ruwe} $>$ 1.4. We find no evidence for a skewed parallax distribution in either the full sample ($p$ = 0.45), the $\varpi$/$\sigma_{\varpi}$ $\leq$ 20 sub-sample ($p$ = 0.16), or the $\varpi$/$\sigma_{\varpi}$ $>$ 20 sub-sample ($p$ = 0.09). The parallax distribution also shows that the majority of the field stars are located closer than $\varpi$ $>$ 0.65 mas and steeply drops off for stars beyond $\varpi$ $<$ 0.6 mas. Background stars most likely go undetected because of the severe extinction by the natal molecular cloud. We find no indication of contamination of fore and background stars.

Following the advice of \citet{BailerJones2020}, we have set up a joint likelihood to estimate the distance directly from the parallax. The joint likelihood is similar to that used in \citet{CantatGaudin2018}, with the only difference being the adopted parallax zero-point offset. We have already applied the \citet{Lindegren2021b} parallax zero-point offset to the parallax in Section~\ref{sec:gaiadata}; it amounts to a mean offset of $-$0.039 mas (standard deviation of 0.008 mas) for members with 5-parameter, and $-$0.051 mas (standard deviation of 0.009 mas) for members with 6-parameter astrometric solutions. The likelihood is given by

\begin{align}
    P(d\ |\ \varpi, \sigma_{\varpi}) &= \prod_{i=1} P(\varpi_{i}\ |\ d, \sigma_{\varpi_{i}}) \nonumber\\&= \prod_{i=1}\frac{1}{\sqrt{2\pi\sigma_{\varpi_{i}}^{2}}}\mathrm{exp}\left(-\frac{(\varpi_{i} - \frac{1}{d})^{2}}{2\sigma_{\varpi_{i}}^{2}}\right).
\end{align}
Here, $P(d\ |\ \varpi, \sigma_{\varpi})$ is the unnormalised probability distribution for the distance $d$ in kpc, given the individual parallaxes $\varpi_{i}$ in mas, and adopted Gaussian parallax uncertainties $\sigma_{\varpi_{i}}$ in mas. Also noted in \citet{CantatGaudin2018}, we neglect correlations between measurements of all stars, and assume that all stars are at the same distance. We determine the distance with the mode of the unnormalised probability distribution. The positive and negative 1$\sigma$ error are determined from the 16$^{\rm{th}}$ and 84$^{\rm{th}}$ percentile, respectively. We determine the distance and parallax to be 1706 $\pm$ 7 pc and 0.587 $\pm$ 0.003 mas (symmetric distributions), respectively. The determined values for the centre, proper motion, parallax and distance of NGC6611 agree with other determinations in the literature \citep[see e.g.][]{CantatGaudin2018,Kuhn2019,MaizApellaniz2022}. Our distance is consistent with the distance of $1681^{+9}_{-10}$ pc determined in \citet[]{CantatGaudin2018} considering the uncertainties in distance and parallax zero-point offset in \textit{Gaia} DR2 and EDR3, and the distance determination of 1697$^{+31}_{-30}$ pc in \citet{MaizApellaniz2022} who include estimates for external uncertainties.

We have collected all known spectral types of stars in the field of NGC6611 determined from optical spectra in the literature \citep{Hillenbrand1993,Evans2005,Wolff2007,Martayan2008}. In most cases, this also allowed the (systemic) radial velocities to be collected. This gives a total of 166 spectral types and 146 radial velocities. Out of the 146 stars with known radial velocities, 34 were excluded based on their poor astrometry. The remaining 66 stars were removed because of their deviating parallax and/or proper motion. We have cross-matched the members of NGC6611 with these spectral types and radial velocities. Focusing on the radial velocities of the members, we show them in the bottom-left panel of Figure~\ref{fig:members_astrometry}. This results in 46 members with a known radial velocity.

In order to estimate the radial velocity of NGC6611, we do not need to analyse the full sample. The radial velocities in the bottom-left panel of Figure~\ref{fig:members_astrometry} display significant spread even accounting for uncertainties (partially because of binarity). To determine the radial velocity of NGC6611 we turn to the early type B stars, the B0 to B4 stars. We do not use the O stars, which we treat in Section~\ref{sec:res_Ostarkin}. We exclude known binary stars \citep[such as W472;][]{Walker1961,Evans2005}. We also remove stars with radial velocity uncertainties larger than 15 km s$^{-1}$. We use a similar approach as \citet{Evans2005} and determine for the B0-4 stars the mean of the radial velocity and exclude stars which deviate more than 2$\sigma$ from the mean. While this method can be repeated multiple times, only 1 iteration was needed to converge. We show the remaining 17 B0-4 stars in the bottom-right panel of Figure~\ref{fig:members_astrometry}. The mean and standard deviation of the radial velocity is 4.7 $\pm$ 3.5 km s$^{-1}$ shown with the dashed black line and shaded region respectively. This is consistent with the mean radial velocity of NGC6611 in \citet{Evans2005} of 10 $\pm$ 8 km s$^{-1}$ within the uncertainties.

\begin{figure*}
\centering
\begin{minipage}{.49\textwidth}
\centering
\includegraphics[width=1.0\columnwidth]{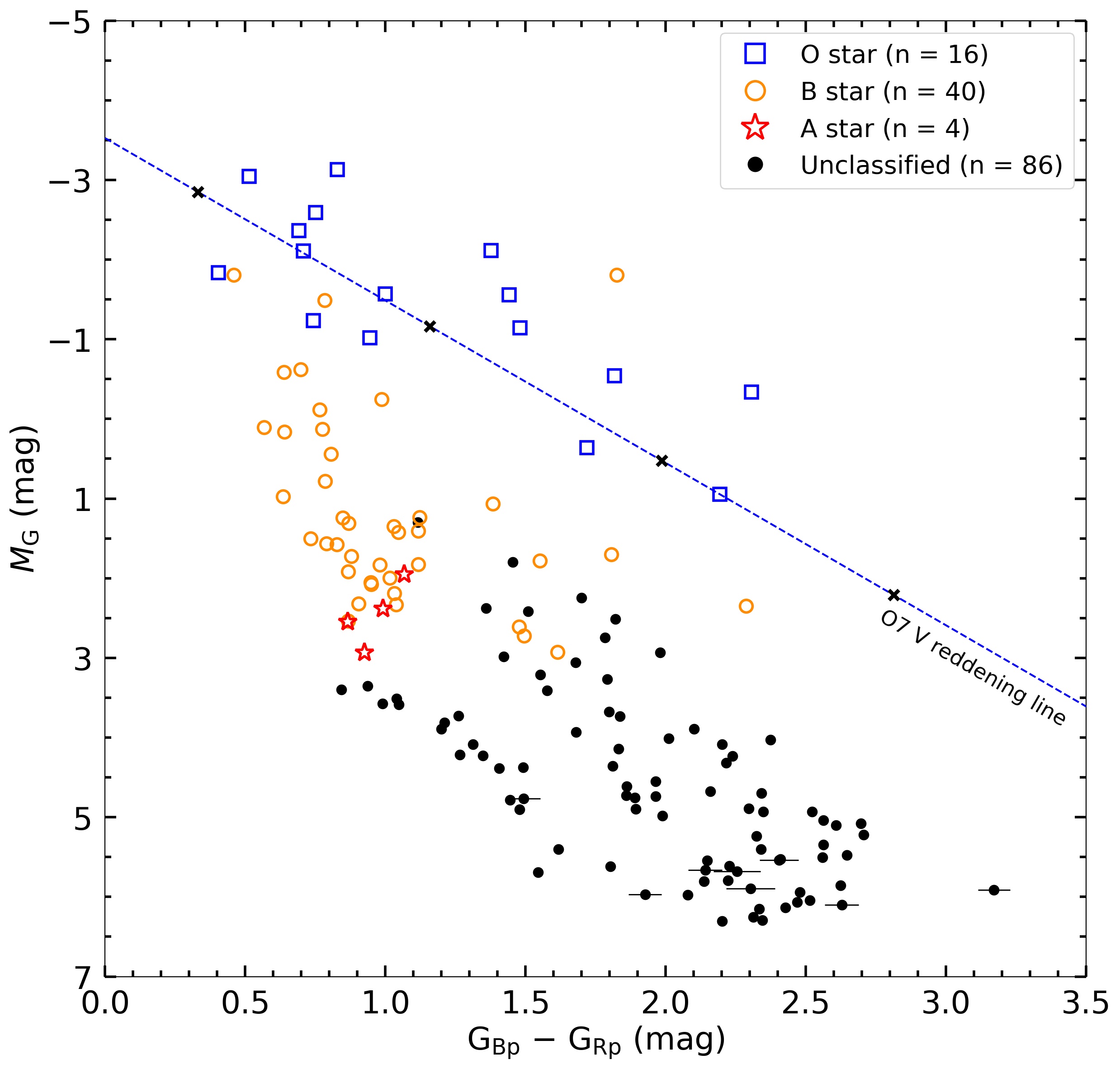}
\end{minipage}
\begin{minipage}{.49\textwidth}
\centering
\includegraphics[width=1.0\columnwidth]{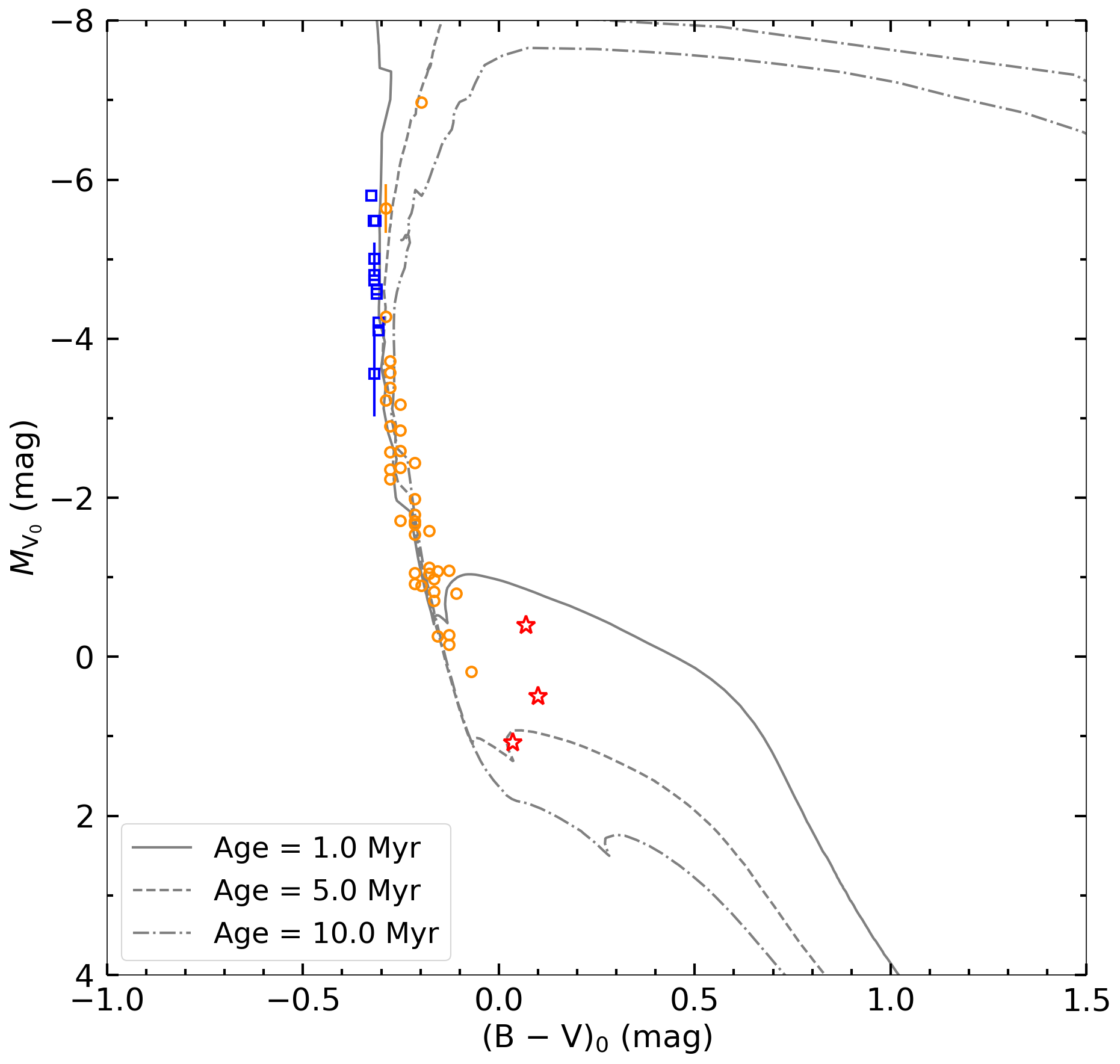}
\end{minipage}
\caption{Colour - absolute magnitude diagram (CAMD) for the average cluster distance for NGC6611. Errors are smaller than the respective symbol if not shown. \textit{Left}: The \textit{Gaia} CAMD of the members with known spectral types. The O stars are shown with the open blue squares, B stars with the open orange circles, A stars with the open red stars and unclassified members in black. The two runaway O stars discussed in Section \ref{sec:res_runaway} are included. We show a reddening line for an O7 V star with $R_{\rm{V}}$ = 3.5, the black crosses indicate $A_{\rm{V}}$ = 2.0, 4.0, 6.0 and 8.0 mag from left to right. \textit{Right}: The extinction corrected M$_{\rm{V_{0}}}$ for the average cluster distance \citep[$R_{\rm{V}}$ = 3.56;][]{Kumar2004} against the intrinsic colour (B $-$ V)$_{0}$ from the \citet{Guarcello2007a,Guarcello2007b} BVI photometry. The stars are marked and coloured similarly as in the left panel. Stars with an uncertain or no spectral type are not shown. We show three \textsc{parsec} + \textsc{colibri} isochrones with ages of 1.0, 5.0 and 10.0 Myr with the solid, dashed and dash-dotted grey lines, respectively.}
\label{fig:HR}
\end{figure*}

\section{Two stellar populations in NGC6611}
\subsection{\textit{Gaia} photometry}
\label{sec:Gaiaphot}
We have constructed the colour - absolute magnitude diagram (CAMD) of the members of NGC6611 with the \textit{Gaia} photometry using the determined cluster distance for all members. We display the \textit{Gaia} CAMD in the left panel of Figure~\ref{fig:HR}, where the different spectral types (O, B or A) are indicated. We show the reddening line for the O7 V((f)) star W222 with $R_{\rm{V}}$ = 3.5 and $A_{\rm{V}}$ = 6.49 mag and for visualisation we show $A_{\rm{V}}$ of 2.0, 4.0, 6.0 and 8.0 mag with the black crosses \citep{MaizApellaniz2018}. This $R_{\rm{V}}$ is similar to the average of $\sim$ 3.56 in NGC6611 \citep{Kumar2004}. The O stars are distributed parallel to this line, likely due to variable extinction. As mentioned, we have included all O stars, including the three distant runaway O stars for completeness.

The B stars are almost all fainter than $M_{\rm{G}}$ $=$ --1 mag. The three B stars that are brighter constitute a B0.5 V + B1: system, a B0.5 V star \citep{Evans2005} and a B2.5 I star \citep{Hillenbrand1993}. The B2.5 I star BD--13$^{\circ}$ 4912, located at (G$_{\rm{Bp}}$ $-$ G$_{\rm{Rp}}$) $\sim$ 2.0 mag and $M_{\rm{G}}$ $\sim$ --2 mag, stands out specifically as it is the only star classified as a supergiant among all the collected spectral types. The \textit{Gaia} astrometry of BD--13$^{\circ}$ 4912 is in excellent agreement with that of NGC6611. The B stars fainter than $M_{\rm{G}}$ $=$ --1 mag show a clear over-density around (G$_{\rm{Bp}}$ $-$ G$_{\rm{Rp}}$) $\sim$ 1.0 mag and $M_{\rm{G}}$ $\sim$ 2 mag. The membership of NGC6611 includes 4 A stars, located around (G$_{\rm{Bp}}$ $-$ G$_{\rm{Rp}}$) $\sim$ 1.0 mag and $M_{\rm{G}}$ $\sim$ 2.5 mag. Three of these are early type A stars, while the fourth has an uncertain spectral type (Ae). 

The spectral types of the members are incomplete and completely missing below $M _{\rm{G}}$ $\sim$ 1 and 3 mag, respectively. Among the members with known spectral type, we identify one outlier: W611. This star is classified as a K0 V star in \citet{Evans2005} and is located around (G$_{\rm{Bp}}$ $-$ G$_{\rm{Rp}}$) $\sim$ 2.0 mag and $M _{\rm{G}}$ $\sim$ 0 mag among the O stars. The parallax of W611 is only marginally consistent at 3$\sigma$ but would still place it too far away to be consistent with a K0 V star. The 3$\sigma$ lower limit on the parallax places it at $\sim$ 1.4 kpc, which would still make it 5 to 6 magnitudes brighter than a typical K0 V star \citep{Peacut2013}. \citet{deWinter1997} classified W611 as a G8 III star, with the luminosity classification based on the apparent brightness. We deem W611 to be a foreground star and have therefore excluded it from the members. Other than this one outlier, we find no evidence for significant contamination with field stars. Another possibility is that W611 is a pre-main-sequence star. This would imply that this star has formed more recently than all other stars in NGC6611.

Towards the fainter end ($M_{\rm{G}} \lesssim 2$ mag) of the CAMD, two groupings are visible. A `blue group' is visible starting at (G$_{\rm{Bp}}$ $-$ G$_{\rm{Rp}}$) $\sim$ 1.0 mag and $M _{\rm{G}}$ $\sim$ 3.5 mag and a `red group' is visible starting at (G$_{\rm{Bp}}$ $-$ G$_{\rm{Rp}}$) $\sim$ 1.5 mag and $M _{\rm{G}}$ $\sim$ 2.5 mag. They are clearly separated by a gap. As we know the distance to these stars, there are two key quantities left in explaining their positions in the CAMD; extinction and age. The clear gap between these two tracks argues against variable extinction, although it can explain the spread of the stars within each of the tracks.

\subsection{BV photometry}
\label{sec:BVphot}
Before drawing further conclusions from the CAMD, we take a look at the (variable) extinction in NGC6611. To do this, we have manually cross-matched our members with the \citet{Guarcello2007a,Guarcello2007b} Johnson BVI photometry obtained with the Wide Field Instrument on the 2.2m telescope of the European Southern Observatory. The cross-matching provided satisfactory counterparts for all but two members: W186 and \textit{Gaia} EDR3 4146405584125976576. These two stars have not been included in further analysis of the BVI photometry. This does not affect our results. With the BV photometry and known spectral types, we can determine the colour excess $E$(B -- V) and extinction in the V-band $A_{\rm{V}}$ with
\begin{align}
    A_{\rm{V}} = R_{\rm{V}}\ E(\rm{B}-\rm{V}) &= R_{\rm{V}}\ [(\rm{B}-\rm{V}) - (\rm{B}-\rm{V})_{0}], 
\end{align}
where $R_{\rm{V}}$ is the slope of the extinction curve and (B -- V)$_{0}$ is the intrinsic colour of the star. We adopt (B -- V)$_{0}$ for luminosity class V from \citet{Peacut2013} and use these intrinsic colours for all luminosity classes. This is an excellent assumption for O stars \citep{Martins2006}, and only results in deviations of $\sim$ $10^{-2}$ mag for the B and A stars. We have not estimated the extinction for the Ae star W112 and the B star W433, as the spectral types are uncertain. We assume the intrinsic colour to be that of the earliest-type star (assumed to be the brightest) in the case of a binary or triple system. These assumptions contribute to the overall uncertainties, however, the adopted spectral type and $R_{\rm{V}}$ are likely to be more significant sources of uncertainty. $R_{\rm{V}}$ is typically assumed to be 3.1 \citep{Cardelli1989}, but can be significantly higher for \HII regions \citep[for NGC6611 see e.g.][]{Kumar2004,MaizApellaniz2018}. We will adopt $R_{\rm{V}}$ = 3.56 from \citet{Kumar2004}, which provides reasonable isochrone fits in Section~\ref{sec:isochrone_fit}, while an $R_{\rm{V}}$ of 3.1 does not.

For the stars with known spectral sub-types, we show the CAMD with the average cluster distance and the extinction corrected $M_{\rm{V_{0}}}$ against the intrinsic colour (B -- V)$_{0}$ in the right panel of Figure~\ref{fig:HR}, using similar colours and markers as in the left panel. The spread in colour which is visible in the left panel has now almost completely disappeared. Almost all of the O and B stars now clearly trace out a main-sequence, which is expected as the (B -- V)$_{0}$ is nearly equal for the O and B stars, with only $M_{\rm{V_{0}}}$ varying. The intrinsically brightest B2.5 I star BD--13$^{\circ}$ 4912 stands out, along with the three A stars which may not have reached the main-sequence. 

For context, we show the stellar isochrones from the Padova and Trieste Stellar Evolution Code (\textsc{parsec}) (v1.2S) + \textsc{colibri} (S\_37 + S\_35 + PR16) models\footnote{\url{http://stev.oapd.inaf.it/cmd}} \citep{Bressan2012,Chen2014,Tang2014,Chen2015,Marigo2017,Pastorelli2019,Pastorelli2020} for ages of 1.0, 5.0 and 10.0 Myr. At the faint-end, the isochrones show the location of stars still on the pre-main-sequence. The transition from the main-sequence to the pre-main-sequence is visible as the isochrone experiences a `knee' and becomes intrinsically redder. For older isochrones, this transition moves towards fainter and redder magnitudes, as the pre-main-sequence lifetime increases for decreasing mass. This gives us a way to determine the age of a young massive cluster \citep{Guo2021}. No isochrone is able to perfectly fit all stars at first glance. Fitting the A stars, the main-sequence and the intrinsically brightest B2.5 I star is already creating problems.

So far, we have only used the members with known spectral types. Including the members without spectral types, which are mostly fainter than $M_{\rm{G}}$ $\sim$ 1 mag could give additional information to better constrain the (variable) extinction and age of NGC6611.

\subsection{Isochrone fitting}
\label{sec:isochrone_fit}
\begin{figure}
\centering
\centering
\includegraphics[width=1.0\columnwidth]{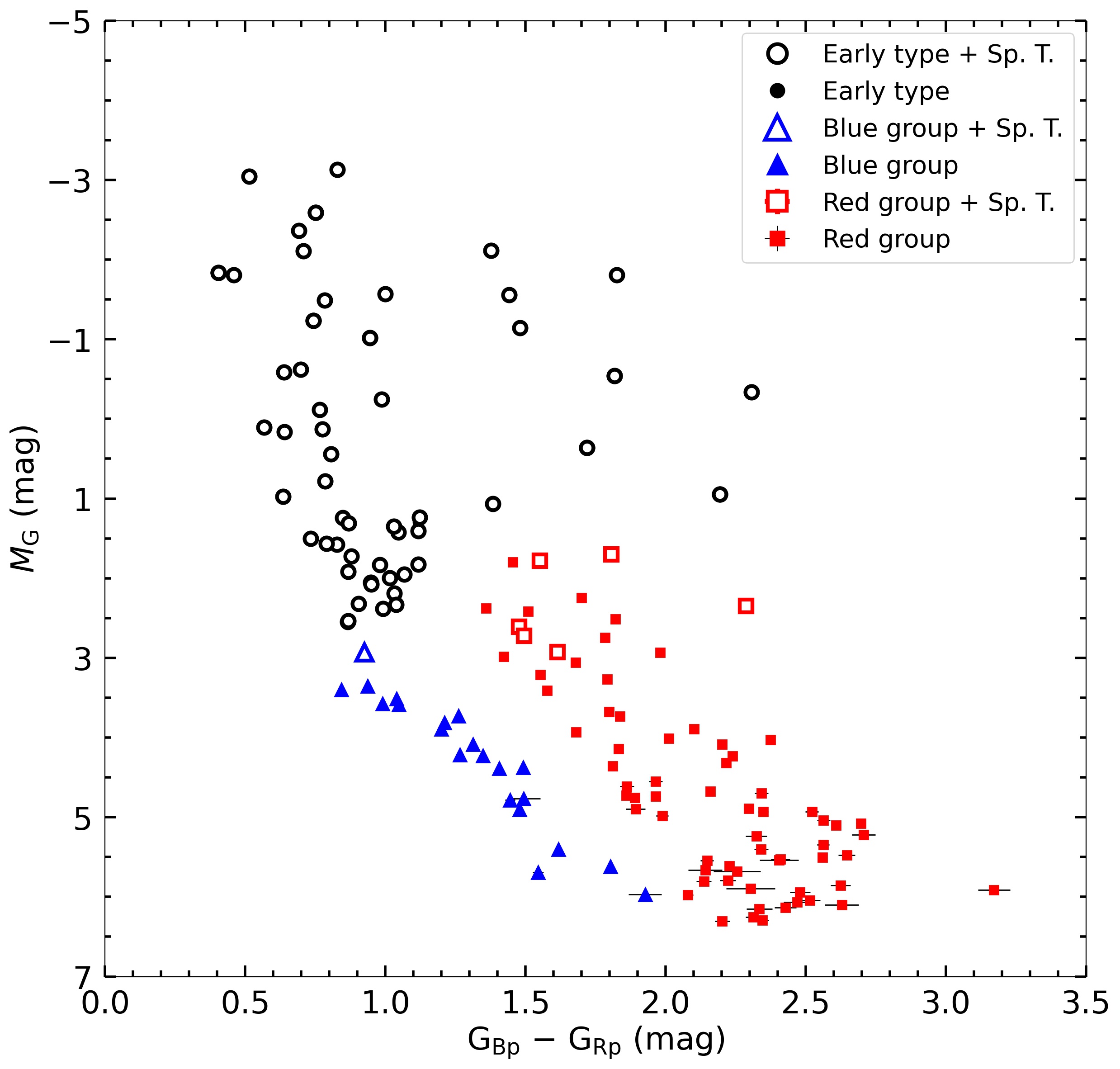}
\caption{\textit{Gaia} colour - absolute magnitude diagram. We have separated the early type stars, the blue and red group shown with the black circles, blue triangles and red squares, respectively. Stars with a known spectral type are shown with an open marker.}
\label{fig:HR_young_old}
\end{figure}

\begin{figure*}
\centering
\begin{minipage}{.49\textwidth}
\centering
\includegraphics[width=1.0\columnwidth]{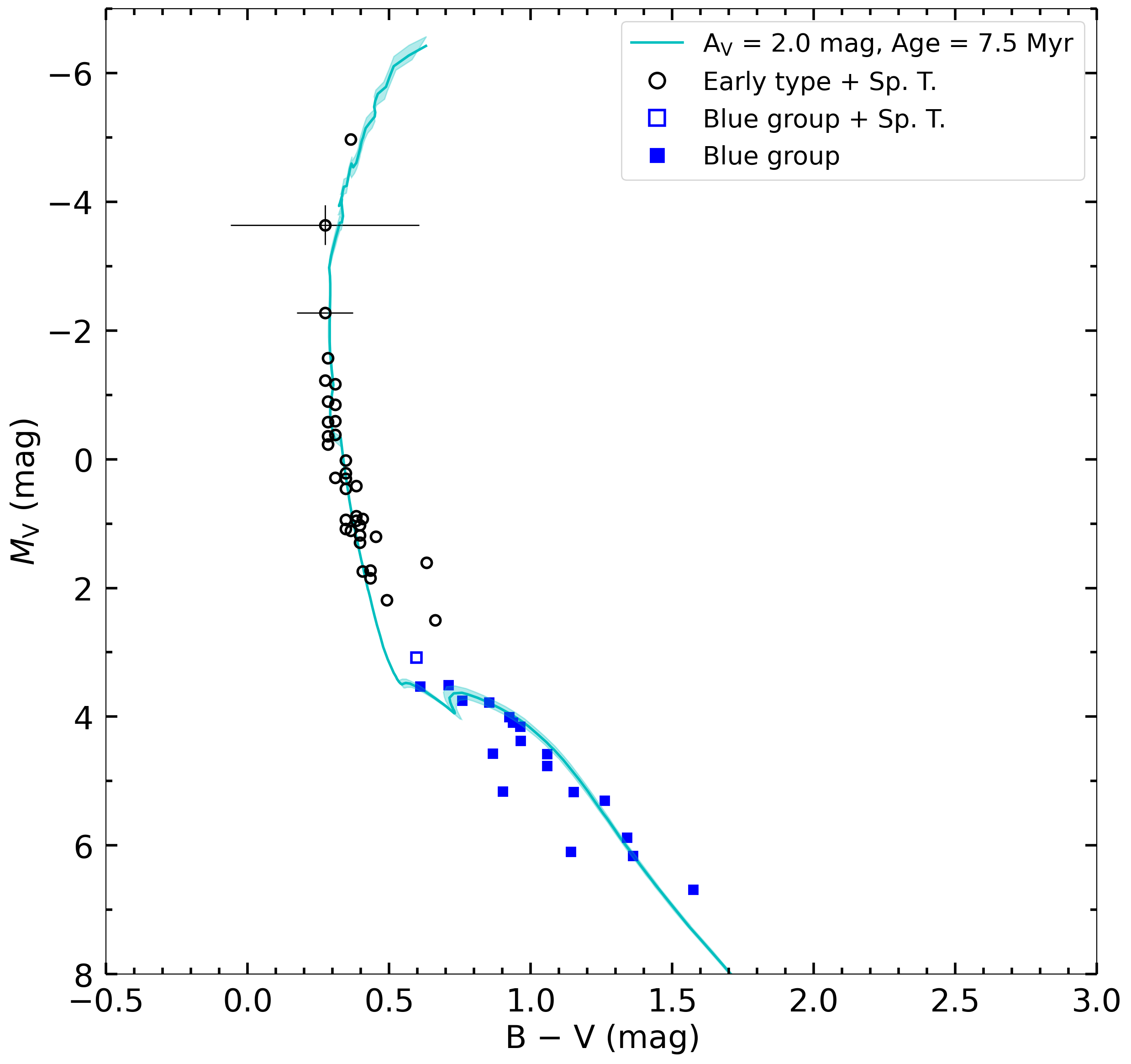}
\end{minipage}
\begin{minipage}{.49\textwidth}
\centering
\includegraphics[width=1.0\columnwidth]{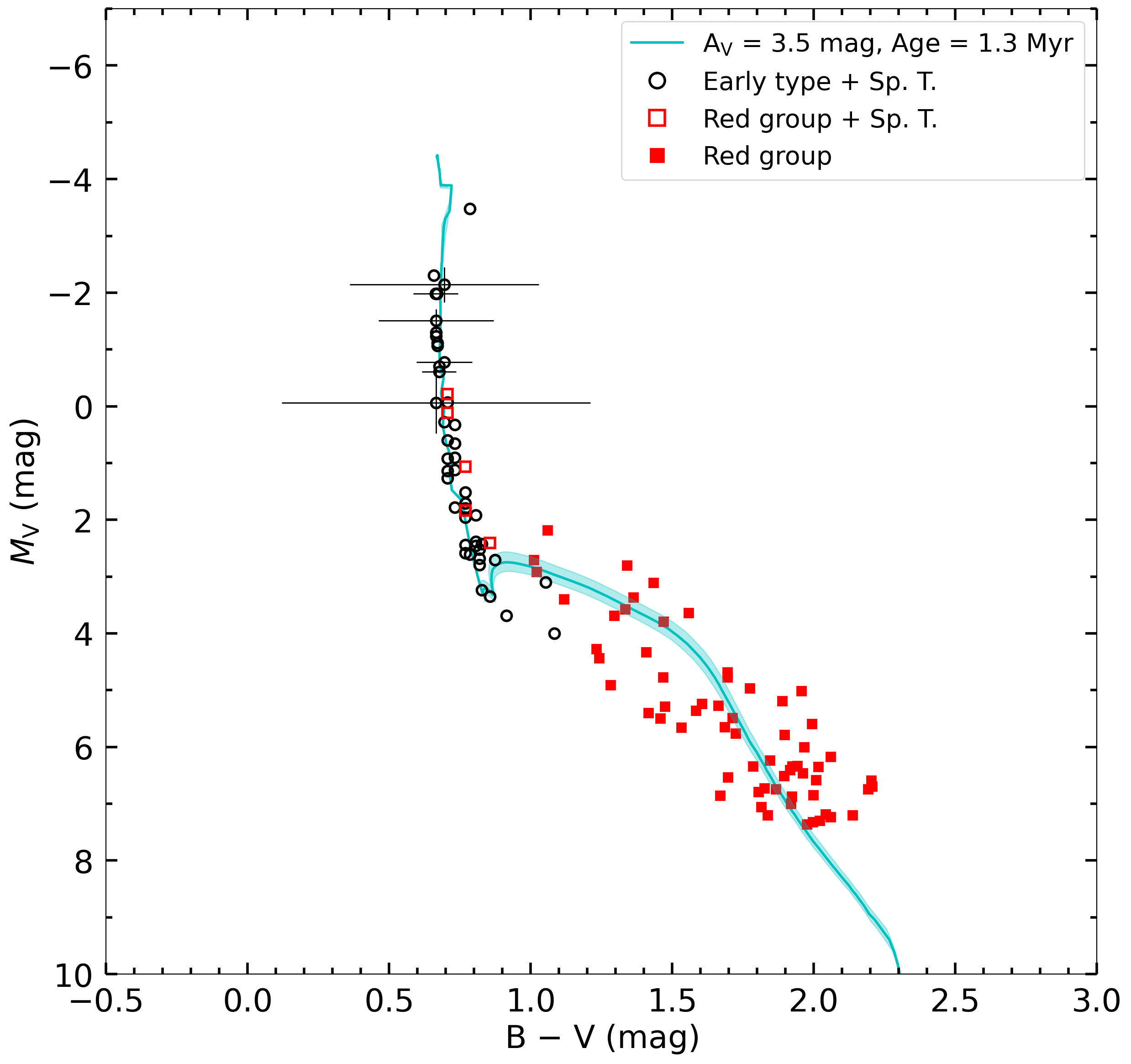}
\end{minipage}
\vskip\baselineskip
\begin{minipage}{.49\textwidth}
\centering
\includegraphics[width=1.0\columnwidth]{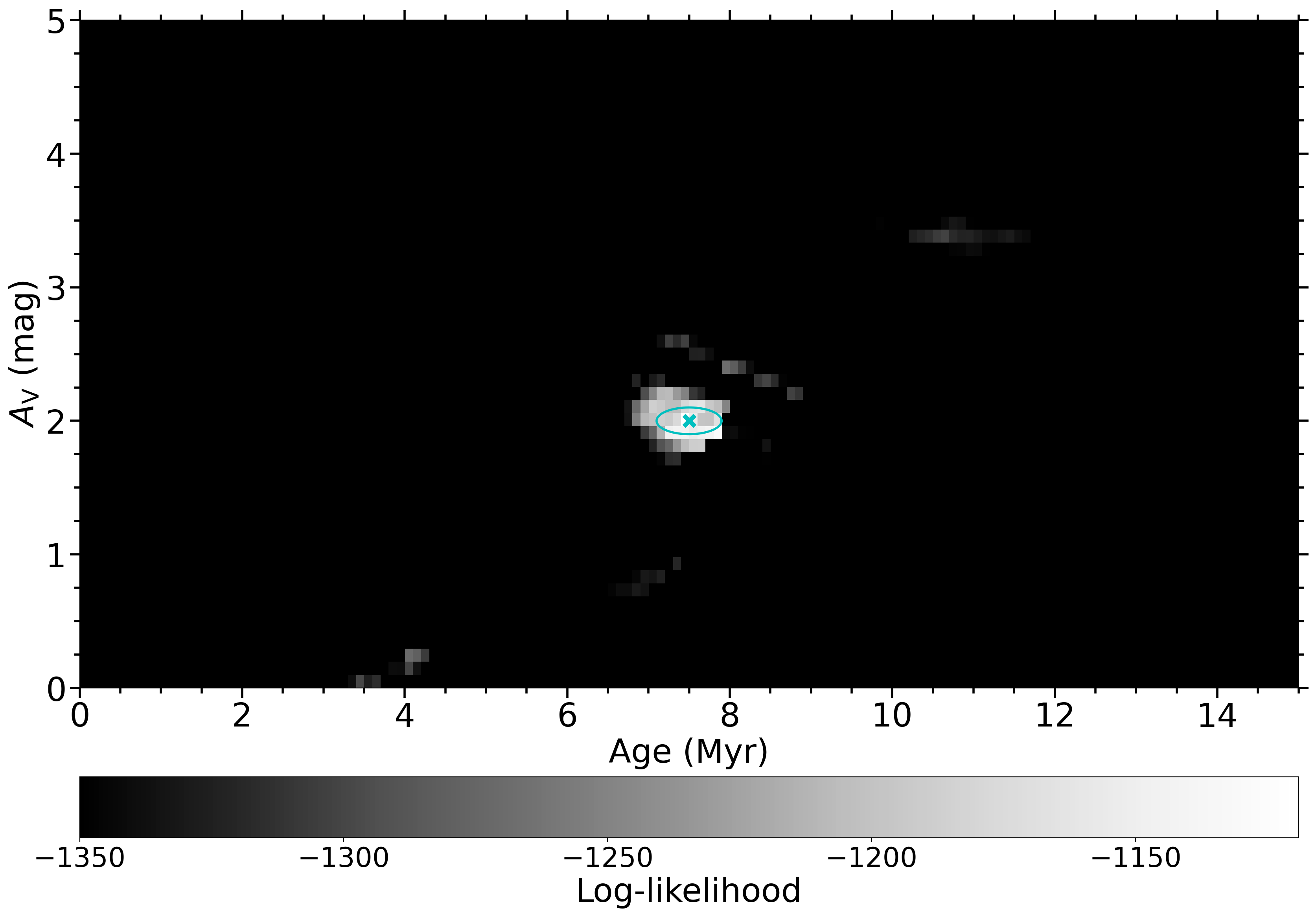}
\end{minipage}
\begin{minipage}{.49\textwidth}
\centering
\includegraphics[width=1.0\columnwidth]{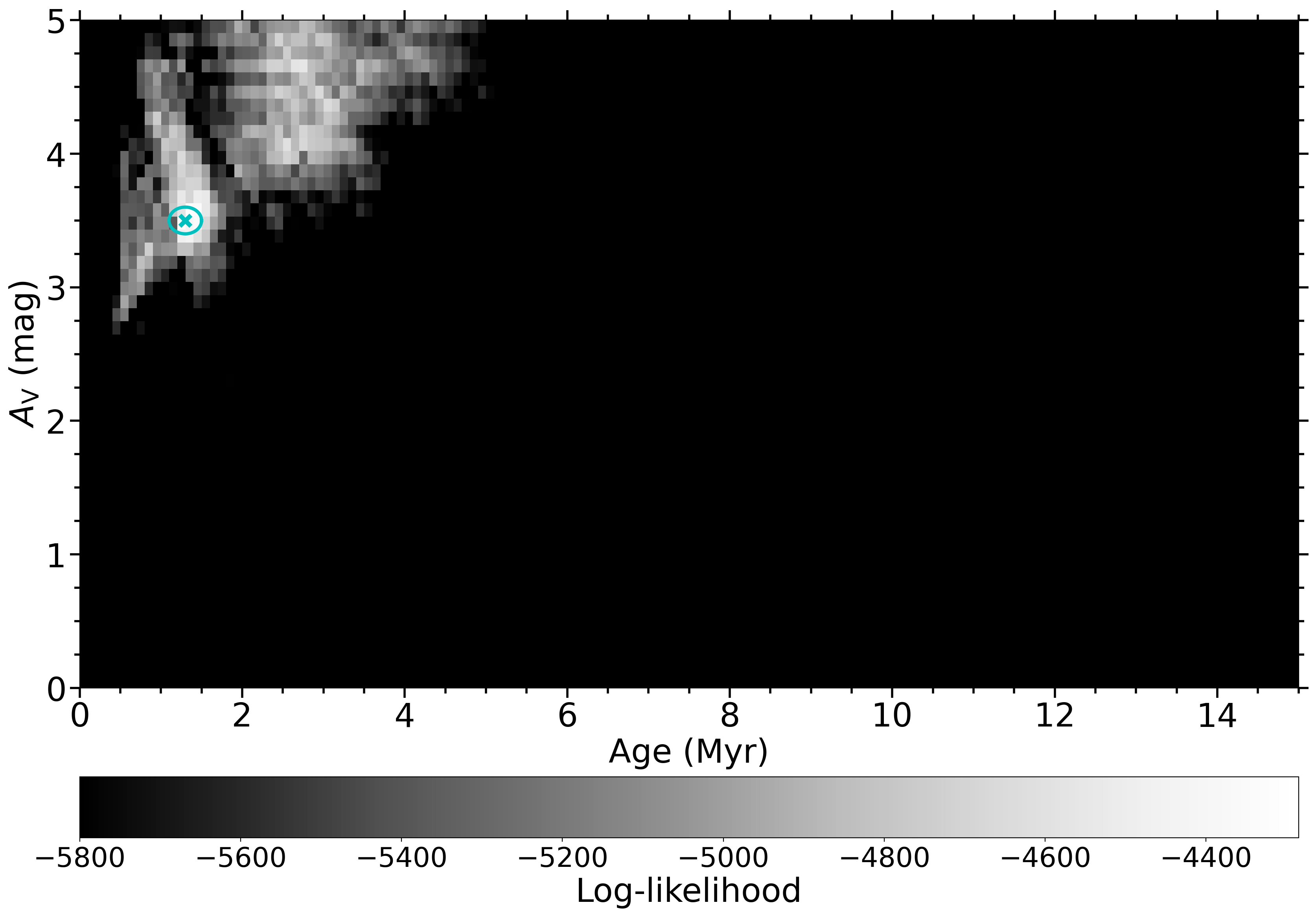}
\end{minipage}
    \caption{Isochrone fitting results of NGC6611. \textit{Top-left:} Best-fit isochrone with $A_{\rm{V}}$ = 2.0 $\pm$ 0.1 mag and age of 7.5 $\pm$ 0.4 Myr shown with the solid cyan line to the early type stars and the blue group (with and without spectral type). An open symbol represents a known spectral type, while a closed symbol does not. The uncertainty on the age for $A_{\rm{V}}$ = 2.0 mag is shown with the cyan shaded region. The stars with known spectral type are corrected for extinction according to the best-fit $A_{\rm{V}}$ with $R_{\rm{V}}$ = 3.56 \citep{Kumar2004}. \textit{Bottom-left:} The log-likelihood for each isochrone in the searched $A_{\rm{V}}$-age grid for the stars indicated in the top-left figure. Each point in the grid is coloured according to its (unnormalised) likelihood value. The log-likelihood gives the quality of the fit, with smaller negative values (in lighter grey) representing better fits. The cyan cross and circle indicate the best-fit $A_{\rm{V}}$ and age with 1$\sigma$ uncertainty range. \textit{Top-right:} Similar to the top-left figure, but now for the early type stars and the red group (with and without spectral type). The solid cyan line gives the best-fit isochrone with $A_{\rm{V}}$ = 3.5 $\pm$ 0.1 mag and age of 1.3 $\pm$ 0.2 Myr. The uncertainty on the age for $A_{\rm{V}}$ = 3.5 mag is shown with the cyan shaded region. \textit{Bottom-right:} Similar to the bottom-left figure, but now for the top-right figure.}
    \label{fig:Isochrones}
\end{figure*}

To estimate the age of NGC6611, we have set up a log-likelihood function similar to \citet{Jorgensen2005} estimating the probability of obtaining an observed set of stars given an isochrone. The log-likelihood is given by
\begin{align}
\label{eq:isochrone_loglik}
    \mathcal{L} \propto \sum_{i=1} \mathrm{log} \left( \sum_{j=1} \frac{1}{2\pi\sigma_{x_{i}}\sigma_{y_{i}}} \mathrm{exp} \left[-\frac{(x_{i} - x_{j})^{2}}{2\sigma_{x_{i}}} - \frac{(y_{i} - y_{j})^{2}}{2\sigma_{y_{i}}} \right] \right),
\end{align}
where in our case $x$ is equal to (B -- V), and $y$ is equal to $M_{\rm{V}}$. We have set up a grid using the \textsc{parsec} + \textsc{colibri} stellar isochrones varying the extinction $A_{\rm{V}}$ between 0.0 and 5.0 mag in steps of 0.1 mag \citep[with $R_{\rm{V}} = 3.56$;][]{Kumar2004} and the age between 0.1 and 15 Myr in steps of 0.1 Myr. For each of the isochrones in the grid, we can calculate the log-likelihood with Equation~\ref{eq:isochrone_loglik} in order to find the best-fitting isochrone in the entire grid. The best-fit isochrone will have the largest value for the log-likelihood as this curve deviates the least from the fit members. We are dealing with a large number of stars with `small' uncertainties and a large number of points on the isochrones. The log-likelihood value will therefore quickly have large negative values, even in log-space.

For each of the isochrones, we choose an $A_{\rm{V}}$. Stars with known spectral type and therefore known $A_{\rm{V}}$ need to be corrected to the $A_{\rm{V}}$ adopted in each isochrone. For example, when fitting an isochrone with $A_{\rm{V}}$ = 2.6 mag, we would place the stars with known spectral type in the CAMD as if they also had $A_{\rm{V}}$ = 2.6 mag by linear inter- or extrapolation.

It is clear from the right panel in Figure~\ref{fig:HR} that we can not accurately constrain the age based on the early type main-sequence stars alone. The members without spectral type display, as mentioned, two clear groups (see the left panel in Figure~\ref{fig:HR}). We have divided the \textit{Gaia} CAMD in three different groups, the early type stars, a blue group and a red group. The location and distribution of these three groups is shown in Figure~\ref{fig:HR_young_old}, where the red and blue groups have their respective colours and are not corrected for extinction. If the spectral type is known, we show the star as an open symbol. To constrain the average extinction and age, we will fit two sets of data comprised of the three groups in Figure~\ref{fig:HR_young_old}. The first set of data is the combination of the early type stars and the blue group (with and without spectral type). As stated above, the early type stars are assigned an $A_{\rm{V}}$ similar to the isochrone in order to make the main-sequence consistent with the blue group. We have excluded the O stars for this set of data, which we will show is a necessary assumption. The second set of data is the combination of the early type stars and the red group (with and without spectral type). Again, the early type stars are assigned an $A_{\rm{V}}$ similar to the isochrone to make the main-sequence consistent with the red group.

The results for the best-fit age and mean $A_{\rm{V}}$ applied to the isochrones are shown in Figure~\ref{fig:Isochrones}. For the first set of data, the early type stars and the blue group, the best fit isochrone has an average $A_{\rm{V}}$ = 2.0 mag and age = 7.5 Myr. This best fit isochrone is shown in the top-left panel, with the stars with known spectral type shifted to the same extinction (stars without spectral type are not shifted). The log-likelihoods of the isochrones are shown in the bottom-left panel. The cyan cross indicates the location of the best-fit isochrone. Similarly, the early type stars and the red group are shown at the right hand side of Figure~\ref{fig:Isochrones}. The best-fit isochrone to this set of data has $A_{\rm{V}}$ = 3.5 mag and age = 1.3 Myr. In this method, we are primarily fitting the mean $A_{\rm{V}}$ and age of the pre-main-sequence stars. Since the early type stars have mostly known spectral type, these stars will always align with the isochrone. However, these early type stars are still valuable to the fit as they dictate where the main-sequence should be with respect to the pre-main-sequence. Fitting only the blue or red group allows too much freedom for the isochrones since we would only be spanning 4 to 5 magnitudes in $M_{\rm{G}}$ (i.e. the early type stars give constraints to the isochrone fitting).

The colour-mapped isochrone grids show more than one local maximum. Specifically, the second set of data including the red group shows seemingly `good fits' with $A_{\rm{V}}$ between 4 and 5 mag and an age of $\sim$ 3 Myr in the bottom-right panel. These isochrones do not accurately fit the transition from pre-main-sequence to main-sequence (the `knee' in the isochrones) but place this around (B -- V) $\sim$ 1.25 mag and $M_{\rm{V}}$ $\sim$ 5 mag. The colour-map is in log-likelihood space, making fits with a log-likelihood, for example, of 100 less than the best-fit seem acceptable even though the probability difference is a factor 10$^{100}$ (due to the small uncertainties on the accurate \textit{Gaia} photometry).

While the average extinction and age of the two sets of data are clearly different, we need to consider if this is significant. If the log-likelihood value of a fit decreases by 1, the p-value decreases by a factor 10. We would determine unrealistically small uncertainties if we were to use the log-likelihood directly. Instead, we determine the uncertainty on the average extinction and age through bootstrapping. Assuming that 10\% of the members are outliers or field stars (since we initially chose a membership probability of 0.9), we randomly select new sets of data containing 90\% of the original sets of data, without replacement. We determine the best-fit extinction and age and re-do the bootstrapping and fitting 1000 times. With a 1000 best-fit extinctions and ages (for each set of data), we determine the uncertainty on the extinction and age with the 16$^{\rm{th}}$ and 84$^{\rm{th}}$ percentiles. This gives for the set of data with the blue group $A_{\rm{V}}$ = 2.0 $\pm$ 0.1 mag and an age of 7.5 $\pm$ 0.4 Myr. For the set of data with the red group we have $A_{\rm{V}}$ = 3.5 $\pm$ 0.1 mag and an age of 1.3 $\pm$ 0.2 Myr. We show the best-fit values with the solid cyan line and cross in the top and bottom panels of Figure~\ref{fig:Isochrones}, respectively. The uncertainty on the age (assuming the best-fit extinction) is shown with the shaded cyan region and circle in the top and bottom panels, respectively.

It becomes obvious now that the blue and red group have a significantly different mean $A_{\rm{V}}$ and age, so we should consider them as two distinct populations of stars. However, this distinction is not as easily made for the early type stars. If we assume an age of 7 to 8 Myr for the older population of stars, the O stars earlier than O7.5-8.5 should already have gone supernova or are considerably evolved. The presence of an O3.5 V((f+)), an O4 V((f+)) and an O6.5 V((f)) star thus argues that these early type stars are part of the younger population of stars \citep{Sana2009}. If these early type O stars are part of the younger population, we could expect many of the late type O stars and B stars to be so as well on the basis of the initial mass function. All O stars have been excluded in the fitting of the older population. One major assumption in the isochrone fitting is that the (B -- V) spread within the blue and red group individually is caused by extinction, which is also observed for the O and B stars. The determined best-fit extinction including uncertainty should only be indicative of the mean extinction value in that group. This is not necessarily the case for the determined ages, as we expect the blue group to be formed in a single star formation episode and the red group to be formed in a different star formation episode, each lasting approximately several 100 kyr. If prolonged star formation took place over several Myr, the observed main-sequence(s) and the two pre-main-sequence tracks, clearly separated by a gap, would be difficult to fit within this scenario.

\section{Astrometry of the two populations}
Up to this point, we analysed the astrometric properties of all members of NGC6611 together. There may be differences in the spatial distribution, proper motion, or even parallax between the young and old population (red and blue group) defined in Figure~\ref{fig:HR_young_old}.

\begin{figure*}
\centering
\begin{minipage}{.49\textwidth}
\centering
\includegraphics[width=1.0\columnwidth]{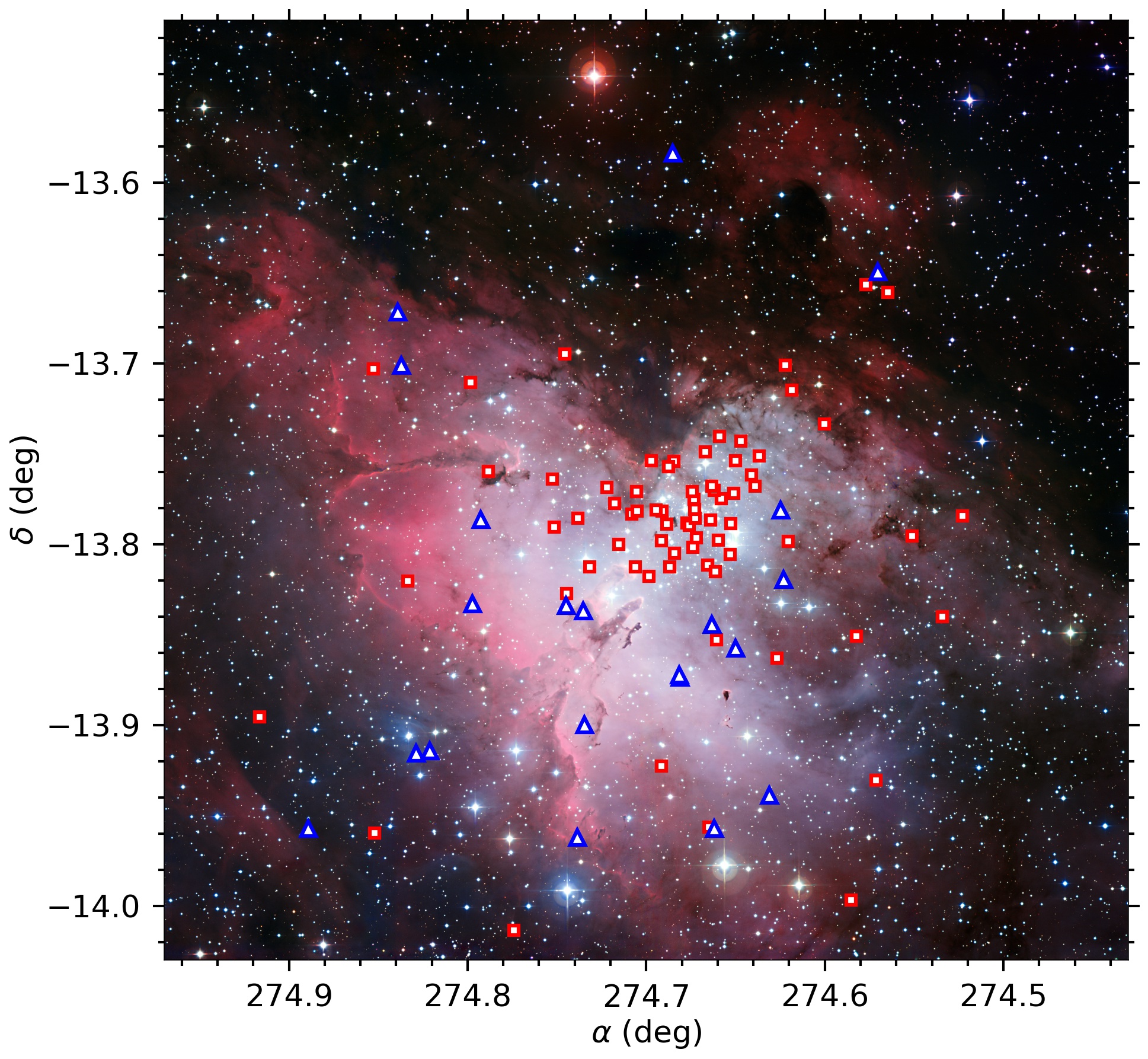}
\end{minipage}
\begin{minipage}{.49\textwidth}
\centering
\includegraphics[width=1.0\columnwidth]{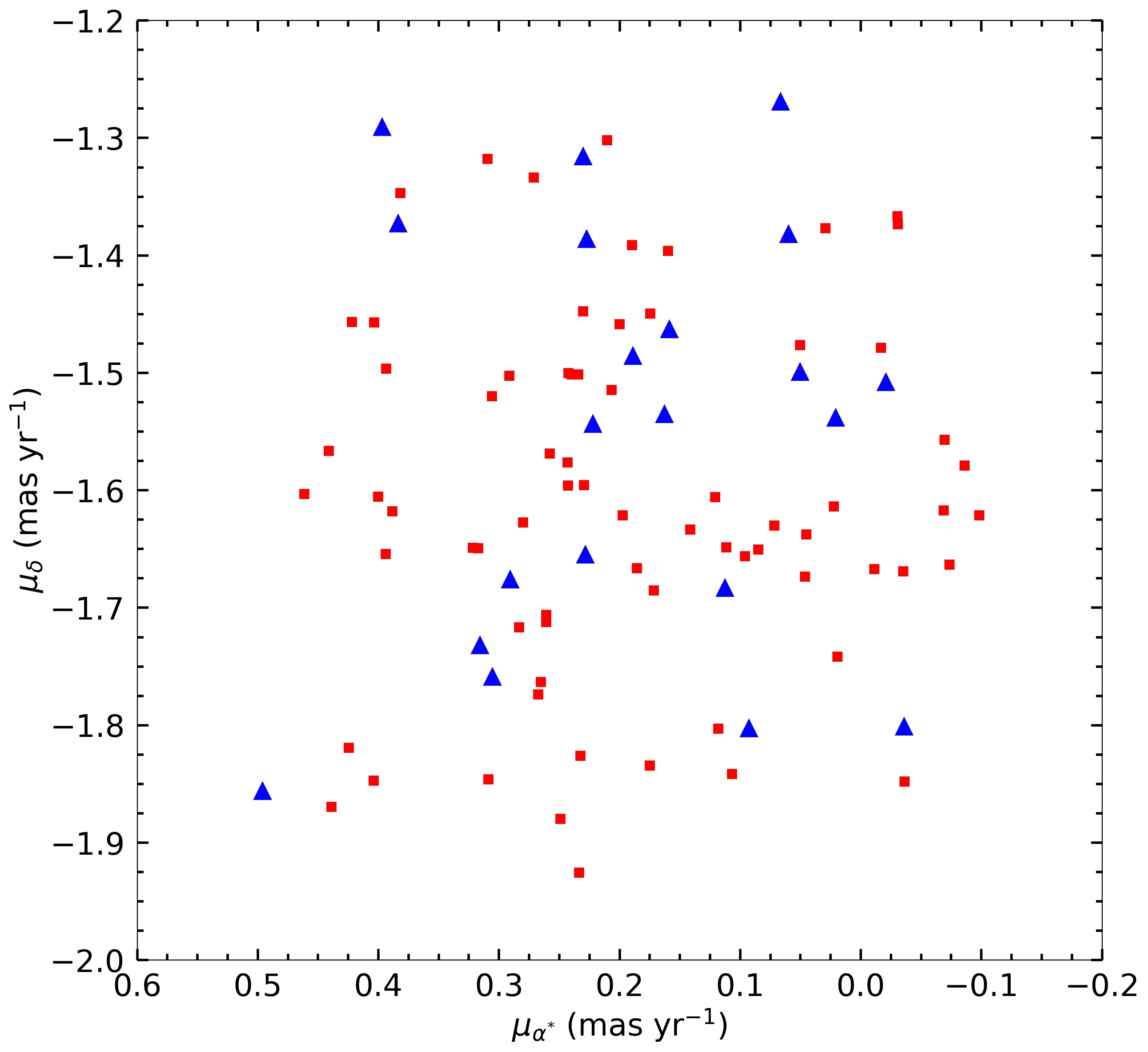}
\end{minipage}
\begin{minipage}{.49\textwidth}
\centering
\includegraphics[width=1.0\columnwidth]{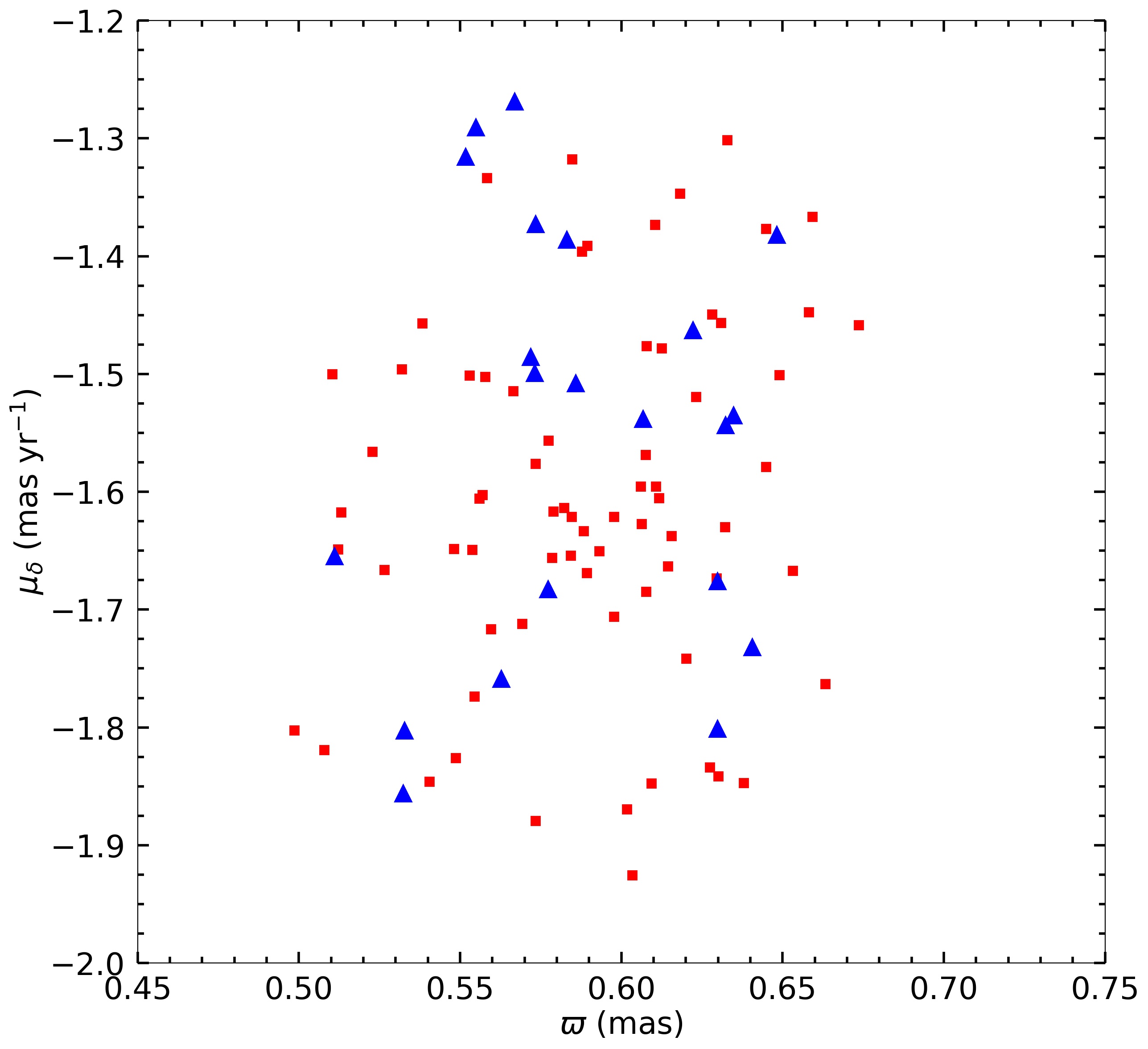}
\end{minipage}
\begin{minipage}{.49\textwidth}
\centering
\includegraphics[width=1.0\columnwidth]{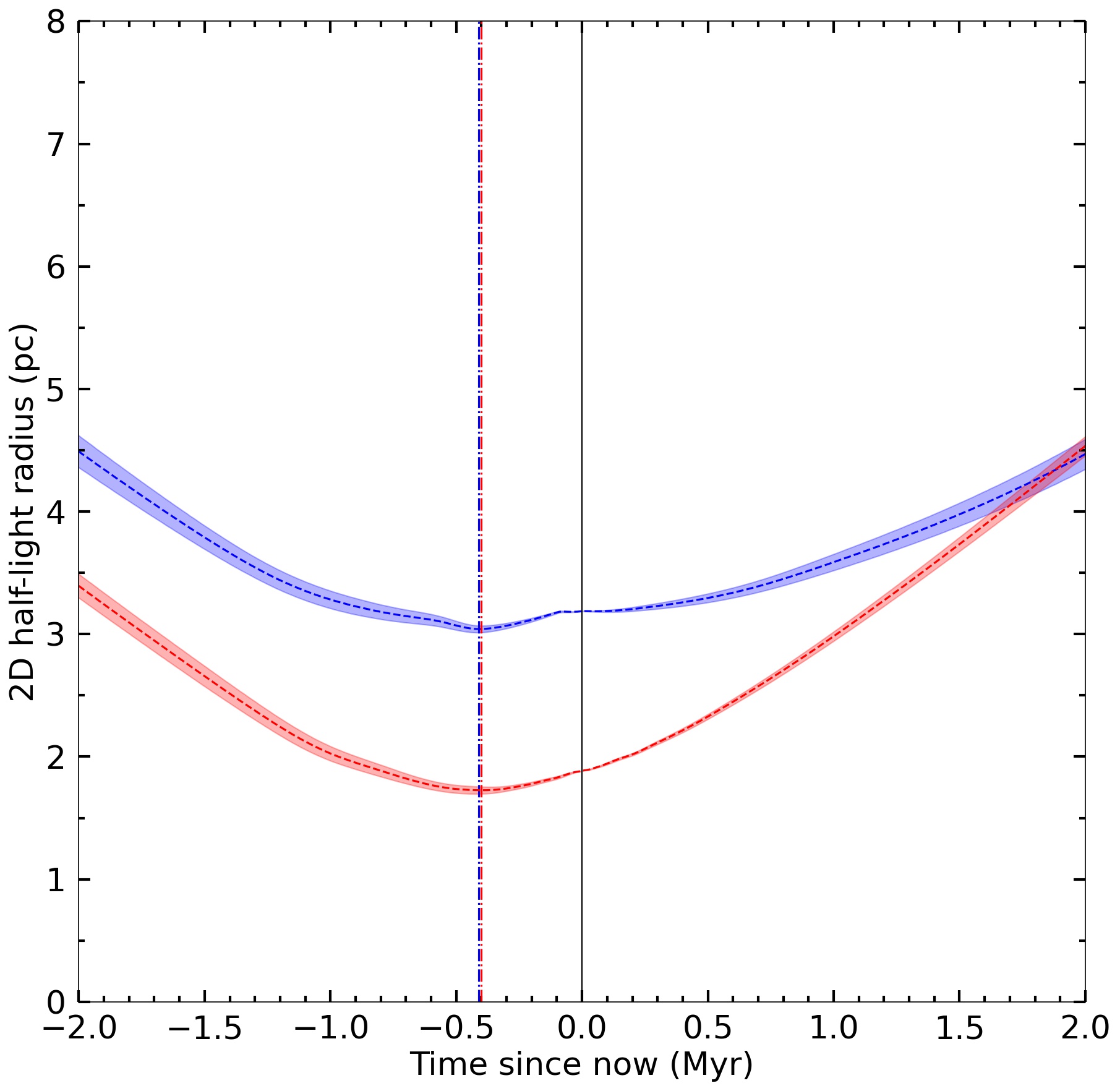}
\end{minipage}
\caption{Distribution of the young and old population (the red and blue group, respectively), coloured and marked similarly as in Figure~\ref{fig:HR_young_old}. \textit{Top-left}: B, V and R composite mosaic image of NGC6611 (ESO, press release 0926) showing the spatial distribution. \textit{Top-right}: The proper motion distribution. \textit{Bottom-left}: The parallax distribution. \textit{Bottom-right}: The 2D half-light radius for each population as a function of time. The dashed line indicates the mean radius and the coloured region indicates 1$\sigma$ uncertainty. The dash dotted lines give the time of minimum radius for each population. The young population shows a minimum --0.40 Myr ago, while the old population shows a minimum --0.41 Myr ago.}
\label{fig:redblue_astrometry}
\end{figure*}

The first thing to investigate is the spatial distribution of the two populations, which we show in the top-left panel of Figure~\ref{fig:redblue_astrometry}, coloured and marked similarly as in Figure~\ref{fig:HR_young_old}. We can see that the young (red) population is clearly concentrated towards the centre of NGC6611, while the old (blue) population is spatially extended. The majority of the old population is scattered south of the centre of NGC6611; however, it is unclear whether this is physical or an observational bias (e.g. stars with relatively low extinction are found south of the centre of NGC6611).

We also show the proper motion and parallax distribution in Figure~\ref{fig:redblue_astrometry}. Both the proper motion and parallax show no clear differences. A two-sided t-test and a Kolmogorov-Smirnov test between the blue and red group for the two proper motion components and the parallax gives p-values consistent within 2$\sigma$ for all cases. We note that biases could possibly come into play in the membership selection, since the younger population has $\sim$ 3 times more stars than the older population ($\sim$ 5 times when including the early type stars). We could therefore select only stars belonging to the older population with a proper motion and parallax consistent with the young population and exclude outliers. The older population can not be explained with the contamination of field stars in our members. While the spatial distribution of this population might be considered consistent with a field star population, both the proper motion and the parallax are in excellent agreement with that of the younger population. As a consistency check, we have selected stars associated with the older population from the 2878 candidate sources in Section~\ref{sec:upmask} by their position in the CAMD which include many field stars. Still, a clear over-density of stars is present at the determined proper motion and parallax of NGC6611. The older population is also redder (or brighter) than the field stars.

The older population may be spatially extended now, but could have been more concentrated in the past, similar to the younger population. To investigate this, we trace back in time the stars belonging to the young and old population separately. We perform 10000 Monte-Carlo simulations randomly drawing the proper motions for each star from a multivariate normal distribution with means equal to the observed values, with the uncertainties given by the uncertainties and correlations. We trace back the young and old population --2 Myr back and forward in time by propagating their proper motions. In increments of 10 kyr we determine the 
2D half-light radius for each population, given by the mean as the 50$^{\rm{th}}$ percentile and 1$\sigma$ uncertainty as the 16$^{\rm{th}}$ and 84$^{\rm{th}}$ percentile (without correcting for extinction). We show the result in the bottom-right panel of Figure~\ref{fig:redblue_astrometry}. The young population has a minimum 2D half-light radius of 1.7 pc --0.40 Myr ago, while the old population has a minimum radius of 3.0 pc --0.41 Myr ago. Both populations were smaller in the past and are currently already expanding. The position of the stars and thus the radius are likely extrapolated too far back and forward in time before and after --1 and 1 Myr, respectively.

\section{Runaway stars}
\label{sec:res_runaway}
\begin{figure*}
\begin{minipage}{.49\textwidth}
\centering
\includegraphics[width=1.0\columnwidth]{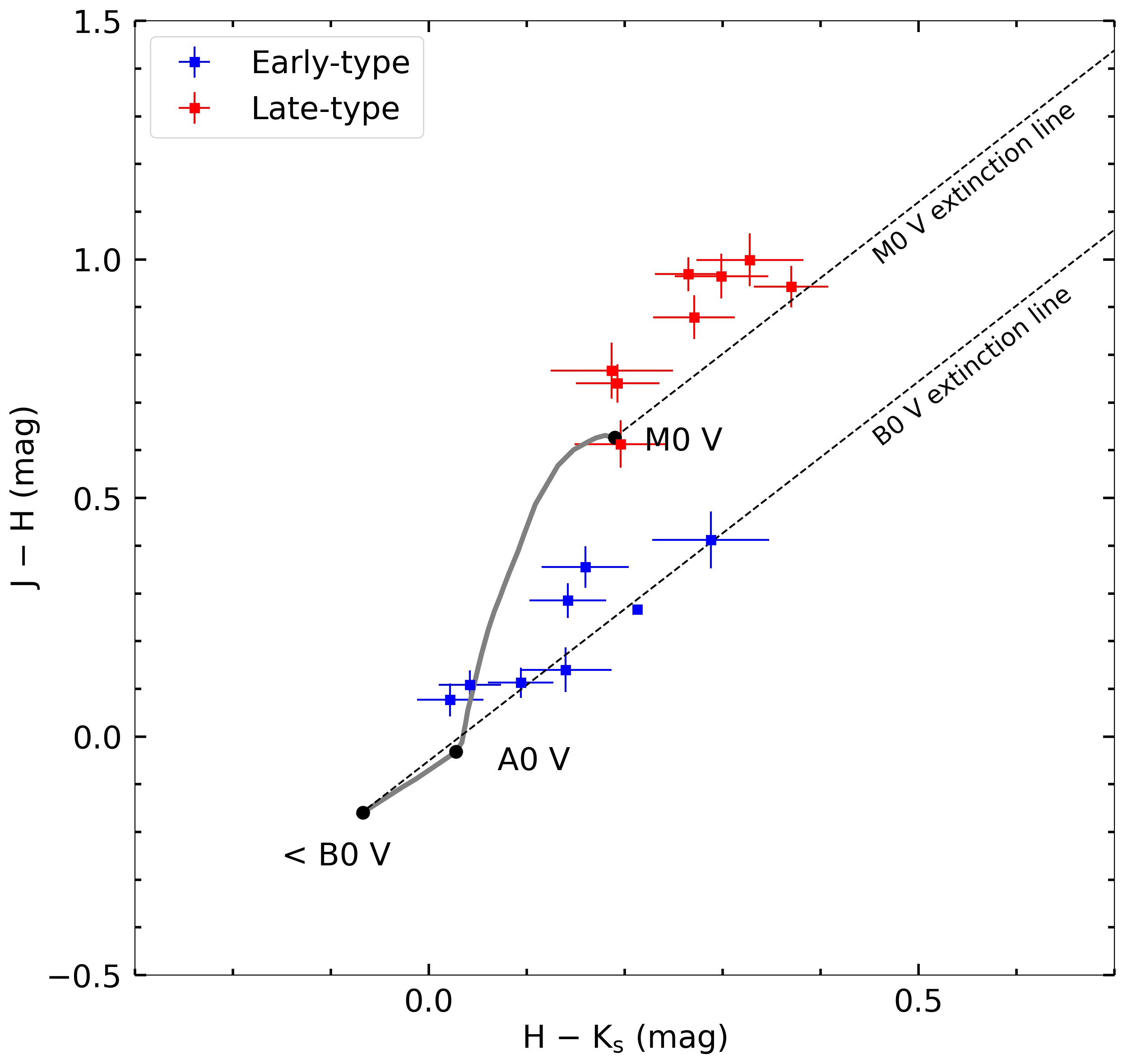}
\end{minipage}
\begin{minipage}{.49\textwidth}
\centering
\includegraphics[width=1.0\columnwidth]{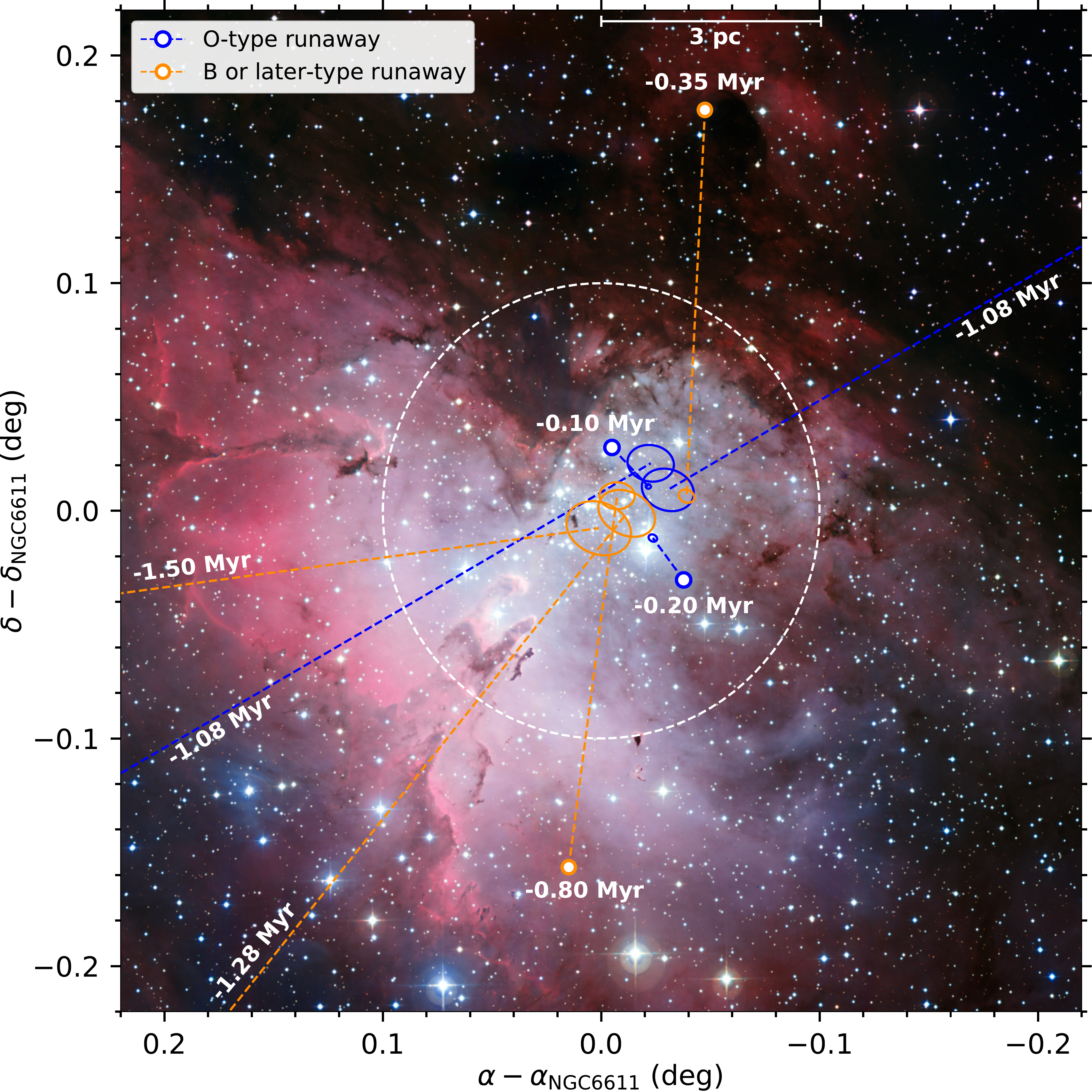}
\end{minipage}
\caption{Properties of the runaway candidates. \textit{Left}: \textit{2MASS} near-infrared colour-colour diagram for the runaway candidates. The early type runaways originally from NGC6611 in the last 3 Myr are shown in blue. Late type stars are highlighted in red and are deemed unlikely to come from NGC6611. We show the location of the zero-age main-sequence in grey \citep{Peacut2013}. The reddening line for a B0\ V (top) and M0 V star (bottom) are shown with the dashed black lines for $R_{\rm{V}}$ = 3.1 \citep{Cardelli1989}. \textit{Right}: B, V and R composite mosaic image of NGC6611 (ESO, press release 0926). We highlight the adopted search cluster search radius, with a radius of 0.1 deg ($\sim$ 3 pc at a distance of 1.7 kpc), with the white circle. The trajectories of the O star runaways are shown in blue, while the B or later-type runaways are shown in orange. The ellipse shapes give the uncertainty on the location for the given trace-back time, which places the runaways near the centre of NGC6611. The kinematic properties of these runaways are summarised in Table~\ref{tab:Ostars_runaways}.}
\label{fig:runaway_JHK_image}
\end{figure*}

We have now discussed the astrometric parameters and the isochronal age(s) of NGC6611. Dynamical interactions in young massive clusters are expected to produce runaway stars \citep{Gvaramadze2012}. The young population has an age of 1.3 $\pm$ 0.2 Myr, so core-collapse supernovae should not have happened yet. Runaways should then be produced through dynamical interactions. With the accurate \textit{Gaia} EDR3 proper motions and parallaxes, we should be able to find runaways that trace back to NGC6611.

We search for runaway stars coming from NGC6611 by performing a cone-search in \textit{Gaia} EDR3 centred on the cluster with ($\alpha_{\rm{NGC6611}}$, $\delta_{\rm{NGC6611}}$) = (274.67 deg, --13.78 deg) with a radius of 15 deg. These stars are subject to the same corrections and filters described in Section~\ref{sec:gaiadata}. Moreover, we require these stars to have $\varpi/\sigma_{\varpi} > 10$ and $\varpi$ to be consistent within 3$\sigma$ from the parallax of NGC6611. This assumption may exclude runaways moving in the radial direction, for which the parallax could now significantly deviate. Unfortunately in \textit{Gaia} EDR3 few radial velocities are available, so there is no viable method for including these. The determined radial velocity for NGC6611 is 4.7 $\pm$ 3.5 km s$^{-1}$ and its radial motion has a minimal effect on the analysis. Next, we have only included stars brighter than $M_{\rm{G}} <$ 1 mag, where we have directly used their (accurate) parallax. We have tried to adjust this cut-off magnitude and found that for fainter potential runaways it becomes increasingly more difficult to distinguish between actual runaways and field stars aligning by chance. As we will show, the adopted cut-off magnitude allows us to separate the early type (O, B or A) stars expected from NGC6611 from bright late type stars (e.g. red giants) in the field at this magnitude.

We identify runaways by tracing back in time their position and checking whether this position aligns with the position of NGC6611 at the same time. Stars consistent to be runaways from NGC6611 should satisfy at some point in time
\begin{align*}
\label{eq:runaway}
    r_{\rm{sep}}(t) = \sqrt{\alpha_{\rm{sep}}^{2}(t) + \delta_{\rm{sep}}^{2}(t)} < r_{\rm{NGC6611}},
\end{align*}
where $\alpha_{\rm{sep}}$, $\delta_{\rm{sep}}$ and $r_{\rm{sep}}$ are the separation between the centre of NGC6611 and the runaway in right ascension, declination and on-sky projection respectively. $r_{\rm{NGC6611}}$ is the cluster radius and will be chosen sufficiently small to ensure that the small-angle approximation holds. $\alpha_{\rm{sep}}$ and $\delta_{\rm{sep}}$ are given by \begin{align*}
    \alpha_{\rm{sep}}(t) &= (\alpha_{\rm{}} + \frac{t \cdot \mu_{\alpha^{*},\rm{}}}{3.6 \times 10^{6} \cdot \rm{cos}(\delta_{\rm{}})}) \\
    &\quad - (\alpha_{\rm{NGC6611}} + \frac{t \cdot \mu_{\alpha^{*},\rm{NGC6611}}}{3.6 \times 10^{6} \cdot \rm{cos}(\delta_{\rm{NGC6611}})})\ \rm{deg}, \\
    \delta_{\rm{sep}}(t) &= (\delta_{\rm{}} + \frac{t \cdot \mu_{\delta,\rm{}}}{3.6 \times 10^{6}}) \\
    &\quad - (\delta_{\rm{NGC6611}} + \frac{t \cdot \mu_{\delta,\rm{NGC6611}}}{3.6 \times 10^{6}})\ \rm{deg}.
\end{align*}
Here, $t$ is the time in years and should be negative for runaways coming from NGC6611 and $\mu_{\alpha^{*}} \equiv \mu_{\alpha}$cos($\delta$). We will take a conservative approach and use a lower limit of --3 Myr for $t$, considering the age of 1.3 $\pm$ 0.2 Myr. We adopt a cluster radius of $r_{\rm{NGC6611}}$ = 0.1 deg. A last filter is applied by only including runaways moving outwards with relatively high transverse velocities with $\Delta v_{\rm{T}} > 3$ km s$^{-1}$ and $\Delta v_{\rm{out}} > 0$ km s$^{-1}$ with respect to NGC6611, where $\Delta v_{\rm{T}}$ and $\Delta v_{\rm{out}}$ are, respectively, the transverse velocity and the outwards velocity component relative to the centre of NGC6611. 

Runaways are typically defined to have a relative velocity $>$ 30 km s$^{-1}$. However, we are limited to a 2D approach as the radial velocities are missing. In the case of the binary supernova ejection scenario, walkaways with a relative velocity $<$ 30 km s$^{-1}$ are predicted \citep{Renzo2019}. But these definitions are not clear for the dynamical ejection scenario, as this is dependent on the cluster escape velocity. For example, the young massive cluster R136 in the large Magellanic Cloud with a mass of $\sim$ 5 $\times$ 10$^{4}$ M$_{\odot}$ and a radius of $\sim$ 1 pc has an escape velocity of $\sim$ 20 km s$^{-1}$ \citep{Schneider2018}. A conservative estimate of the current 3D escape velocity of NGC6611 is $\sim$ 4 km s$^{-1}$ at a distance of 2 pc from the centre (see Section~\ref{sec:dis_ngc6611_params}). If runaways are isotropically ejected, we expect the projected $\Delta v_{\rm{T}}$ to be on average $\gtrsim$ 3 km s$^{-1}$, which we adopt as our lower limit. \citet{Kuhn2019} select members from X-ray, optical and infrared studies and show that stars in NGC6611 have a median $\Delta v_{\rm{out}}$ of 0.90 $\pm$ 0.23 km s$^{-1}$ and NGC6611 is therefore on average expanding. Our lower limit on $\Delta v_{\rm{T}}$ is therefore sufficiently large to exclude members of NGC6611. Note that dynamical interactions may contribute to the expansion of a cluster like NGC6611. While $\Delta v_{\rm{T}}$ may not be equal to $\Delta v_{\rm{out}}$ depending on the direction of motion, our use of $\Delta v_{\rm{T}}$ instead of $\Delta v_{\rm{out}}$ does not affect our results. 

We find a total of 16 candidate runaways coming from NGC6611 in the last 3 Myr. The first thing to consider is whether these are `true' runaways or field stars aligning with the position of NGC6611 by chance. As mentioned, we assume the runaways to be early type stars ($M_{\rm{G}} < 1$ mag). To investigate this, we can look at their \textit{2MASS} JHK$_{\rm{s}}$ photometry. We show the (J -- H) - (H -- K$_{\rm{s}}$) colour-colour diagram in the left panel of Figure~\ref{fig:runaway_JHK_image}. The position of the zero-age main sequence is shown with the grey track. The early type stars should be located in the bottom-left below (J -- H) and (H -- K$_{\rm{s}}$) $\lesssim$ 0 mag. Extinction will cause stars to move towards the top-right. For example, we show the reddening line (with $R_{\rm{V}} = 3.1$) of a B0 V (bottom) and M0 V star (top) with a dashed black line. Since stars with a spectral type of $\sim$ A0 V or earlier are more or less expected to have overlapping reddening lines, the early type runaway stars should lie roughly on the B0 V track. This is the case for 8 of the candidate runaways. The other 8 clearly lie above the M0 V extinction track making them inconsistent with the early type stars, but consistent with late type red giant stars. The late type stars also have seemingly random kinematic properties, coming from random positions within the cluster (i.e. not from the centre of NGC6611). The late type stars also have randomly distributed kinematic ages. For this reason, we consider these 8 stars as field stars aligning by chance and have excluded them. 

We remain with 8 early type runaways, which we consider as our final runaway sample. These 8 runaways include 3 O stars and 2 B stars with a spectral classification. Based on the \textit{Gaia} and \textit{2MASS} magnitude and distance from their individual parallax, we estimate the spectral type of an early O star, an early B star and a late B/A star for the 3 unclassified runaways. Two runaway O stars, BD$-$14$^{\circ}$ 5040 and UCAC2 27149134 were also found by \citet{Gvaramadze2008} and are known to produce bowshocks, which we show in Appendix~\ref{sec:Bowshocks}.

\begin{figure}
\centering
\includegraphics[width=1.0\columnwidth]{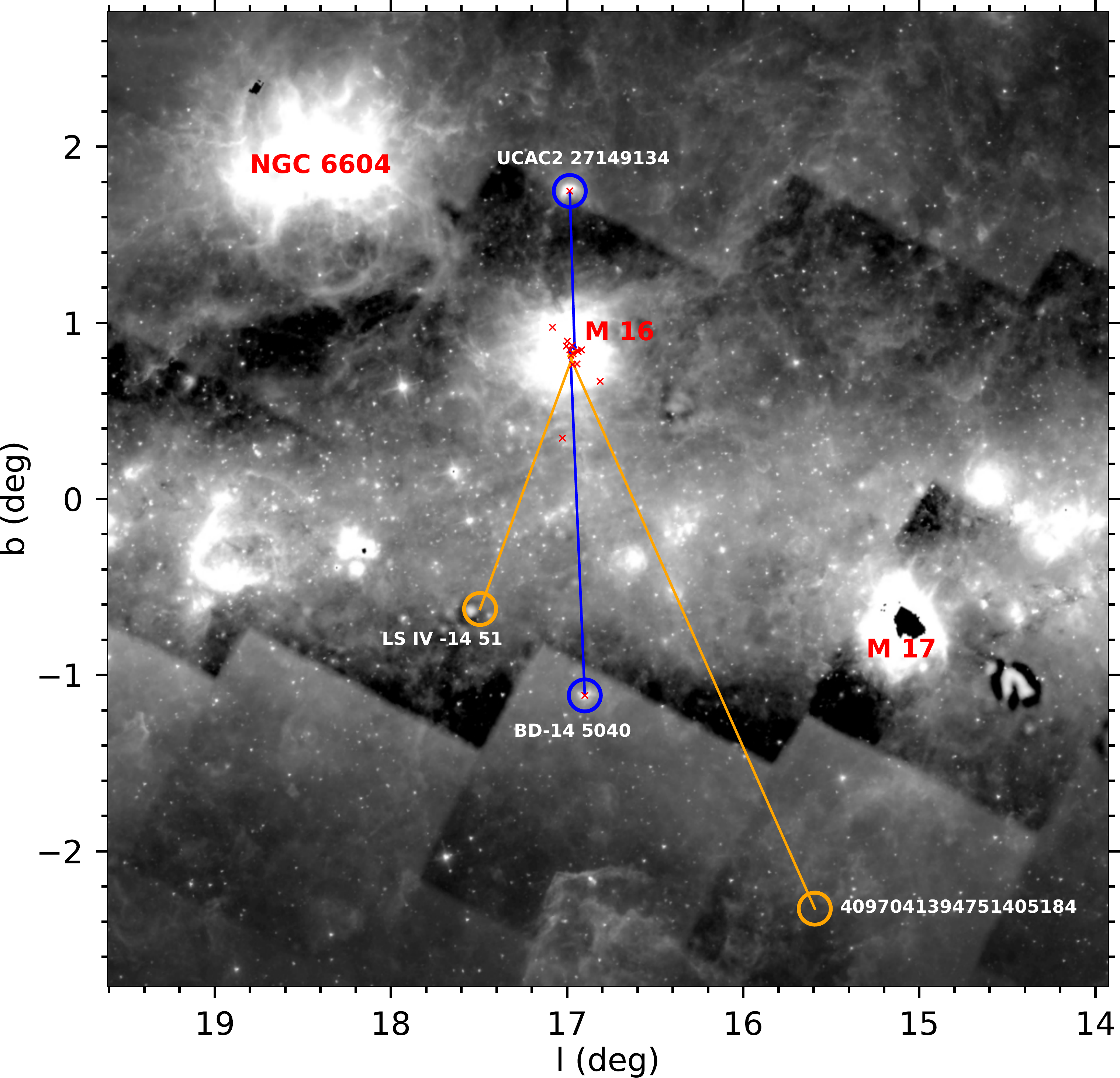}
\caption{\textit{AllWISE} W4 (22 $\mu$m) grey-scale image of M16 and the surrounding region in Galactic coordinates. We show the current position of the runaways found outside M16 also shown in Figure~\ref{fig:runaway_JHK_image} adopting similar colours. We trace back the motion of these runaways \textit{relative} to the cluster. UCAC2 27149134 and BD--14$^{\circ}$ 5040 have a clear bowshock which we show in Appendix~\ref{sec:Bowshocks}. We show for context the position of the O stars with the red crosses and indicate the position of NGC6604 and M17.}
\label{fig:runaway_allwise}
\end{figure}

We show the trajectories of the 8 runaways with respect to NGC6611 in the right panel of Figure~\ref{fig:runaway_JHK_image}, with a trace-back time that places them near the centre of NGC6611. We show for each of the runaways their 2$\sigma$ error ellipse on the position calculated from the position, proper motion and corresponding covariance matrix. Figure~\ref{fig:runaway_allwise} shows the runaways far outside NGC6611 in the \textit{AllWISE} W4 (22 $\mu$m) image in Galactic coordinates, adopting similar colouring as before. The two bowshock producing O stars BD$-$14$^{\circ}$ 5040 and UCAC2 27149134 are shown with the blue tracks and are consistent with coming from the same position in NGC6611 $\sim$ 1.08 Myr ago and move in almost exactly opposite directions. Two other runaways found far outside of M16 are shown in orange. The found early type runaways have kinematic ages $\lesssim$ 1.6 Myr, consistent with the age of the young population. Of the 8 runaways found, 4 have $\Delta v_{\rm{T}} >$ 20 km s$^{-1}$ and 4 have $\Delta v_{\rm{T}} <$ 20 km s$^{-1}$. We summarise the kinematic properties of the runaways in Table~\ref{tab:Ostars_runaways} and give the identifier, spectral type and kinematic properties. We note that $\Delta v_{\rm{T}}$ for the runaways is determined with the individual parallax.

We have increased the trace-back time to find runaways consistent with coming from the older population with an age of 7.5 $\pm$ 0.4 Myr. However, we find no clear candidates coming from the older population in the last 10 Myr. There are several limitations to this, as the Galactic potential may not be ignored over such long time scales, and the centre and mean proper motion of the older population may be different.

\section{The O stars in NGC6611}
\label{sec:res_Ostarkin}
\begin{table*}
\centering
\caption{Identifier, spectral type, and kinematic properties of the runaways and O stars originating from NGC6611. We first give the runaway O stars and B or later type runaways. We also give the O stars which could have had dynamical interactions in the past, as well as the remaining O stars. Each sub-table is sorted by $|\Delta v_{\rm{T}}|$.}
\label{tab:Ostars_runaways}
\begin{tabular}{l l l l l l l}
\hline
\hline
Identifier & Spectral type & |$\Delta v_{\rm{T}}$| & $v_{\rm{R}}$ & $t_{\rm{kin}}$ & $P_{\rm{orb}}$ & Ref.\\
- & - & km s$^{-1}$ & km s$^{-1}$ & Myr & days & - \\
\hline
\multicolumn{6}{c}{Runaway O stars}\\
\hline \vspace{1mm}
BD--14$^{\circ}$ 5040\tablefootmark{a} & O5.5 V(n)((f)) & 53 $\pm$ 1 & - & 1.04 -- 1.10 & - & 1 \\ \vspace{1mm}
UCAC2 27149134\tablefootmark{a} & O3-4 V\tablefootmark{c} & 24 $\pm$ 1 & - & 1.05 -- 1.19 & - & - \\ \vspace{1mm}
HD 168504 & O7.5 V(n)z & 8.7 $\pm$ 0.8 & 5 $\pm$ 5 & 1.41 -- 1.88 & - & 1 \\\vspace{1mm}
BD--13$^{\circ}$ 4927 & O7 II(f) & 6.8 $\pm$ 0.4 & 20 $\pm$ 4 & 0.00 -- 0.27 & - & 2 \\ \vspace{1mm}
HD 168183 & O9.5 III + B3-5 V/III & 4.6 $\pm$ 0.3 & 19.2 $\pm$ 1.0 & 0.83 -- 2.02 & 4 & 2 \\ \vspace{1mm}
LS IV --13 14 & O9 V & 3.4 $\pm$ 0.2 & 19 $\pm$ 5 & 0.00 -- 0.87 & - & 3 \\ \vspace{1mm}
W161 & O8.5 V & 2.6 $\pm$ 0.2 & --6 $\pm$ 5 & 0.32 -- 1.37 & - & 3 \\ \vspace{1mm}
HD 168137\tablefootmark{b} & O7 V + O8 V & 2.8 $\pm$ 0.3 & 26.3 $\pm$ 9.2 & 0.17 -- 1.29 & 1836 & 2,4 \\
\hline
\multicolumn{6}{c}{B or later-type runaway stars}\\
\hline \vspace{1mm}
4097041394751405184 & B/A \tablefootmark{c} & 78 $\pm$ 3 & - & 1.26 -- 1.31 & - & - \\ \vspace{1mm}
LS IV --14 51 & B0-2 V\tablefootmark{c} & 30 $\pm$ 1 & - & 1.45 -- 1.57 & - & - \\ \vspace{1mm}
W583 & B9 III & 14.1 $\pm$ 0.6 & --1 $\pm$ 5 & 0.30 -- 0.45 & - & 3\\ \vspace{1mm}
W290 & B2.5 V & 6.0 $\pm$ 0.3 & 8 $\pm$ 5 & 0.51 -- 1.04 & - & 3 \\
\hline
\hline
\multicolumn{6}{c}{Candidate dynamically interacted O stars}\\
\hline \vspace{1mm}
BD--13$^{\circ}$ 4929 & O7 V + (B0.5 V + B0.5 V) & 5.5 $\pm$ 0.5 & 15 $\pm$ 2 & - & undef. + (4?) & 2 \\ \vspace{1mm}
W222 & O7 V((f)) & 1.5 $\pm$ 0.2 & 16 $\pm$ 5 & - & - & 3 \\ \vspace{1mm}
HD 168075 & O6.5 V((f)) + B0-1 V & 1.0 $\pm$ 0.2 & 17 $\pm$ 10 & - & 44 & 2,5 \\ \vspace{1mm}
HD 168076 & O3.5 V((f+)) + O7.5 V & - & 10 $\pm$ 2 & - & $\gtrsim 10^{5}$? & 2 \\
\hline
\multicolumn{6}{c}{Other O stars}\\
\hline \vspace{1mm}
W584\tablefootmark{b} & O9 V & 1.2 $\pm$ 0.2 & 6 $\pm$ 5 & - & - & 3 \\ \vspace{1mm}
BD--13$^{\circ}$ 4930 & O9.5 Vp & 0.7 $\pm$ 0.2 & 5 $\pm$ 1 & - & - & 2 \\ \vspace{1mm}
BD--13$^{\circ}$ 4928 & O9.5 V & 0.3 $\pm$ 0.2 & 17 $\pm$ 15 & - & - & 2 \\ \vspace{1mm}
BD--13$^{\circ}$ 4923 & O4 V((f+)) + O7.5 V & - & 3.4 $\pm$ 5.3 & - & 13 & 2 \\
\hline
\hline
\end{tabular}
\tablefoottext{a}{Visible bowshock in the \textit{AllWISE} photometry}
\tablefoottext{b}{Candidate bowshock in the \textit{Spitzer} photometry}
\tablefoottext{c}{Spectral type estimated from photometry}
\tablebib{(1) \citet{MaizApellaniz2016}; (2) \citet{Sana2009,Sana2012}; (3) \citet{Evans2005}; (4) \citet{LeBouquin2017}; (5) \citet{Barba2010}.}
\end{table*}

\begin{figure*}
\centering
\begin{minipage}{.49\textwidth}
\centering
\includegraphics[width=1.0\columnwidth]{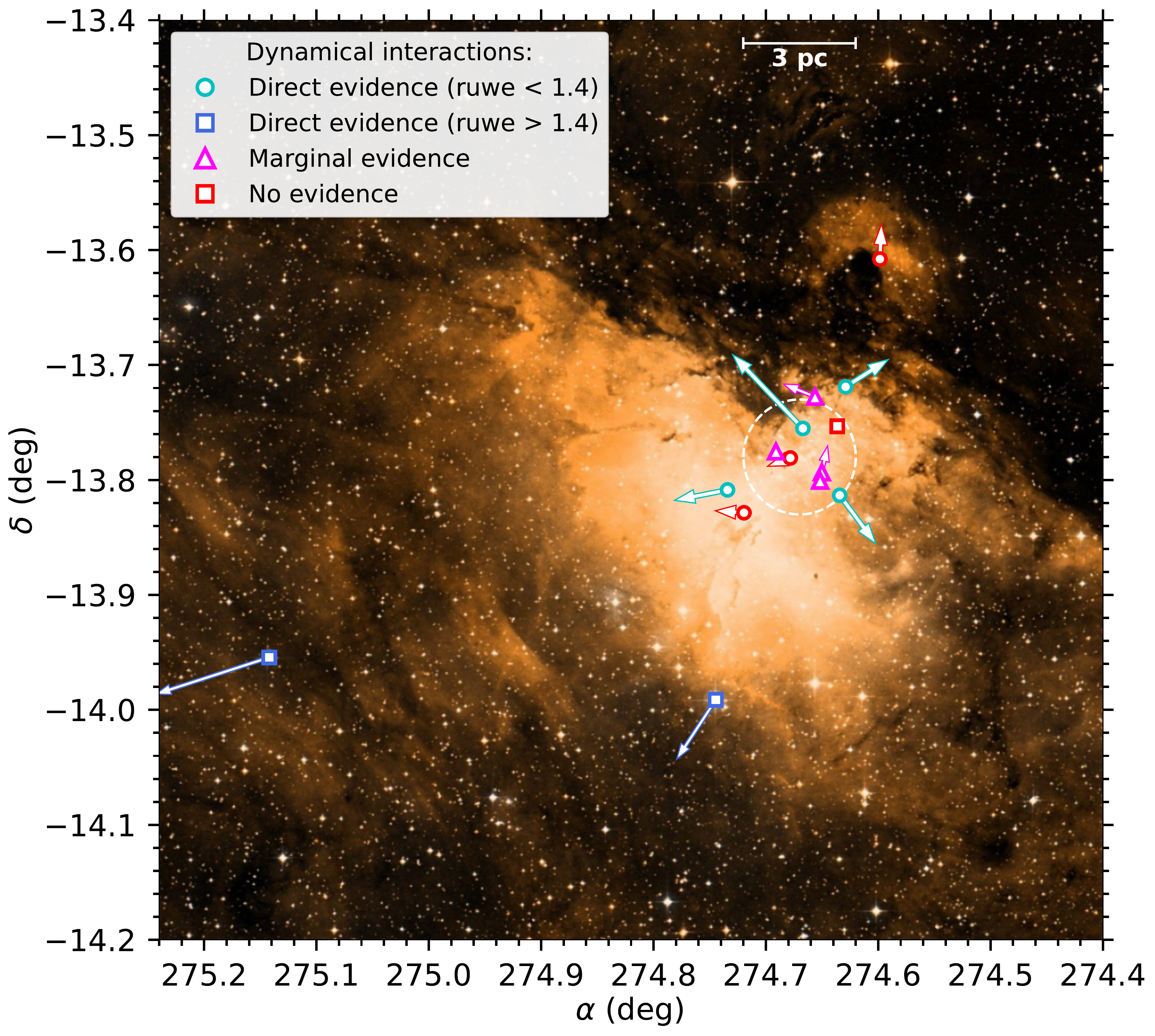}
\end{minipage}
\begin{minipage}{.49\textwidth}
\centering
\includegraphics[width=1.0\columnwidth]{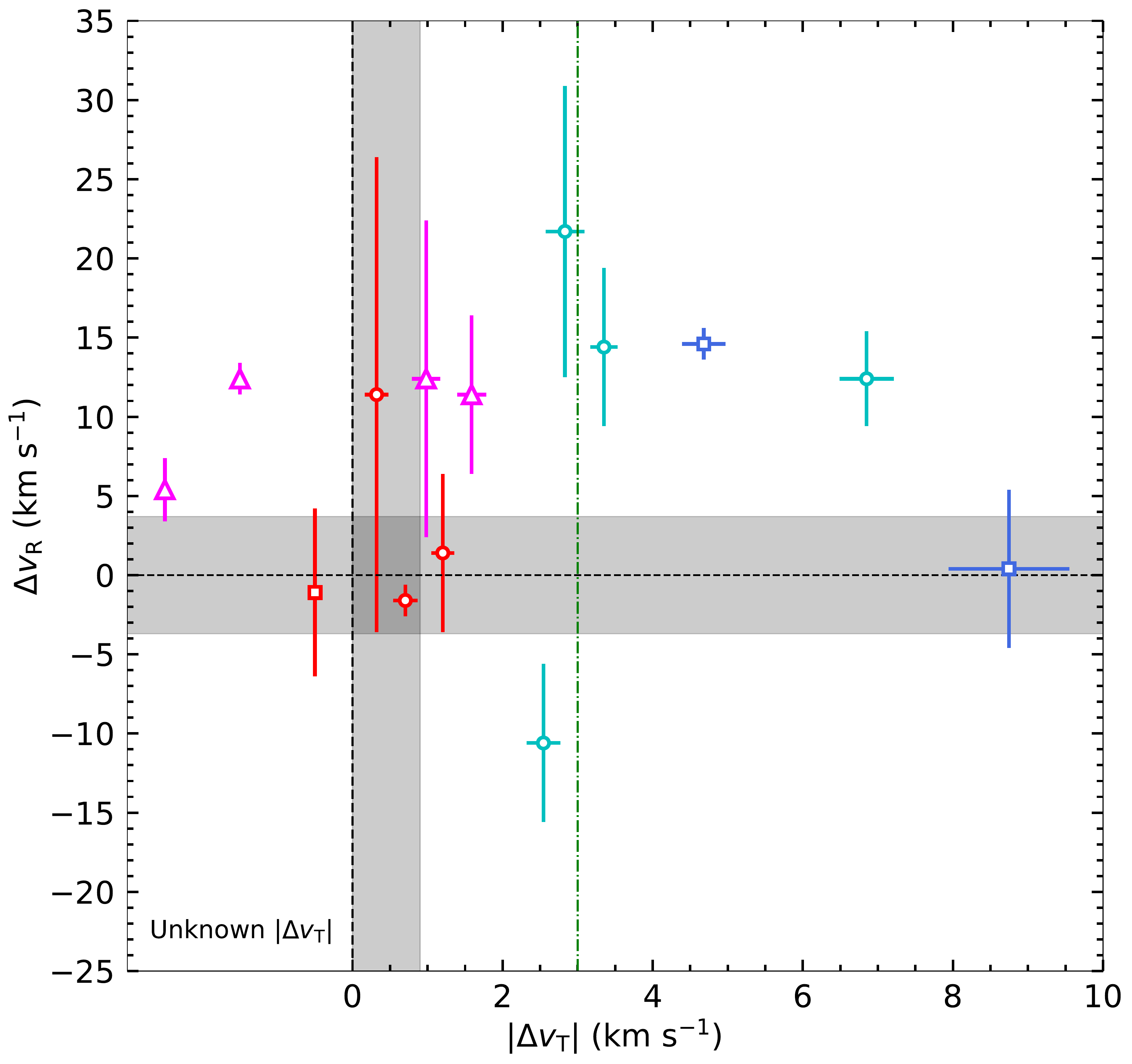}
\end{minipage}
\caption{Kinematic properties of the O stars in the vicinity of NGC6611. \textit{Left:} Position and relative proper motion of the O stars near NGC6611 overlapped on the DSS2 B, R and I colour image. The O stars shown in blue and cyan have velocities comparable or greater than the escape velocity and move away from the centre of NGC6611. The O stars shown in magenta do not have this clear transverse motion away from the centre of NGC6611, but could have had dynamical interactions in the past (see Section~\ref{sec:res_NGC6611_center}. The O stars in red have no evidence for dynamical interactions in the past. The angular distance equivalent to 3 pc is given for a distance of 1.70 kpc. We highlight the centre of NGC6611 with the dashed white circle (r = 0.05 deg). \textit{Right}: The relative (systemic) radial velocity $\Delta v_{\rm{R}}$ as a function of the absolute relative transverse velocity |$\Delta v_{\rm{T}}$| for the O stars in NGC6611, coloured similarly to the \textit{left} panel. The O stars for which we have not used the \textit{Gaia} astrometry (due to \texttt{ruwe} $>$ 1.4) and thus have unknown |$\Delta v_{\rm{T}}$| are shown on the left. The uncertainty on the radial velocity of NGC6611 determined from the B0 to B4 stars is shown with the horizontal black shaded region. The median outward velocity determined by \citet{Kuhn2019} for NGC6611 is shown with the vertical black shaded region.}
\label{fig:dyn_int}
\end{figure*}

\begin{figure}
\centering
\includegraphics[width=1.0\columnwidth]{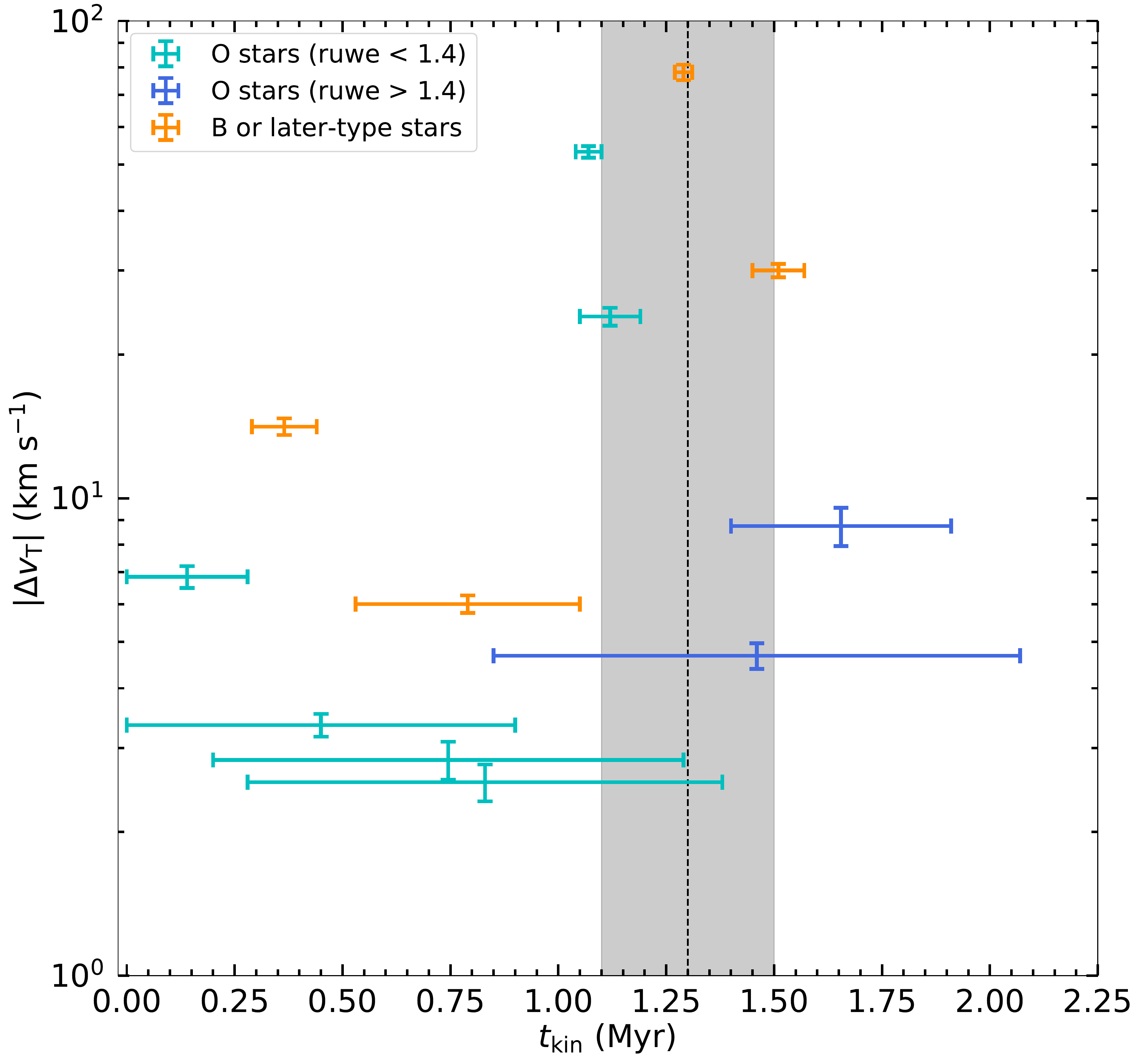}
\caption{Relative transverse velocity as a function of the determined range in kinematic age. We show the best-fit age and uncertainty determined with isochrone fitting with the black dashed line and shaded region respectively. The runaway O stars are shown in blue and cyan, while the runaway B or later-type stars are shown in orange. The O stars are coloured similarly as in Figure~\ref{fig:dyn_int}.}
\label{fig:vT_tkin}
\end{figure}

For a more complete view on the importance of dynamical interactions for massive stars in NGC6611, we turn to the O stars. The majority of the O stars in NGC6611 have been identified and spectrally classified \citep{Evans2005,Sana2009,Sota2011,Sota2014,MaizApellaniz2016}. The binary properties for part of the O stars have been well studied through radial velocity variations in optical spectra \citep{Sana2009,Sana2012}. Systemic radial velocities have been determined for several binary O star systems, allowing us to study the transverse and radial motion for most of the single and binary O star systems. The systemic radial velocities have been collected from the literature where possible. We list the O stars, along with the found runaways, in Table~\ref{tab:Ostars_runaways} and give their spectral type. We also give the absolute $\Delta v_{\rm{T}}$ and calculate the uncertainty using the individual parallax and assuming Gaussian uncertainties on the proper motion of the individual system and NGC6611.

The relative transverse motion ($\Delta v_{\rm{T}}$) of the O stars in the vicinity of NGC6611 is shown in the left panel of Figure~\ref{fig:dyn_int}. The length of each arrow is proportional to the $\Delta v_{\rm{T}}$ of the star. Two runaway O stars are missing as they are outside this image. In the right panel of Figure~\ref{fig:dyn_int} we show the relative radial velocity $\Delta v_{\rm{R}}$ as a function of their $|\Delta v_{\rm{T}}|$ with respect to the average of NGC6611, coloured and marked similarly as in the left panel. The grey dashed lines and shaded regions indicate respectively the mean and standard deviation on $\Delta v_{\rm{R}}$ determined in Section~\ref{sec:ngc6611_members}. For $|\Delta v_{\rm{T}}|$, we indicate the median outward velocity from \citet{Kuhn2019} with the grey shaded region. The runaway lower limit on $|\Delta v_{\rm{T}}|$ of 3 km s$^{-1}$ is shown with the green dash-dotted line and is comparable to the escape velocity for a typical runaway. The three O star systems BD--13$^{\circ}$ 4923, BD--13$^{\circ}$ 4929 and HD 168076, have poor quality astrometry for which we only show the (systemic) radial velocity on the left side. We neglect the expansion velocity in the radial direction, since the uncertainty on the average cluster radial velocity is dominated by the formal error of 3.7 km s$^{-1}$, compared to the average expansion velocity of $\sim$ 0.9 km s$^{-1}$ in the transverse direction.

Several other O stars near the centre of NGC6611 show a receding motion away from the centre of NGC6611, highlighted in cyan. This includes the two runaway O stars already identified in Section~\ref{sec:res_runaway}, BD--13$^{\circ} 4927$ and LS IV --13 14. On top of this, two other O star systems show a similar receding motion. These are the O7 V + O8 V binary system HD 168137 with an highly eccentric orbit \citep[$e$ = 0.902 $\pm$ 0.058;][]{Sana2009,LeBouquin2017} and the O8.5 V star W161. Both O stars might only be considered to move away marginally significant (2.6 $\pm$ 0.2 km s$^{-1}$ and 2.8 $\pm$ 0.3 km s$^{-1}$) compared to the 3 km s$^{-1}$ adopted. Complementing the transverse velocities, the (systemic) radial velocities also indicate differential motion at the 2$\sigma$ level. Specifically HD 168137 could turn out to be a runaway moving mostly in the radial direction, if the systemic radial velocity can be constrained better. This could for example be visible as a bowshock in the infrared with \textit{Spitzer}, where strong emission is seen $\sim$ 25$^{\prime\prime}$ from HD 168137 \citep[see Appendix \ref{sec:Bowshocks};][]{Pilbratt1998,Flagey2011}. We have therefore included both O star systems as accelerated stars.

HD 168504 and HD 168183 have \textsc{ruwe} $>$ 1.4 (1.670 and 1.415 respectively) and were therefore excluded from the runaway search. These two systems have parallaxes consistent (within 2 and 1$\sigma$, respectively) with that of NGC6611 (see Figure~\ref{fig:members_astrometry}). This might indicate that the astrometric solution and thus the proper motion may not be of poor quality \citep[see e.g.][]{MaizApellaniz2021}. HD 168504 is an O7.5V(n)z star; HD 168183 is an O9.5 III star in an eclipsing binary system with a mid-type B star \citep{Sana2009, MaizApellaniz2016}. These two systems are shown with the blue squares in Figure~\ref{fig:dyn_int} and have $|\Delta v_{\rm{T}}|$ $>$ 4 km s$^{-1}$. While we could attribute this significantly deviating $\Delta v_{\rm{T}}$ to a poor astrometric solution, the systemic radial velocity of HD 168504 is significantly deviating ($>$ 3$\sigma$) from the average of NGC6611. HD 168504 and HD 168183 are also located several tenths of degrees from the centre of NGC6611 and show a clear receding motion. The combination of their location, transverse and radial velocities make these two systems clear outliers with respect to the cluster average and we have included them as runaways.

This conclusion is not easily made for BD--13$^{\circ}$ 4929. While the parallax is also consistent within 1$\sigma$ from NGC6611 with \texttt{ruwe} $>$ 1.4, the transverse velocity does not clearly point away from the centre of NGC6611 but instead from the northern part of the nebula (not shown). BD--13$^{\circ}$ 4929 is a hierarchical triple system where the observed O7 V star is in a wider orbit around two early type B stars in an inner orbit. The average radial velocity of BD--13$^{\circ}$ 4929 is 17 $\pm$ 1 km s$^{-1}$ (determined from the O star); this is not the systemic radial velocity and could vary over a longer time scale \citep{Sana2009,Sana2012}. We therefore include BD--13$^{\circ}$ 4929 as a `candidate dynamically interacted O star' in Table~\ref{tab:Ostars_runaways}. For these O stars, we can not draw conclusions on whether these stars have dynamically interacted, but this remains a possibility. See for example Section~\ref{sec:dis_bully_binary} on why HD 168076 could have participated in dynamical interactions.

Other O star systems show no evidence for significantly deviating transverse or (systemic) radial velocities. These systems are BD--13$^{\circ}$ 4923, BD--13$^{\circ}$ 4928, BD--13$^{\circ}$ 4930, HD 168075, HD 168076, W222 and W584, for which both the (systemic) radial velocity and $|\Delta v_{\rm{T}}|$ are consistent (or unknown) with NGC6611. These systems are highlighted in red and magenta. \citet{Sana2009} report systemic radial velocities for BD--13$^{\circ}$ 4923 of 30.7 $\pm$ 4.3 and 3.4 $\pm$ 5.3 km s$^{-1}$ for the primary O4 V((f+)) star and secondary O7.5 V star respectively. A significant difference is present between these two velocities, while they should be equal. The spectral line He II $\lambda$4686 used for the orbital solution can be affected by a stellar wind for early type O stars \citep{Sana2013}. Here, we will assume the 3.4 $\pm$ 5.3 km s$^{-1}$ belonging to the O7.5 V star to be the systemic radial velocity, which is consistent with the average radial velocity of NGC6611. The binary orbit for HD 168075 is solved, but the systemic radial velocity is not reported in the literature. We will adopt a conservative 17 $\pm$ 10 km s$^{-1}$ based on figure 1 in \citet{Barba2010}. W584 is reported to have a bowshock in the infrared as observed with \textit{Spitzer} \citep{Guarcello2010b}. However, the transverse and radial velocity are in reasonable agreement with the average of NGC6611 and we have not included this star as a runaway.

Next, we will quantify the time scale over which the O stars and runaways have been accelerated. In order to do this, we trace back the runaways and determine how long ago these stars were near the centre of NGC6611. Specifically, we calculate the range in time the runaways spent within a radius $r_{\rm{NGC6611}}$ = 0.05 deg from the centre coordinates ($\alpha_{\rm{NGC6611}}$, $\delta_{\rm{NGC6611}}$) = (274.67 deg, --13.78 deg). This radius is the equivalent of 1.5 pc at at distance of $\sim$ 1.70 kpc, but we note that this radius has been conservatively chosen and that the closest approach of almost all runaway stars is within 1.0 pc. We show this radius with the white dashed circle in the left panel of Figure~\ref{fig:dyn_int}. The only exception is HD 168183, for which we use a radius of 0.08 deg.

Since both the proper motion of the star and NGC6611 have uncertainties, we resort to Monte-Carlo simulations. We neglect the radial motion and uncertainty on the initial position. We assume the proper motion of the star and NGC6611 to follow Gaussian distributions with means equal to the observed proper motions and standard deviations equal to the observed proper motion errors (as determined in Section~\ref{sec:ngc6611_members} for NGC6611). In each iteration, we draw proper motions following these Gaussian distributions for the star and NGC6611. We consequently calculate the times when the star enters and leaves the circle with $r_{\rm{NGC6611}}$ = 0.05 deg for each iteration, giving two respective distributions over all iterations. We determine the time when the star enters and leaves this circle as the 14$^{\rm{th}}$ and 86$^{\rm{th}}$ percentile over their respective distribution from 1000 iterations, which respectively give the lower and upper bound on the kinematic age $t_{\rm{kin}}$. We list the range in $t_{\rm{kin}}$ for each accelerated star in Table~\ref{tab:Ostars_runaways}.

To visualise the kinematic ages, we show them as a function of $|\Delta v_{\rm{T}}|$ for each star in Figure~\ref{fig:vT_tkin} coloured similarly as in Figure~\ref{fig:dyn_int}. The orange data indicate the B or later-type runaways. The kinematic ages are better constrained for stars with higher relative transverse velocities, as they spent less time in NGC6611 when traced back, as expected. We also show the age and uncertainty of the young population, 1.3 $\pm$ 0.2 Myr, with the dashed black line and shaded region, respectively. The kinematic ages of the runaways with $|\Delta v_{\rm{T}}|$ $\gtrsim$ 20 km s$^{-1}$ are in excellent agreement with this isochronal age. The kinematic ages of the accelerated stars with $|\Delta v_{\rm{T}}|$ $\lesssim$ 20 km s$^{-1}$ span a larger range between $\sim$ 0.0 and 2.0 Myr. This shows that runaways might not have been ejected instantaneously and that ejections could even happen one Myr after the formation of the cluster.

\section{Discussion}
\label{sec:discussion}
We have studied the young massive cluster NGC6611 in the Eagle Nebula (M16). NGC6611 hosts a rich early type stellar population and formed at least 19 O stars. From the CAMD, we find evidence for two populations of stars. The younger population has an average $A_{\rm{V}}$ = 3.5 $\pm$ 0.1 mag and an age of 1.3 $\pm$ 0.2 Myr. The older population has an average $A_{\rm{V}}$ = 2.0 $\pm$ 0.1 mag and an age of 7.5 $\pm$ 0.4 Myr. These two populations show a clear difference in spatial distribution, with the younger population concentrated towards the centre of NGC6611, while the older population is spatially more extended. 

We have searched for runaways with $|\Delta v_{\rm{T}}| >$ 3 km s$^{-1}$ and find 4 O stars and 4 B or later type stars which have been ejected in the last 1.6 Myr. Dynamical interactions between stars in the younger population are the prime mechanism to explain these runaways. We have studied the kinematics of the O stars in NGC6611 and show how at least nine out of 19 O stars have velocities comparable to or greater than the escape velocity of the cluster ($\sim$ 3 km s$^{-1}$). These runaway O stars can all be traced back to the centre region of NGC6611 and have a kinematic age between 0.0 and 2.0 Myr, which is within the uncertainties less than or comparable to the age of the younger population.

\begin{table}
\centering
\caption{Physical parameters of NGC6611.}
\label{tab:NGC6611params}
\begin{tabular}{l l}
\hline
\hline \vspace{1mm}
Right Ascension ($\alpha$) & 274.67 $\pm$ 0.01 deg \\  \vspace{1mm}
Declination ($\delta$) & --13.78 $\pm$ 0.01 deg \\ \vspace{1mm}
Proper motion ($\mu_{\alpha^{*}}$) & 0.21 $\pm$ 0.01 mas yr$^{-1}$ \\ \vspace{1mm}
Proper motion ($\mu_{\delta}$) & --1.59 $\pm$ 0.01 mas yr$^{-1}$ \\ \vspace{1mm}
Radial velocity ($v_{\rm{R}}$) & 4.7 $\pm$ 3.5 km s$^{-1}$\\ \vspace{1mm}
1D velocity dispersion$^{(1)}$ ($\sigma_{\rm{1D}}$) & 1.8 $\pm$ 0.2 km s$^{-1}$ \\ \vspace{1mm}
Median outward velocity$^{(1)}$ ($v_{\rm{out}}$) & 0.90 $\pm$ 0.23 km s$^{-1}$ \\ \vspace{1mm}
Parallax ($\varpi$) & 0.587 $\pm$ 0.003 mas \\ \vspace{1mm}
Distance ($d$) & 1.706 $\pm$ 0.008 kpc \\ 
\hline
\multicolumn{2}{c}{Young population}\\
\hline \vspace{1mm}
Mass & 5-10 $\times$ $10^{3}$ M$_{\odot}$\\ \vspace{1mm}
Half-mass radius ($r_{\rm{hm}}$) & 1.5-2.0 pc \\ \vspace{1mm}
Age & 1.3 $\pm$ 0.2 Myr \\ \vspace{1mm}
Extinction ($A_{\rm{V}}$) & 3.5 $\pm$ 0.1 mag \\ 
\hline
\multicolumn{2}{c}{Old population} \\
\hline \vspace{1mm}
Mass & - \\ \vspace{1mm}
Half-mass radius & 2.5-3.0 pc \\ \vspace{1mm}
Age & 7.6 $\pm$ 0.4 Myr \\ \vspace{1mm}
Extinction ($A_{\rm{V}}$) & 2.0 $\pm$ 0.1 mag \\
\hline
\hline
\end{tabular}
\tablebib{(1) \citet{Kuhn2019}.}
\end{table}

\subsection{Physical parameters of NGC6611}
\label{sec:dis_ngc6611_params}
Before further discussing our results, we will consider some of the key physical parameters of NGC6611. These parameters include the astrometry, (initial) mass function, mass, radius, stellar multiplicity and age.

We summarise the astrometric parameters of NGC6611 in Table~\ref{tab:NGC6611params}, where we give the centre coordinates, proper motion, distance and radial velocity determined in Section~\ref{sec:ngc6611_members}. We also list $\sigma_{\rm{1D}}$ and the median $v_{\rm{out}}$ adopted from \citet{Kuhn2019}. 

We will assume a Salpeter \citep{Salpeter1955} or Kroupa \citep{Kroupa2001} distribution for the initial masses. The Kroupa initial mass function is also used to define the stellar occupation along the isochrones. With the initial mass function, we can roughly estimate the mass of NGC6611. If we assume the sample of 19 O stars (with a conservative M $>$ 18 M$_{\odot}$) to be mostly complete, we determine the mass of NGC6611 to be $\sim$ 8.9 $\times$ $10^{3}$ or 5.4 $\times$ $10^{3}$ M$_{\odot}$ for a Salpeter and Kroupa initial mass function, respectively, between 0.08 and 100 M$_{\odot}$. This mass is still uncertain as we are dealing with small-number statistics, possibly excluding unclassified stars and undetected O stars. If the 7 O stars which have not been investigated for the prescence of a companion have an O star companion, we determine a mass of $\sim$ 7-12 $\times$ $10^{3}$ M$_{\odot}$. Similarly, including all spectrally classified B0-2 stars consistent with the distance of NGC6611 yields a mass that agrees with the aforementioned values \citep[assuming a 50\% binary fraction;][]{Banyard2021}. The mass of NGC6611 is estimated in the literature at $\sim$ 2 to 25 $\times$ $10^{3}$ M$_{\odot}$, taking into account the $\sim$ 170 early type B stars \citep{Bonatto2006,Wolff2007,PortegiesZwart2010}. We note that if the mass is 20 $\times$ $10^{3}$ M$_{\odot}$, we would expect to find $\gtrsim$ 67 O stars, for which we find no evidence. We will consider a mass in the range of 5-10 $\times$ $10^{3}$ M$_{\odot}$, which is mostly sensitive to the adopted initial mass function. Estimating the mass for the older population is problematic. We have found 21 stars (22 with the B2.5 I) belonging to the old population, which brings large uncertainties due to small number statistics. On top of this, the older population is spatially more extended, which could result in a bias against identification of cluster membership. In summary, we can not give a reliable estimate for the mass of the older population and leave this for future work.

Several definitions exist for the `radius' of a cluster, including the half-mass radius, effective radius, virial radius and core radius \citep{PortegiesZwart2010}. We will use the half-mass radius $r_{\rm{hm}}$ here. Using all spectrally classified stars consistent with the distance of NGC6611 we determine the current projected $r_{\rm{hm}}(\rm{proj})$ to be $\sim$ 2.0 pc from a weighted average, where the weights are taken to be the mass from \citet{Peacut2013}. Using different sub-samples, such as the O stars or B stars, gives similar results in the range of 2-2.5 pc for $r_{\rm{hm}}(\rm{proj})$. Estimates for the half-light radius, taking into account the young population of pre-main-sequence stars, similarly give results in the range of $r_{\rm{hm}}(\rm{proj})$ of 2-2.5 pc. This is the projected 2D radius. Assuming for simplicity a Plummer density distribution \citep{Plummer1911} this corresponds to a 3D  $r_{\rm{hm}}$ of 0.77$r_{\rm{hm}}(\rm{proj})$ $\sim$ 1.5-2.0 pc.

Most massive stars are in binaries or higher-order systems. This is also the case for NGC6611, where 5 out of 9 systems studied in depth for the prescence for a companion are binary or higher-order systems \citep{Sana2009,Sana2014}. We should therefore consider binary fraction percentages between 50--100\%. While Table~\ref{tab:Ostars_runaways} contains 10 spectrally classified `single' O stars, 7 of these have not been investigated for the prescence of a companion and could still hide one. HD 168504 is an O7.5V(n)z star and is presumed to be single, although the \texttt{ruwe} $>$ 1.4 gives a first indication of a possible binary system \citep{MaizApellaniz2016,Belokurov2020}. \textit{Gaia} indicates that W161 is separated from another source by only $\sim$ 3$^{\prime\prime}$. W161 could possibly be a binary with the radial velocity measured by \citet{Evans2005} and \citet{Martayan2008} deviating by 16 km s$^{-1}$, although consistent within 2$\sigma$ considering the uncertainties. W161 is also relatively bright in the near-infrared for an O8.5 V star, with a \textit{2MASS} K$_{\rm{s}}$ magnitude brighter than the O7 V + O8 V system HD 168137, despite being more reddened. Similarly W584 located in the north-west part of the nebula also shows another source separated by $\sim$ 2$^{\prime\prime}$ and is only $\sim$ 2.5 mag fainter in the G-band. In summary, the binary properties of the O stars in NGC6611 have been well studied for the apparently brightest O stars, mostly located near the centre of NGC6611, but remains incomplete when considering the late type and runaway O stars. Thus, in order to properly determine the high-mass end of the IMF, the OB runaways have to be accounted for.

\subsection{The ages of the two populations of NGC6611}
The CAMD has revealed two populations of stars, specifically for stars fainter than $M_{G} \gtrsim$ 3 mag. Not only are the estimated ages of the two populations different with 1.3 $\pm$ 0.2 Myr compared to 7.5 $\pm$ 0.4 Myr for the young and old population, also the average extinction is different. The younger and older population require on average 3.5 $\pm$ 0.1 mag and 2.0 $\pm$ 0.1 mag extinction in the V-band. This could imply that the older population is located in front of the younger population. If so, the difference in distance would be small ($\lesssim$ 20 pc), as we find no deviations in the parallax, not taking into account that we have partially selected members from the parallax.  

Another difference between these two population is their spatial distribution. The young population has a half-light radius of $\sim$ 2 pc, while the old population has a half-light radius of $\sim$ 3.5 pc, corresponding to a 3D $r_{\rm{hm}}$ (assuming the half-light and half-mass radius are similar) of 1.5 and 2.7 pc, respectively. The fact that the size of the older population is larger than that of the younger population agrees with the expectation that the clusters are still expanding.

Assuming a cluster mass of 7.5 $\times$ 10$^{3}$ M$_{\sun}$ and a mean stellar mass of 0.5 M$_{\sun}$ we derive half-mass relaxation times on the order of $\sim$ 100 Myr, the time it takes for a cluster to reach virial equilibrium. This is much longer than the age of both populations, so the cluster is not yet in hydrodynamical equilibrium (unless the initial $r_{\rm{hm}}$ of the cluster was much smaller than the present radius and on the order of $\sim$ 0.2 pc). From Figure~\ref{fig:redblue_astrometry} the increase in the projected half light radius is about $\sim$ 0.8 pc Myr$^{-1}$ and 0.3 pc Myr$^{-1}$ for the young and old population respectively, corresponding to an expansion of $\sim$ 0.8 and 0.3 km s$^{-1}$, respectively. This expansion velocity for the young population is consistent with that measured by \citet{Kuhn2019} of 0.90 $\pm$ 0.23 km s$^{-1}$. We can also estimate the dynamical timescale $t_{\rm{dyn}}$, the time it takes a typical star to cross the cluster. With a cluster mass of 7.5 $\times$ 10$^{3}$ M$_{\odot}$ and 3D $r_{\rm{hm}}$ of 1.5 and 2.7 pc, we find $t_{\rm{dyn}}$ $\sim$ 0.4 and $\sim$ 0.7 Myr for the young and old population respectively. The $t_{\rm{dyn}}$ of 0.4 Myr for the young population is similar to the time of minimum radius found in Figure~\ref{fig:redblue_astrometry}. This indicates that we should not blindly trust the radius given in Figure~\ref{fig:redblue_astrometry} at a time of --0.4 Myr or earlier. The motions of the stars could have changed direction or crossed the cluster on longer timescales. The $t_{\rm{dyn}}$ of 0.7 Myr for the old population differs from the time of minimum radius (0.4 Myr), but this is not surprising considering the uncertainties involved as mentioned in Section~\ref{sec:dis_ngc6611_params}. The calculation for $t_{\rm{dyn}}$ also assumes a virialised cluster, which may not be the case here.

The age of the young population in NGC6611 can also be derived from the astrometry. The kinematics of the runaways and O stars show that dynamical interactions happened over the last $\sim$ 2 Myr. The four runaways with $| \Delta v_{\rm{T}}| >$ 20 km s$^{-1}$ specifically have been kicked out 1.0 to 1.6 Myr ago. This is in excellent agreement with the isochronal age of the young population. We expect the dynamical interactions to take place right after star formation \citep{Fujii2011}.

For the stars not included in the blue or red group, only one star shows evidence of belonging to the old population. The B2.5 supergiant BD--13$^{\circ}$ 4912 shows a clear deviation from the young population in Figure~\ref{fig:Isochrones}. BD--13$^{\circ}$ 4912 has been used as an argument in favour of an older age or noncoevality for NGC6611 as a whole, with its evolutionary status implying an age of $\sim$ 6 Myr \citep{Hillenbrand1993,Gvaramadze2008}. Similar to \citet{Hillenbrand1993}, we find no evidence for other bright stars belonging to the old population. The O stars earlier than O7.5-8.5 should have already experienced their supernova or are considerably evolved if they were to belong to the old population. The O and B stars are almost all spectrally classified as dwarfs and show a clear spatial clustering towards the centre of NGC6611, similar to the young population. We therefore deem it justified to consider the majority of the early type stars as part of the young population.

The presence of an older population in NGC6611 has been suggested before. \citet{Guarcello2007a,Guarcello2007b,Guarcello2009,Guarcello2010b} report an over-density of blue stars in the optical which also display a near-infrared excess indicative of circumstellar discs. These `blue stars with excess' (similar to our `blue group' but located at $M_{\rm{G}}$ $\sim$ 7 mag) were initially attributed to stars in the young population possessing a circumstellar disc. When viewing these stars at a certain inclination, the scattering of light from the disc could make these stars appear bluer \citep{Guarcello2010b,Bonito2013}. However, \citet{DeMarchi2013} show that only a few percent of the stars in the young population are subject to the aforementioned effect and that most of the blue stars with excess indeed make up an older population with an uncertain age of 8 to 32 Myr. Other evidence supporting the presence of an older population comes from the X-ray luminosity, lithium absorption lines and their optical magnitudes \citep{DeMarchi2013}. \citet{DeMarchi2013} also find a significantly different spatial distribution for the older population, consistent with our spatially extended distribution for the blue group in Figure~\ref{fig:redblue_astrometry}.

Several theories exist for the formation of a cluster with multiple populations of stars. For example, the Orion nebula cluster is now observed to have three different populations of stars differing in age by $\sim$ 0.5 to 1 Myr each \citep{Beccari2017,Jerabkova2019}. If the first generation of stars (observed as the oldest) formed several OB stars, these stars could halt star formation through their ionising photons and strong stellar winds. If these OB stars are all ejected by dynamical interactions, the remaining gas in the molecular cloud can be allowed to collapse again and form a second generation, also including O stars \citep{Kroupa2018}. This is likely not the case for NGC6611, since it would require the two populations to differ in age by $\lesssim$ 1 Myr, while we find $\gtrsim$ 6 Myr. The younger population also appears to have formed significantly more (O) stars than the older population, while this is opposite to what is observed in the Orion Nebula Cluster \citep{Jerabkova2019}.

Another scenario is that a nearby older population of stars was captured by the younger population as it was experiencing cloud-collapse \citep{PflammAltenBurg2007}. Stars present in the field, which could have been formed by earlier star formation, could be gravitationally bound to the molecular cloud that formed NGC6611 if their paths cross. The Sagittarius Spiral Arm has seen star formation sites (see Figure~\ref{fig:runaway_allwise}), with M17, M16 and NGC6604 visible as the giant \HII regions to mention a few. The capture of an older less massive cluster should therefore not be excluded. Similarly, if the older population passed by a different molecular cloud, a supernova could have triggered the molecular cloud to collapse and form the younger population \textcolor{blue}{(Guo et al. in prep)}. The difference in ages between the young and old population of $\sim$ 6 to 7 Myr would be consistent with the core-collapse supernova of a massive star initially belonging to the old population. More detailed analysis is needed on the older population before conclusions can be drawn on the formation mechanism of the younger population.

\subsection{Dynamical interactions in NGC6611}
We now turn to the kinematics of the NGC6611 runaways and O stars. We have found a total of 9 O and 4 B or later-type stars with velocities comparable to or greater than the escape velocity for NGC6611 ($|\Delta v_{\rm{T}}|$ $\gtrsim$ 3 km s$^{-1}$). Since the young population is $<$ 2 Myr old, we do not expect core-collapse supernova to have happened yet, making the dynamical ejection scenario the only feasible mechanism. The runaways coming from the centre of NGC6611 are in line with this assumption as dynamical interactions are expected to be most prominent in the densest regions of young star clusters \citep{Fujii2011}. Out of the remaining 10 O stars, 5 O stars could have had dynamical interactions in the past based on their kinematic properties, but this remains inconclusive. The 5 O stars for which we find no evidence for acceleration or dynamical interactions contain three late type O stars (O9 V, O9.5 Vp and O9.5 V) and an O4 V((f+)) + O7.5 V binary system for which we have poor quality astrometry. If we assume that the runaway O stars were created through dynamical interactions, we should consider that roughly $\gtrsim$ 50\% of the O stars underwent dynamical interactions.

It is unclear if these percentages are to be expected for clusters such as NGC6611. Simulations and models of young massive clusters have been performed to study the physical properties of runaways and the dynamical evolution of clusters \citep{Fujii2011,Oh2012,Fujii2014,Oh2015,Oh2016}. We have found two runaways with $M$ $\gtrsim$ 15 M$_{\odot}$, which is consistent with the results of \citet{Fujii2011}. For example, their figure 3 predicts a runaway fraction of 0.05--0.2 for a cluster mass of 6 $\times$ $10^{3}$ M$_{\odot}$ compared to our runaway fraction of $\sim$ 0.1 ($v_{\rm{T}}$ $\gtrsim$ 20 km s$^{-1}$). Our results are also in agreement with their figure 6, where we find that runaways with relatively low $|\Delta v_{\rm{T}}|$ are more common than runaways with relatively high $|\Delta v_{\rm{T}}|$. This is similar to what has been found in \citet{Oh2015} and \citet{Oh2016}, where a large fraction of the O stars could be ejected at low velocity ($\lesssim$ 10 km s$^{-1}$) under certain conditions. 

This creates a consistent picture so far; however, one important parameter in these models is the initial half-mass radius $r_{\rm{hm}}$. The initial $r_{\rm{hm}}$ typically assumed in these models is in the range of 0.1-0.5 pc. NGC6611 is a relatively dispersed cluster for which we determine a present-day $r_{\rm{hm}}$ $\simeq$ 1.5-2 pc. We visualise radii of 0.1, 0.5 and 1.5 pc in Appendix \ref{sec:res_NGC6611_center}. Even when considering the expansion of a cluster like NGC6611 and dynamical interactions contributing to this, it is difficult to reconcile this with a small initial $r_{\rm{hm}}$. The O stars can not be traced back to such a small radius, which is similar for the fainter stars. A radius of 0.1-0.5 pc is more typical of dense clusters such as Westerlund 2 and NGC 3603 and it is difficult to fit NGC6611 among these clusters \citep[see e.g.][]{Pfalzner2009,PortegiesZwart2010,Pfalzner2013}.

What so far has not been considered in models of dynamical evolution of clusters is that the massive binary system population initially had wider orbits than what is currently observed. \citet{Ramirez2021} find that the massive binaries harden over time, implying that their orbits shrink as part of the (late) formation process. The time scale over which these binaries harden is $\sim$ 1 to 2 Myr. If the massive binaries indeed form in wider orbits, it would increase the cross-section for dynamical interactions. The orbital period distribution considered in the simulations of young massive clusters often follows the \citet{Sana2012} distribution, which has been derived for clusters that are 2-5 Myr old. 

Westerlund 2, one of the clusters included in \citet{Ramirez2021}, shows a rich history of high velocity ejections ($\gtrsim$ 25 km s$^{-1}$) in the last $\sim$ 1 Myr, with likely 7 O stars being thrown out \citep{Drew2018}. NGC 3603 shows a similar behaviour, with 11 O star ejections at runaway velocity \citep{Drew2019}. Since both have estimated ages of $\lesssim$ 2 Myr, only the dynamical ejection scenario is expected to contribute to the production of runaways. For Westerlund 2, the percentage of O stars observed as runaways is $\sim$ 25\% \citep{Drew2018}. Considering that we see more low velocity runaway O stars, Westerlund 2 and NGC 3603 may have also produced a significant number of low velocity runaway O stars, depending on their respective escape velocity.

\subsection{Bully binary in NGC6611}
\label{sec:dis_bully_binary}
HD 168076 is the brightest visual star in NGC6611/M16 and is spectrally classified as an O4 IV(f) star \citep{Sota2011,MaizApellaniz2016}. \citet{Bosch1999} show a bright companion separated 0.15$^{\prime\prime}$ from the primary star. \citet{Sana2014} also report the prescence of a secondary at this separation. \citet{Sana2009} studied the multiplicity properties of HD 168076 with high-resolution optical spectra and find no significant radial velocity variations in line with a large binary orbital period. They also assigned a tentative spectral classification of O7.5 V / O9 III to the secondary.

This makes HD 168076 a massive binary system possibly with a wide binary orbit. Such a massive system could be responsible for a major part of the ejections of other O stars, the so-called `bully binary' scenario \citep{Fujii2011}. Unfortunately, the astrometry of HD 168076 is of poor quality with \texttt{ruwe} = 8.382, and makes it impossible for us to trace its position back in time. If the lower limit on the binary orbital period of $P_{\rm{orb}}$ $\gtrsim$ $10^{5}$ days reported by \citet{Sana2009} is correct, it would suggest that HD 168076 even now has a high cross-section for dynamical interactions \citep[see e.g. figure 2 in][]{Fujii2011}. A recent role of HD 168076 in dynamical ejections could possibly be seen in the acceleration of the two O stars BD$-$13$^{\circ}$ 4927 and LS IV --13 14, which we show in Appendix \ref{sec:res_NGC6611_center}. If HD 168076 was the culprit of the accelerations, $t_{\rm{kin}}$ would be $\lesssim$ 0.3 Myr for both O stars. More investigation is needed on the 3D velocities of HD 168076, LS IV --13 14 and BD$-$13$^{\circ}$ 4927 to draw further conclusions, so it remains inconclusive as to what role HD 168076 played in accelerating the O stars.

\subsection{Runaways}
The two runaway O stars BD--14$^{\circ}$ 5040 and UCAC2 27149134 were also discussed in \citet{Gvaramadze2008}, initially discovered by their bowshock. Relative to NGC6611 the two O stars move in exactly opposite directions and are consistent with coming from the same position near the centre of NGC6611. This could suggest that one of the binaries initially formed in NGC6611 has been disrupted through binary-binary interactions. 

This is similar to the well-known example of the two runaway stars $\mu$ Col and AE Aur which are now thought to have been ejected $\sim$ 2.5 Myr ago in or near the vicinity of the Orion Nebula Cluster \citep{Blaauw1954,Hoogerwerf2001}. \citet{Gualandris2004} showed that a binary-binary interaction involving $\mu$ Col, AE Aur and the binary system $\iota$ Ori could lead to the ejection of the two runaway stars. 

BD--14$^{\circ}$ 5040 and UCAC2 27149134 may have been subject to a similar binary-binary interaction, though we find no evidence for other runaways consistent with coming from this event. We could therefore assume the other stars involved to remain in the vicinity of NGC6611, similar to the case of $\iota$ Ori.

Since we know the spectral type for BD--14$^{\circ}$ 5040, an O5.5 V star, a rough estimate of the mass is $\sim$ 34 $M_{\odot}$ \citep{Martins2005}. To conserve momentum among the two stars, UCAC2 27149134 needs to have a mass $\sim$ 74 $M_{\odot}$. We note that including radial velocities in this calculation would only change the resulting mass by $\sim$ 10\% assuming BD--14$^{\circ}$ 5040 moves towards us with -100 km s$^{-1}$. Even then, a conservative lower limit on the mass would be 60-65 $M_{\odot}$ for UCAC2 27149134.

Similar to \citet{Gvaramadze2008}, we can roughly estimate the spectral type from the \textit{2MASS} JHK$_{\rm{s}}$ magnitudes and the \textit{Gaia} EDR3 parallaxes. UCAC2 27149134 has K$_{\rm{s}}$ = 7.396 $\pm$ 0.023 mag, (J -- H) = 0.41 $\pm$ 0.06 mag and (H -- K$_{\rm{s}}$) = 0.288 $\pm$ 0.06 mag. The distance to UCAC2 27149134 is 1750$^{+75}_{-60}$ pc from the individual parallax. With $R_{\rm{V}}$ = 3.1, we estimate $A_{\rm{V}}$ $\sim$ 5.7 mag. These values agree with \citet{Gvaramadze2008}, however \citet{Martins2006} do not provide a photometric calibration for the \textit{2MASS} passbands, which is noticeable when comparing to \citet{Peacut2013}. For UCAC2 27149134, we estimate a photometric spectral type of O3-4 V, with the dwarf luminosity classification based on the prescence of an O3.5 V((f+)) and O4 V((f+)) star in NGC6611. A spectral type of O3 V would give a rough estimate of $\sim$ 60 $M_{\odot}$ on the mass, in reach of the 60-65 $M_{\odot}$ `lower limit' mentioned earlier. Spectroscopic follow-up on this source may confirm our hypothesis.

Both UCAC2 27149134 and BD--14$^{\circ}$ 5040 produce bowshocks visible in the \textit{AllWISE} data (see Appendix~\ref{sec:Bowshocks}). A third bowshock producing runaway O star analysed in \citet{Gvaramadze2008}, HD 165319, is spectrally classified as an O9.7 Ib star. This star was identified as a runaway coming from NGC6611, for which we find no evidence. HD 165319 is located at a distance of $\sim$ 1.4 kpc and would be even closer in the past considering its radial velocity of $\sim$ 25 km s$^{-1}$. It is thus inconsistent with being at the same distance as NGC6611. The proper motion of HD 165319 indicates that its projected closest approach is about $\sim$ 0.5 deg away from the centre of NGC6611 and requires a kinematic age of $\sim$ 4 to 5 Myr, again arguing against an origin in NGC6611.

\section{Summary \& conclusion}
\label{sec:conclusion}
We have studied the Eagle Nebula and its young massive cluster NGC6611 in detail with \textit{Gaia} EDR3. We summarise our findings below:
\begin{itemize}
    \item We have confirmed 137 members of NGC6611, at a distance of 1706 $\pm$ 8 pc.
    \item NGC6611 includes two populations of stars. We have determined the mean age and extinction with isochrone fitting. The young population has an age of 1.3 $\pm$ 0.2 Myr and a mean extinction of $A_{\rm{V}}$ = 3.5 $\pm$ 0.1 mag. The old population has an age of 7.5 $\pm$ 0.4 Myr and a mean extinction of $A_{\rm{V}}$ = 2.0 $\pm$ 0.1 mag. Almost all OB stars are most likely belonging to the young population. Only the B2.5 I star BD--13$^{\circ}$ 4912 may be associated with the old population.
    \item We have found 8 runaways ($|\Delta v_{\rm{T}}| >$ 3 km s$^{-1}$) coming from NGC6611, of which 4 are O stars and 4 are B or later-type stars. These runaways have kinematic ages consistent with being ejected from the young population.
    \item Investigating the kinematics of all O stars present in NGC6611 in detail shows that at least 9 out of 19 O stars have velocities comparable to or greater than the escape velocity of the cluster ($\sim$ 3 km s$^{-1}$). With the dynamical interactions as the sole ejection mechanism given the young age of the cluster, we show that the accelerated O stars and runaways have kinematic ages in agreement with the young population.
    \item If massive binaries are initially formed with wider separations than what is currently observed, it will increase the cross-section for dynamical interactions. The percentage of escaping O stars in NGC6611 ($\gtrsim$ 45\%) is difficult to reconcile with n-body simulations of young massive clusters. If these simulations account for an initially wide massive binary systems, it may bring their runaway fraction in line with our results.
\end{itemize}
The number of young massive clusters younger than $\sim$ 1 Myr is extremely limited. Runaway stars can give valuable insight into the early evolution of young massive clusters and massive binaries as shown here. Extending these results to other young massive clusters can give a larger sample-size to study the importance of dynamical interactions and the initial massive binary population.
\begin{acknowledgements}
The data and data products presented in this paper are available at the following DOI: \url{http://doi.org/10.5281/zenodo.6372280}. MS acknowledges support from NOVA. SR acknowledges funding from the European Research Council Horizon 2020 research and innovation programme (Grant No. 833925, project STAREX). This research has made use of NASA’s Astrophysics Data System, the SIMBAD database, operated at CDS, Strasbourg, France, the "Aladin sky atlas" developed at CDS, Strasbourg Observatory, France.

This work has made use of data from the European Space Agency (ESA) mission
{\it Gaia} (\url{https://www.cosmos.esa.int/gaia}), processed by the {\it Gaia}
Data Processing and Analysis Consortium (DPAC,
\url{https://www.cosmos.esa.int/web/gaia/dpac/consortium}). Funding for the DPAC
has been provided by national institutions, in particular the institutions
participating in the {\it Gaia} Multilateral Agreement.
\end{acknowledgements}

\clearpage

\bibliographystyle{aa}
\bibliography{REFERENCES.bib}

\clearpage

\begin{appendix}
\section{Bowshocks of runaways}
\label{sec:Bowshocks}
\begin{figure}[h]
\centering
\includegraphics[width=1.0\columnwidth]{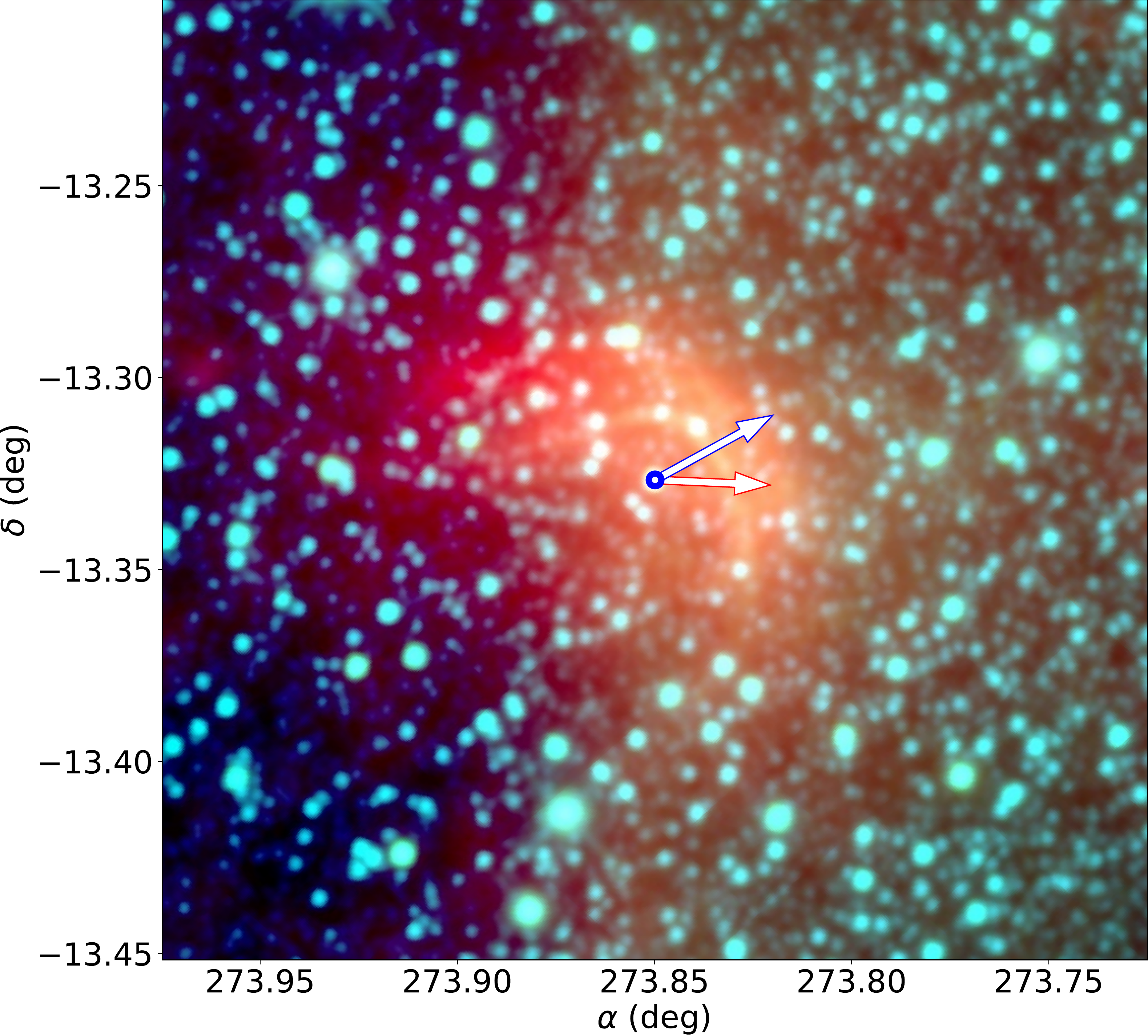}
\caption{\textit{AllWISE} (W1, W2 and W4 as blue, green and red, respectively) false-colour image of the bowshock produced by UCAC2 27149134 (18$^{\prime}$ by 18$^{\prime}$). The blue circle gives the position of the star. The red arrow gives the absolute transverse motion of the star, while the blue arrow gives the transverse motion relative to the cluster.}
\label{fig:bowshock_UCAC}
\end{figure}

\begin{figure}[h]
   \centering
   \includegraphics[width=0.99\linewidth]{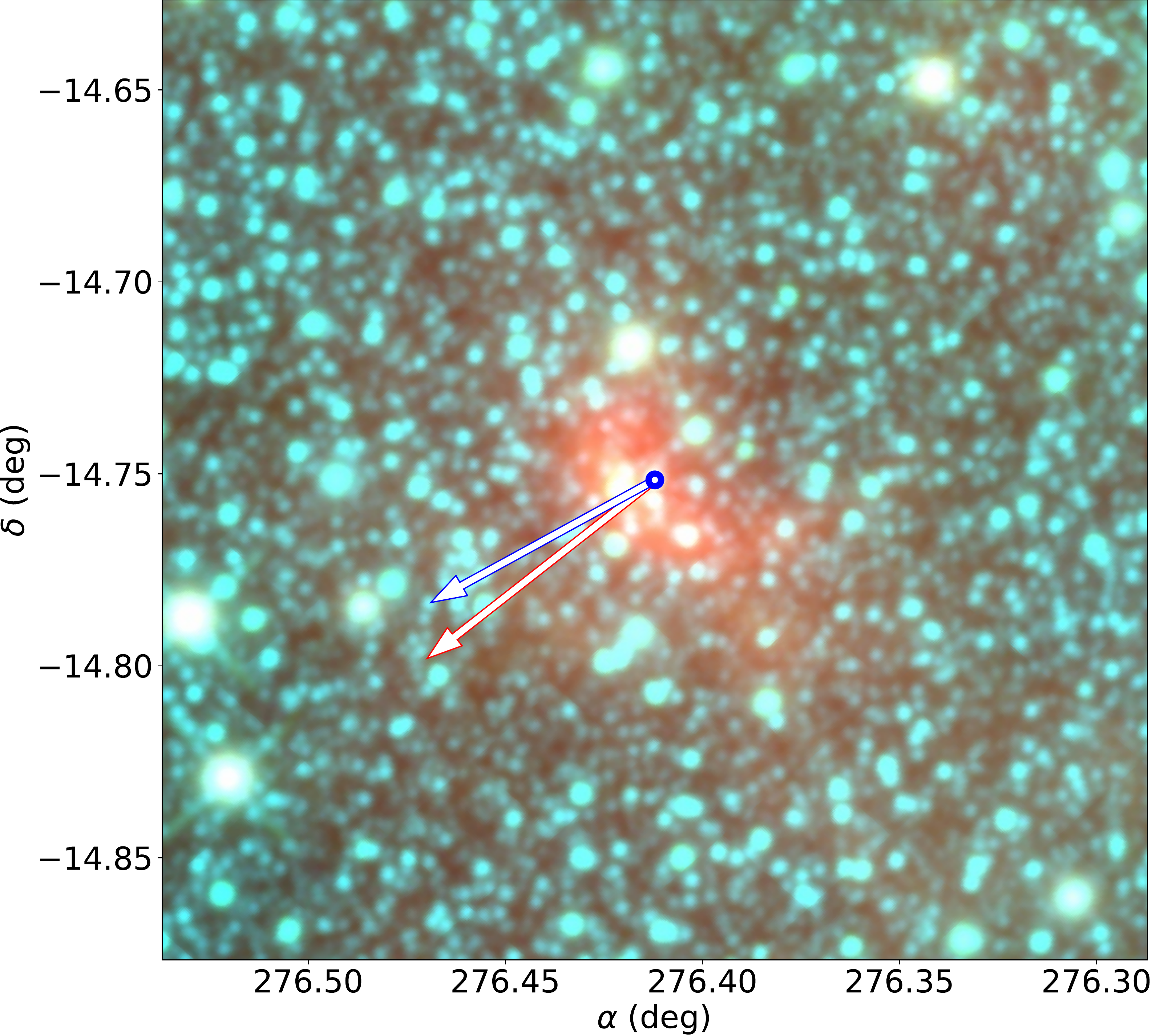}
      \caption{\textit{AllWISE} (W1, W2 and W4 as blue, green and red, respectively) false-colour image of the bowshock produced by BD--14$^{\circ}$ 5040 (18$^{\prime}$ by 18$^{\prime}$). The blue circle gives the position of the star. The red arrow gives the absolute transverse motion of the star, while the blue arrow gives the transverse motion relative to the cluster.}
         \label{fig:bowshock_BD}
\end{figure}

\begin{figure}[h]
   \centering
   \includegraphics[width=0.99\linewidth]{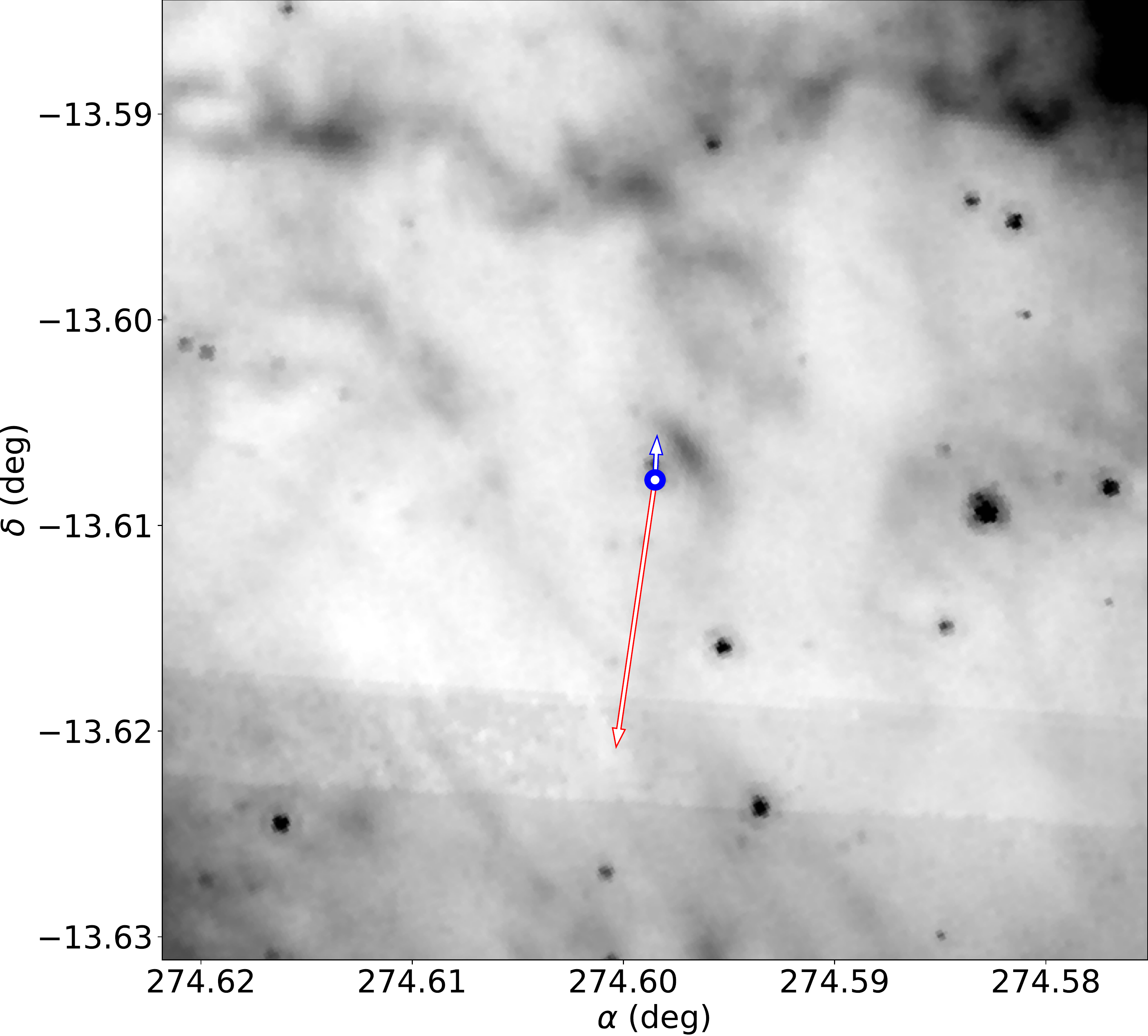}
      \caption{\textit{Spitzer} IRAC 8.0 $\mu$m image of W584 (2.8$^{\prime}$ by 2.8$^{\prime}$)
      The blue circle gives the position of the star. The red arrow gives the absolute transverse motion of the star, while the blue arrow gives the transverse motion relative to the cluster.}
         \label{fig:bowshock_HD168137}
\end{figure}

\begin{figure}[h]
   \centering
   \includegraphics[width=0.99\linewidth]{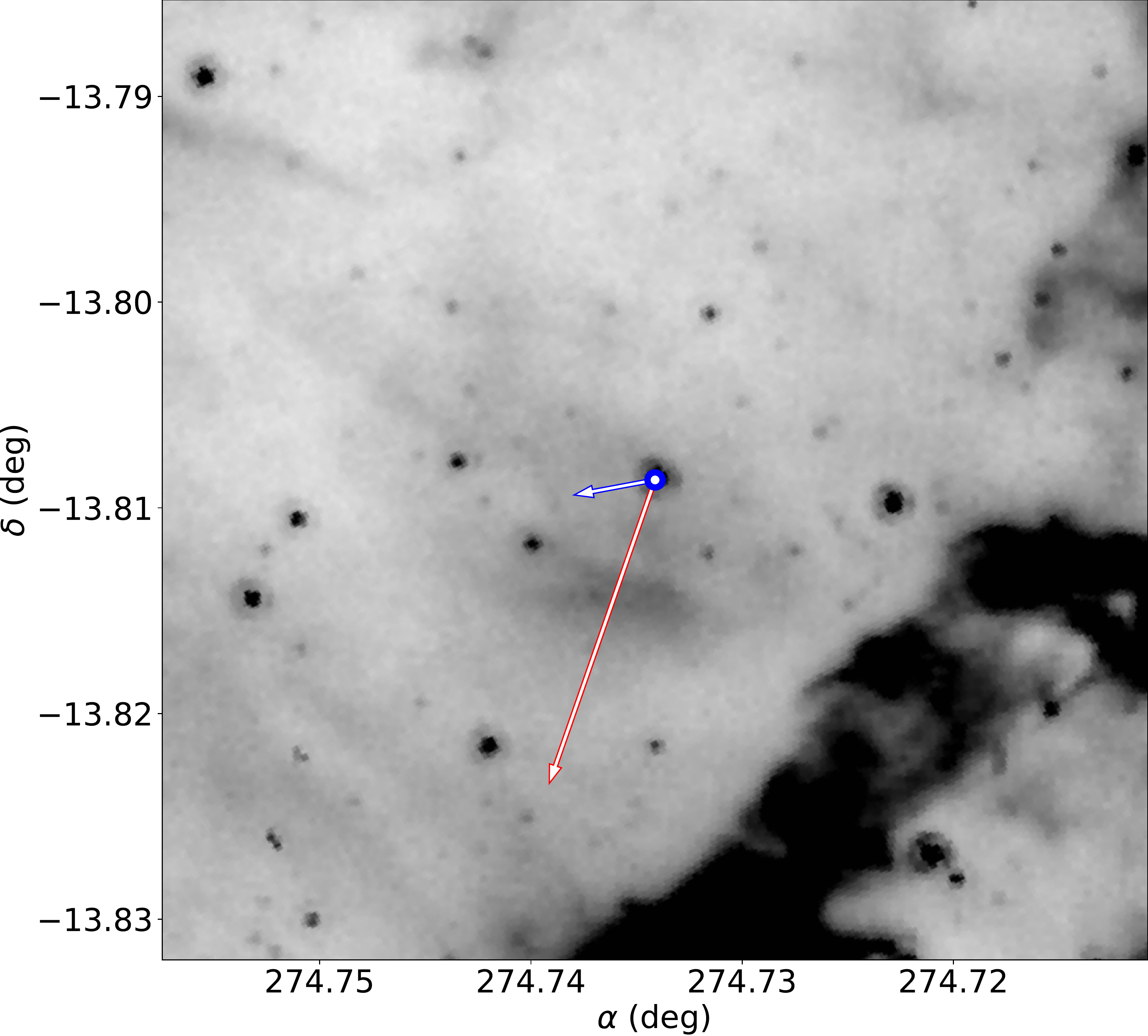}
      \caption{\textit{Spitzer} IRAC 8.0 $\mu$m image of HD 168137 (2.8$^{\prime}$ by 2.8$^{\prime}$)
      The blue circle gives the position of the star. The red arrow gives the absolute transverse motion of the star, while the blue arrow gives the transverse motion relative to the cluster.}
         \label{fig:bowshock_W584}
\end{figure}

The runaway O stars which have visible bowshocks in the \textit{AllWISE} imaging are UCAC2 27149134 shown in Figure~\ref{fig:bowshock_UCAC} and BD--14$^{\circ}$ 5040 in Figure~\ref{fig:bowshock_BD}. W584 and HD 168137 are reported to have possible bowshocks in the \textit{Spitzer} images which we show in Figure~\ref{fig:bowshock_W584} and Figure~\ref{fig:bowshock_HD168137}.

\clearpage

\section{The centre of NGC6611}
\label{sec:res_NGC6611_center}

\begin{figure*}
\centering
\includegraphics[width=1.0\textwidth]{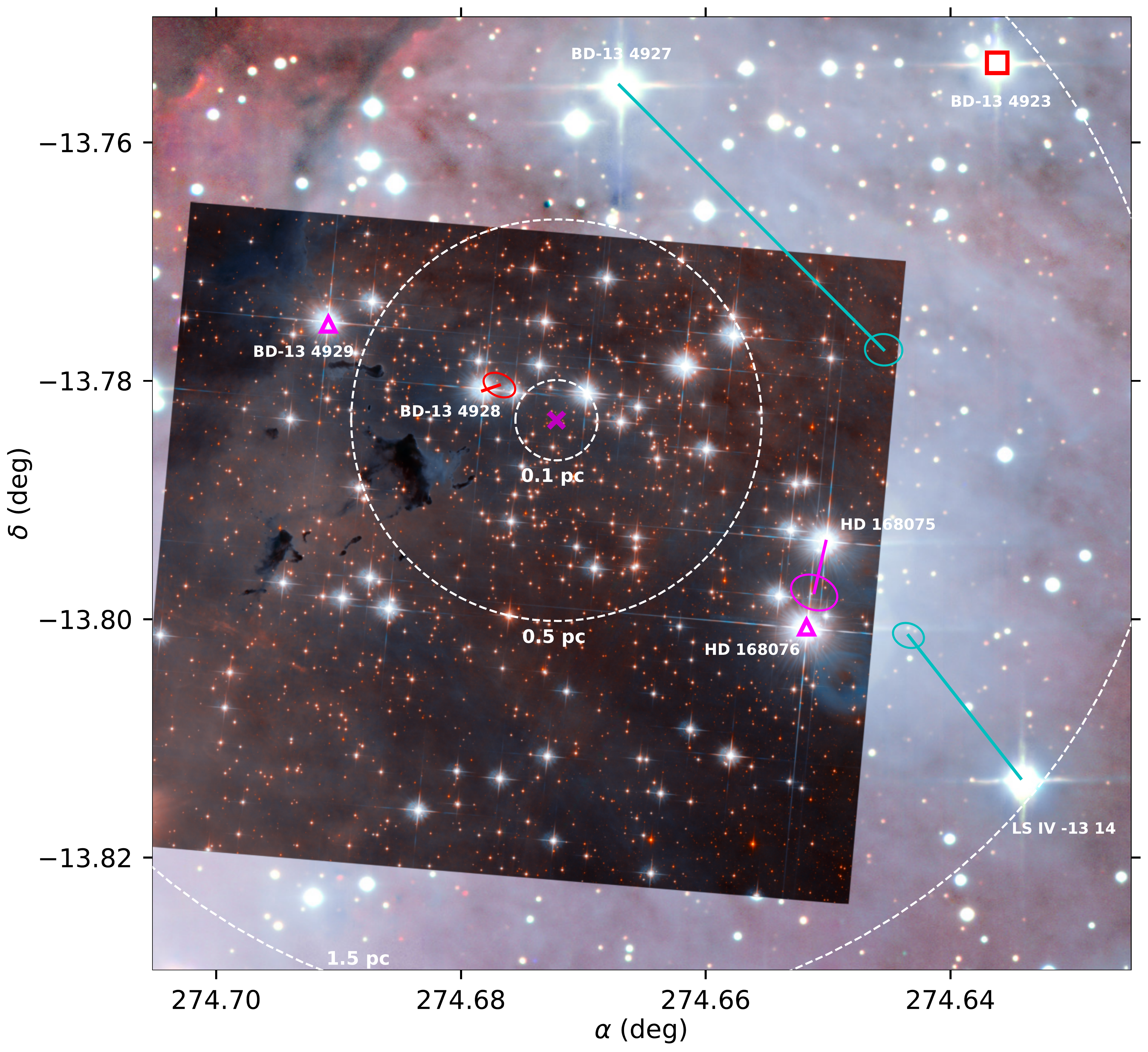}
\caption{B, V and R composite mosaic image of NGC6611 (ESO, press release 0926). We overplot the near-infrared Hubble Space Telescope image of the centre of NGC6611 (ESA/Hubble). The purple cross indicates the centre of NGC6611, with three white dashed circles with radii of 0.1, 0.5 and 1.5 pc at a distance of 1.70 kpc. We show the motion of the O stars and their position at $t$ = --130 kyr relative to NGC6611, coloured similarly as in Figure~\ref{fig:dyn_int}, with the error ellipse giving the uncertainty on the proper motion.}
\label{fig:HST_dynint}
\end{figure*}

The runaway O stars can all be traced back to the centre part of NGC6611. If we assume dynamical interactions to be the cause of their ejection, we might be able to identify the exact location where these interactions took place. This is best done for the O stars most recently ejected, since the uncertainties increase over the time traced back. To do this we show the \textit{Hubble Space Telescope} image of NGC6611 with the Advanced Camera for Surveys image on top of the ESO B, V and R composite mosaic image in Figure~\ref{fig:HST_dynint}. We show the location and relative proper motion traced back for the O stars in this image for $t$ = --0.13 Myr, coloured and marked similarly as in Figure~\ref{fig:dyn_int}.

One O star system stands out, the bright system HD 168076 composed of an O3.5 V((f+)) + O7.5 V. HD 168076 fits the description of a bully binary \citep{Fujii2011}, being able to accelerate other O stars (see Section~\ref{sec:dis_bully_binary} for a discussion). While the proper motion of HD 168076 is unusable, both HD 168075 and LS IV --13 14 are roughly consistent with coming from a similar position. This is also true for the O7 V((f)) star W222, located north of BD--13$^{\circ}$ 4923 and BD--13$^{\circ}$ 4927 and not visible in this image. W222 is consistent with being at the same location as W161 for $t$ = --0.22 Myr. These are the only cases where the positions (with uncertainty) overlap with each other in the last 0.5 Myr. This of course does not confirm that these O stars have dynamically interacted in the past, since we are ignoring the most difficult parameter to constrain; the radial depth of the cluster. We refer to HD 168076, HD 168075 and W222 in Table~\ref{tab:Ostars_runaways} as 'candidate dynamically interacted O stars'.

For other accelerated O star systems it is more difficult to constrain the location of dynamical interaction. BD--13$^{\circ}$ 4927 does not come across any other O star system when traced back. This is not certain as the local standard of rest of the dynamical interaction could be different and the proper motions of the O star systems HD 168076 and BD--13$^{\circ}$ 4923 are unknown. The complete 3D velocities for the O stars in NGC6611 could give more insight on the importance of dynamical interactions in NGC6611. We have also not included the (early type) B stars, for which (systemic) radial velocities are incomplete.

\clearpage
\section{Members of NGC6611}
\onecolumn
\begin{landscape}
\topcaption{Members of NGC6611 with, if known, their spectral, astrometric and photometric data.}
\label{tab:members_all}
\begin{supertabular}{l l l l l l l l l l l l l}
\hline
\textit{Gaia} source\_id & Spectral type & $\alpha$ & $\delta$ & $\varpi$ & $\mu_{\alpha^{*}}$ & $\mu_{\delta}$ & $v_{\rm{R}}$ & $\sigma_{v_{\rm{R}}}$ & (B--V) & V   & E(B--V) & Ref \\
- & - & deg & deg & mas & mas yr$^{-1}$ & mas yr$^{-1}$ & km s$^{-1}$ & km s$^{-1}$ & mag & mag & mag & - \\
\hline
4146599166888812032 & O6.5 V((f)) + B0-1 V & 274.6502 & -13.7935 & 0.629 & 0.181 & -1.481 & 17.0 & 10.0 & 0.54 & 8.752 & 0.86 & 1 \\
4146599682284818176 & O7 V + O8 V & 274.7341 & -13.8086 & 0.566 & 0.55 & -1.668 & 26.3 & 9.2 & 0.41 & 8.942 & - & 1 \\
4146599476126383872 & O9.5 Vp & 274.7195 & -13.8285 & 0.633 & 0.293 & -1.594 & 3.0 & 1.0 & 0.34 & 9.368 & 0.65 & 1 \\
4146594386587183360 & B0.5 V + B1: & 274.7732 & -13.914 & 0.601 & 0.401 & -1.705 & -104.0 & 5.0 & 0.42 & 9.396 & 0.7 & 2 \\
4146598995090108288 & B0.5 V & 274.6248 & -13.8327 & 0.586 & 0.047 & -1.676 & 14.0 & 5.0 & 0.88 & 9.687 & 1.17 & 2 \\
4146600781797073920 & O7 V + (B0.5 V + B0.5 V) & 274.691 & -13.7753 & 0.621 & 0.374 & -2.263 & 17.0 & 1.0 & 0.71 & 9.803 & - & 1 \\
4146604939324925440 & B2.5 I & 274.4627 & -13.8488 & 0.578 & 0.17 & -1.621 & - & - & 1.41 & 9.902 & 1.6 & 4 \\
4146612670266023040 & O4 V((f+)) + O7.5 V & 274.6364 & -13.7533 & 0.256 & -0.433 & -2.124 & 3.5 & 5.3 & 0.98 & 10.007 & - & 1 \\
4146600678717290624 & O9.5 V & 274.6782 & -13.7808 & 0.623 & 0.244 & -1.613 & 16.0 & 15.0 & 0.56 & 10.044 & 0.87 & 1 \\
4146599029451913984 & O9 V & 274.6343 & -13.8134 & 0.583 & -0.04 & -1.932 & 19.0 & 5.0 & 0.73 & 10.296 & 1.04 & 2 \\
4146599819725862912 & B0.5 Vn & 274.7703 & -13.8057 & 0.612 & 0.406 & -1.772 & 20.0 & 5.0 & 0.47 & 10.628 & 0.76 & 2 \\
4146594906281066368 & B1.5 V & 274.8335 & -13.906 & 0.582 & 0.158 & -1.529 & -47.0 & 5.0 & 0.43 & 10.733 & 0.68 & 2 \\
4146612494169472256 & O7 II(f) & 274.6671 & -13.7552 & 0.605 & 0.788 & -0.985 & 17.0 & 3.0 & 0.67 & 11.114 & - & 1 \\
4146612395388089344 & B1 V & 274.6578 & -13.7764 & 0.555 & 0.17 & -1.59 & 8.0 & 5.0 & 0.71 & 11.122 & 0.99 & 2 \\
4146613632338722432 & O8.5 V & 274.629 & -13.719 & 0.586 & -0.058 & -1.43 & -6.0 & 5.0 & 1.24 & 11.215 & 1.55 & 2 \\
4146600330822092160 & B1 V & 274.7112 & -13.8035 & 0.52 & 0.319 & -1.572 & 0.0 & 5.0 & 0.46 & 11.217 & 0.74 & 2 \\
4146599162597495680 & B1 V & 274.6541 & -13.798 & 0.549 & 0.1 & -1.865 & 8.0 & 5.0 & 0.61 & 11.407 & 0.88 & 2 \\
4146594803201839744 & B1.5 V & 274.827 & -13.9278 & 0.539 & 0.11 & -1.642 & 42.0 & 5.0 & 0.31 & 11.436 & 0.56 & 2 \\
4146597105304415488 & B1.5 V & 274.6666 & -13.9093 & 0.575 & 0.171 & -1.307 & 4.0 & 5.0 & 0.5 & 11.474 & 0.76 & 2 \\
4146600575638051712 & B1.5 V & 274.6861 & -13.799 & 0.637 & 0.229 & -1.932 & 14.0 & 5.0 & 0.63 & 11.718 & 0.88 & 2 \\
4146600674426016640 & B1 V & 274.6758 & -13.7811 & 0.516 & 0.173 & -1.579 & 8.0 & 10.0 & 0.59 & 12.037 & 0.87 & 3 \\
4146617819936287488 & O9 V & 274.5985 & -13.6078 & 0.566 & 0.2 & -1.451 & 6.0 & 5.0 & 1.24 & 12.052 & 1.55 & 2 \\
4146593183996341120 & B2 V & 274.8046 & -13.9606 & 0.541 & 0.21 & -1.778 & 6.0 & 10.0 & 0.39 & 12.262 & 0.61 & 3 \\
4146598887718503168 & B3 V & 274.6837 & -13.8157 & 0.578 & 0.177 & -1.741 & 8.0 & 5.0 & 0.65 & 12.529 & 0.83 & 2 \\
4152606172568933504 & B1.5 V & 274.7518 & -13.7114 & 0.541 & 0.405 & -1.317 & 7.0 & 5.0 & 1.04 & 12.594 & 1.29 & 2 \\
4146600575638052608 & B2 V & 274.6877 & -13.7964 & 0.575 & 0.155 & -1.574 & 6.0 & 10.0 & 0.67 & 12.607 & 0.88 & 3 \\
4146600777505252608 & B1 V & 274.6876 & -13.7614 & 0.57 & 0.108 & -1.714 & 2.0 & 10.0 & 0.81 & 12.676 & 1.09 & 3 \\
4146604076033711360 & - & 274.4674 & -13.8843 & 0.625 & 0.063 & -1.464 & - & - & 0.84 & 12.705 & - & - \\
4146600640059737088 & B2 V & 274.66 & -13.7859 & 0.576 & 0.365 & -1.835 & -50.0 & 20.0 & 0.73 & 12.753 & 0.95 & 2 \\
4146600472558819200 & B9 III & 274.7326 & -13.7817 & 0.574 & 0.337 & -1.795 & -32.0 & 20.0 & 0.7 & 12.789 & - & 2 \\
4146594768842190592 & B2.5 V & 274.739 & -13.8701 & 0.575 & 0.029 & -1.4 & 17.0 & 5.0 & 0.51 & 12.793 & 0.71 & 2 \\
4146600777505249408 & B2 Vn & 274.6855 & -13.7634 & 0.583 & 0.025 & -1.668 & 1.0 & 5.0 & 0.81 & 12.828 & 1.03 & 2 \\
4146601022314634368 & B3 V + ? & 274.7696 & -13.7457 & 0.608 & 0.323 & -1.392 & 2.0 & 5.0 & 0.61 & 12.834 & 0.78 & 2 \\
4146598818993604480 & B5 III & 274.6922 & -13.8232 & 0.558 & 0.412 & -1.676 & 23.0 & 5.0 & 0.62 & 12.866 & 0.78 & 2 \\
4146612872133107584 & O7 V((f)) & 274.6562 & -13.7276 & 0.605 & 0.388 & -1.526 & 16.0 & 5.0 & 1.52 & 12.968 & 1.84 & 2 \\
4146600678717295104 & B3 V & 274.6737 & -13.7789 & 0.598 & 0.35 & -1.572 & 5.0 & 5.0 & 0.64 & 13.038 & 0.82 & 2 \\
4146612425456526208 & B2 V & 274.6682 & -13.7717 & 0.56 & 0.321 & -1.343 & 6.0 & 10.0 & 0.81 & 13.259 & 1.02 & 3 \\
4152605214803721216 & B1-3 V & 274.8297 & -13.7312 & 0.542 & 0.416 & -1.656 & 2.0 & 5.0 & 0.64 & 13.285 & 0.85 & 2 \\
4146600399547025280 & B5 III & 274.7049 & -13.8012 & 0.573 & 0.26 & -1.762 & 5.0 & 5.0 & 0.51 & 13.288 & 0.67 & 2 \\
4146598685852419584 & A2 IV & 274.665 & -13.8483 & 0.58 & 0.236 & -1.813 & -4.0 & 10.0 & 0.79 & 13.325 & 0.72 & 3 \\
\end{supertabular}

\renewcommand\thetable{C.1.}
\topcaption{Continued.}
\begin{supertabular}{l l l l l l l l l l l l l}
\hline
\textit{Gaia} source\_id & Spectral type & $\alpha$ & $\delta$ & $\varpi$ & $\mu_{\alpha^{*}}$ & $\mu_{\delta}$ & $v_{\rm{R}}$ & $\sigma_{v_{\rm{R}}}$ & (B--V) & V   & E(B--V) & Ref \\
- & - & deg & deg & mas & mas yr$^{-1}$ & mas yr$^{-1}$ & km s$^{-1}$ & km s$^{-1}$ & mag & mag & mag & - \\
\hline
4146600541278311680 & - & 274.722 & -13.7687 & 0.638 & 0.404 & -1.847 & - & - & 1.06 & 13.347 & - & - \\
4146600536980521344 & B4 V & 274.7209 & -13.7792 & 0.597 & 0.205 & -1.55 & 12.0 & 10.0 & 0.69 & 13.379 & 0.85 & 3 \\
4146600403839355392 & B4 V & 274.6948 & -13.7969 & 0.578 & 0.331 & -1.508 & 4.0 & 10.0 & 0.74 & 13.392 & 0.9 & 3 \\
4146599544847962624 & B7 V & 274.7171 & -13.8248 & 0.63 & -0.112 & -1.589 & 5.0 & 10.0 & 0.58 & 13.403 & 0.71 & 3 \\
4146612803413634688 & B2 V & 274.6589 & -13.7404 & 0.645 & -0.086 & -1.579 & -3.0 & 10.0 & 1.11 & 13.435 & 1.32 & 3 \\
4146612528535774080 & B1 V & 274.6874 & -13.757 & 0.629 & 0.046 & -1.673 & 7.0 & 10.0 & 1.31 & 13.443 & 1.59 & 3 \\
4146600609998380800 & B8 V:e & 274.6942 & -13.7833 & 0.588 & 0.41 & -1.679 & - & - & 0.79 & 13.552 & 0.89 & 4 \\
4146599579205603456 & B7 V & 274.7289 & -13.8127 & 0.55 & 0.433 & -1.706 & 10.0 & 10.0 & 0.62 & 13.666 & 0.75 & 3 \\
4146600644357559680 & B4 V & 274.6656 & -13.7823 & 0.566 & 0.239 & -1.604 & 38.0 & 10.0 & 0.75 & 13.705 & 0.91 & 3 \\
4146600674426021888 & A3 III-II & 274.6787 & -13.7785 & 0.615 & 0.146 & -1.434 & -1.0 & 10.0 & 0.68 & 13.733 & 0.58 & 3 \\
4146598273535559936 & B9 & 274.6513 & -13.8596 & 0.589 & 0.172 & -1.491 & 26.0 & 10.0 & 0.63 & 13.843 & 0.7 & 3 \\
4146598166158712576 & Ae & 274.6051 & -13.8887 & 0.595 & 0.273 & -1.357 & - & - & 0.62 & 13.844 & - & 4 \\
4146610913628015872 & - & 274.6527 & -13.7887 & 0.614 & -0.073 & -1.663 & - & - & 1.01 & 13.869 & - & - \\
4146599579205601920 & - & 274.7316 & -13.8124 & 0.56 & 0.284 & -1.716 & - & - & 1.34 & 13.965 & - & - \\
4146600713077005312 & - & 274.705 & -13.7818 & 0.588 & 0.142 & -1.633 & - & - & 1.02 & 14.078 & - & - \\
4146600609997796864 & B2 V & 274.6886 & -13.789 & 0.608 & 0.172 & -1.685 & 2.0 & 10.0 & 1.1 & 14.178 & 1.31 & 3 \\
4146600678717298304 & - & 274.6732 & -13.7753 & 0.616 & 0.045 & -1.637 & - & - & 1.43 & 14.267 & - & - \\
4146612704625753600 & B7 V & 274.6501 & -13.7537 & 0.598 & 0.261 & -1.706 & 13.0 & 10.0 & 1.06 & 14.314 & 1.19 & 3 \\
4146594871922064000 & A1 V & 274.8212 & -13.9141 & 0.572 & 0.189 & -1.485 & 22.0 & 10.0 & 0.62 & 14.322 & 0.58 & 3 \\
4146613907216640256 & B1: V & 274.6223 & -13.7012 & 0.659 & -0.03 & -1.366 & - & - & 1.69 & 14.445 & 1.97 & 4 \\
4146600335119862912 & - & 274.7153 & -13.7999 & 0.612 & 0.4 & -1.605 & - & - & 1.36 & 14.53 & - & - \\
4152617927906894080 & B & 274.7454 & -13.6948 & 0.585 & 0.31 & -1.317 & - & - & 1.19 & 14.558 & - & 4 \\
4146407748789496320 & - & 274.9165 & -13.8955 & 0.557 & 0.462 & -1.603 & - & - & 1.12 & 14.56 & - & - \\
4146591161069679744 & - & 274.6311 & -13.9387 & 0.63 & -0.036 & -1.801 & - & - & 0.71 & 14.671 & - & - \\
4146588103052936064 & - & 274.6621 & -13.9571 & 0.552 & 0.231 & -1.316 & - & - & 0.61 & 14.694 & - & - \\
4146600884875679232 & - & 274.7524 & -13.764 & 0.584 & 0.394 & -1.654 & - & - & 1.33 & 14.738 & - & - \\
4146612391096772224 & - & 274.6579 & -13.7748 & 0.573 & 0.243 & -1.576 & - & - & 1.56 & 14.8 & - & - \\
4146588103052934784 & - & 274.665 & -13.9566 & 0.611 & 0.23 & -1.595 & - & - & 1.29 & 14.849 & - & - \\
4146596418109622016 & - & 274.7973 & -13.8328 & 0.555 & 0.397 & -1.29 & - & - & 0.76 & 14.91 & - & - \\
4146597822566610944 & - & 274.6812 & -13.8735 & 0.632 & 0.222 & -1.543 & - & - & 0.85 & 14.938 & - & - \\
4146592054422916608 & - & 274.5715 & -13.9306 & 0.582 & 0.022 & -1.614 & - & - & 1.47 & 14.951 & - & - \\
4146598995090119040 & - & 274.6232 & -13.8193 & 0.573 & 0.384 & -1.372 & - & - & 0.92 & 15.163 & - & - \\
4146598685852424576 & - & 274.6631 & -13.8442 & 0.563 & 0.305 & -1.759 & - & - & 0.94 & 15.253 & - & - \\
4146593085214950144 & - & 274.7387 & -13.9622 & 0.583 & 0.227 & -1.386 & - & - & 0.96 & 15.317 & - & - \\
4146612670266024192 & - & 274.6369 & -13.7513 & 0.609 & -0.036 & -1.848 & - & - & 1.23 & 15.436 & - & - \\
4146610844902247424 & - & 274.6204 & -13.7983 & 0.618 & 0.382 & -1.347 & - & - & 1.41 & 15.489 & - & - \\
4146610952279094400 & - & 274.6248 & -13.7811 & 0.567 & 0.067 & -1.269 & - & - & 0.96 & 15.535 & - & - \\
4146590439515163392 & - & 274.5854 & -13.9968 & 0.573 & 0.249 & -1.879 & - & - & 1.24 & 15.594 & - & - \\
4146594562683743744 & - & 274.7346 & -13.8998 & 0.532 & 0.496 & -1.856 & - & - & 0.86 & 15.733 & - & - \\
4152622562164286336 & - & 274.6853 & -13.5837 & 0.648 & 0.06 & -1.382 & - & - & 1.06 & 15.74 & - & - \\
\end{supertabular}

\renewcommand\thetable{C.1.}
\topcaption{Continued.}
\begin{supertabular}{l l l l l l l l l l l l l}
\hline
\textit{Gaia} source\_id & Spectral type & $\alpha$ & $\delta$ & $\varpi$ & $\mu_{\alpha^{*}}$ & $\mu_{\delta}$ & $v_{\rm{R}}$ & $\sigma_{v_{\rm{R}}}$ & (B--V) & V   & E(B--V) & Ref \\
- & - & deg & deg & mas & mas yr$^{-1}$ & mas yr$^{-1}$ & km s$^{-1}$ & km s$^{-1}$ & mag & mag & mag & - \\
\hline
4146600713077013632 & - & 274.7053 & -13.7709 & 0.607 & 0.258 & -1.569 & - & - & 1.69 & 15.848 & - & - \\
4146599304328319360 & - & 274.7447 & -13.8337 & 0.573 & 0.051 & -1.499 & - & - & 1.06 & 15.928 & - & - \\
4146610226426904832 & - & 274.5822 & -13.8508 & 0.532 & 0.394 & -1.496 & - & - & 1.47 & 15.933 & - & - \\
4146600644357558016 & - & 274.6641 & -13.7866 & 0.51 & 0.243 & -1.5 & - & - & 1.69 & 15.936 & - & - \\
4146611429017808128 & - & 274.5512 & -13.7955 & 0.588 & 0.16 & -1.396 & - & - & 1.28 & 16.072 & - & - \\
4146612597255215360 & - & 274.667 & -13.7488 & 0.548 & 0.112 & -1.648 & - & - & 1.77 & 16.129 & - & - \\
4146612425456535424 & - & 274.6742 & -13.7709 & 0.513 & 0.389 & -1.618 & - & - & 1.96 & 16.179 & - & - \\
4152606378727398656 & - & 274.837 & -13.7012 & 0.577 & 0.113 & -1.683 & - & - & 0.9 & 16.321 & - & - \\
4146599269967938816 & - & 274.7353 & -13.8368 & 0.622 & 0.159 & -1.463 & - & - & 1.15 & 16.327 & - & - \\
4146612391096778880 & - & 274.662 & -13.7701 & 0.631 & 0.422 & -1.456 & - & - & 1.89 & 16.355 & - & - \\
4146612528535768960 & - & 274.6844 & -13.754 & 0.628 & 0.175 & -1.449 & - & - & 1.61 & 16.402 & - & - \\
4146599063809568000 & - & 274.6655 & -13.8115 & 0.585 & -0.098 & -1.621 & - & - & 1.66 & 16.432 & - & - \\
4146599300035455744 & - & 274.7445 & -13.8273 & 0.602 & 0.439 & -1.869 & - & - & 1.47 & 16.45 & - & - \\
4146597822566611840 & - & 274.6814 & -13.8721 & 0.635 & 0.163 & -1.535 & - & - & 1.26 & 16.467 & - & - \\
4146594180429043200 & - & 274.6914 & -13.9228 & 0.54 & 0.309 & -1.846 & - & - & 1.58 & 16.52 & - & - \\
4146611605109426432 & - & 274.523 & -13.7842 & 0.554 & 0.317 & -1.649 & - & - & 1.42 & 16.561 & - & - \\
4146612459816239232 & - & 274.651 & -13.772 & 0.593 & 0.085 & -1.65 & - & - & 1.71 & 16.647 & - & - \\
4146596864786208768 & - & 274.8335 & -13.8202 & 0.523 & 0.441 & -1.566 & - & - & 1.46 & 16.656 & - & - \\
4146600365188397568 & - & 274.6915 & -13.798 & 0.674 & 0.2 & -1.458 & - & - & 1.99 & 16.759 & - & - \\
4146598303597678336 & - & 274.6267 & -13.8631 & 0.658 & 0.23 & -1.447 & - & - & 1.68 & 16.814 & - & - \\
4146404896931217536 & - & 274.7741 & -14.0137 & 0.63 & 0.107 & -1.841 & - & - & 1.53 & 16.816 & - & - \\
4146600609997801856 & - & 274.6911 & -13.7816 & 0.556 & 0.121 & -1.606 & - & - & 1.9 & 16.95 & - & - \\
4146617304533144832 & - & 274.5702 & -13.6492 & 0.63 & 0.291 & -1.676 & - & - & 1.34 & 17.039 & - & - \\
4146600575638069376 & - & 274.6758 & -13.7896 & 0.606 & 0.28 & -1.627 & - & - & 1.97 & 17.163 & - & - \\
4152606589193262080 & - & 274.839 & -13.6716 & 0.533 & 0.093 & -1.803 & - & - & 1.14 & 17.26 & - & - \\
4146406821076538880 & - & 274.8894 & -13.9571 & 0.511 & 0.229 & -1.654 & - & - & 1.36 & 17.326 & - & - \\
4146617205751575168 & - & 274.5646 & -13.6605 & 0.549 & 0.233 & -1.826 & - & - & 2.06 & 17.337 & - & - \\
4146600506918571392 & - & 274.7079 & -13.7832 & 0.569 & 0.261 & -1.712 & - & - & 1.84 & 17.4 & - & - \\
4146613597978990080 & - & 274.6184 & -13.7147 & 0.612 & -0.017 & -1.478 & - & - & 1.94 & 17.493 & - & - \\
4146599128236638336 & - & 274.6529 & -13.8057 & 0.527 & 0.186 & -1.666 & - & - & 1.92 & 17.506 & - & - \\
4146598651492680832 & - & 274.6606 & -13.8529 & 0.589 & 0.19 & -1.391 & - & - & 1.78 & 17.507 & - & - \\
4146612429747830144 & - & 274.6633 & -13.768 & 0.579 & 0.096 & -1.656 & - & - & 2.02 & 17.511 & - & - \\
4146600296467811840 & - & 274.7062 & -13.8127 & 0.623 & 0.306 & -1.52 & - & - & 1.91 & 17.567 & - & - \\
4146600640066272512 & - & 274.6726 & -13.7857 & 0.553 & 0.235 & -1.501 & - & - & 1.96 & 17.623 & - & - \\
4146600644357556352 & - & 274.6728 & -13.7807 & 0.606 & 0.243 & -1.596 & - & - & 1.89 & 17.67 & - & - \\
4146600468261266304 & - & 274.7384 & -13.7856 & 0.577 & -0.069 & -1.557 & - & - & 1.7 & 17.695 & - & - \\
4146617240111308544 & - & 274.577 & -13.6563 & 0.633 & 0.21 & -1.302 & - & - & 2.01 & 17.743 & - & - \\
4146600811858666240 & - & 274.6972 & -13.7539 & 0.62 & 0.02 & -1.741 & - & - & 2.2 & 17.749 & - & - \\
4146594867623592704 & - & 274.8288 & -13.9154 & 0.641 & 0.316 & -1.732 & - & - & 1.57 & 17.851 & - & - \\
4146612459816216960 & - & 274.6392 & -13.7679 & 0.627 & 0.175 & -1.834 & - & - & 2.21 & 17.858 & - & - \\
\end{supertabular}

\renewcommand\thetable{C.1.}
\topcaption{Continued.}
\begin{supertabular}{l l l l l l l l l l l l l}
\hline
\textit{Gaia} source\_id & Spectral type & $\alpha$ & $\delta$ & $\varpi$ & $\mu_{\alpha^{*}}$ & $\mu_{\delta}$ & $v_{\rm{R}}$ & $\sigma_{v_{\rm{R}}}$ & (B--V) & V   & E(B--V) & Ref \\
- & - & deg & deg & mas & mas yr$^{-1}$ & mas yr$^{-1}$ & km s$^{-1}$ & km s$^{-1}$ & mag & mag & mag & - \\
\hline
4146598892010860800 & - & 274.6868 & -13.8125 & 0.558 & 0.271 & -1.334 & - & - & 1.82 & 17.889 & - & - \\
4146600609998381056 & - & 274.6945 & -13.7811 & 0.603 & 0.234 & -1.925 & - & - & 1.87 & 17.906 & - & - \\
4146612700334401024 & - & 274.6469 & -13.7429 & 0.589 & -0.035 & -1.669 & - & - & 2.19 & 17.908 & - & - \\
4146600541278903680 & - & 274.7177 & -13.7775 & 0.499 & 0.118 & -1.803 & - & - & 1.8 & 17.956 & - & - \\
4146599162597503360 & - & 274.6596 & -13.7977 & 0.608 & 0.05 & -1.476 & - & - & 2.0 & 18.008 & - & - \\
4152605657177150080 & - & 274.8528 & -13.7031 & 0.653 & -0.011 & -1.667 & - & - & 1.67 & 18.019 & - & - \\
4146599063809568512 & - & 274.6613 & -13.815 & 0.61 & -0.031 & -1.373 & - & - & 1.92 & 18.033 & - & - \\
4146600262109272704 & - & 274.7884 & -13.7598 & 0.649 & 0.24 & -1.501 & - & - & 1.92 & 18.164 & - & - \\
4146599716644558464 & - & 274.7514 & -13.7907 & 0.663 & 0.265 & -1.763 & - & - & 1.81 & 18.22 & - & - \\
4152606039437447936 & - & 274.7982 & -13.7107 & 0.632 & 0.072 & -1.63 & - & - & 2.04 & 18.344 & - & - \\
4146613559321611264 & - & 274.6002 & -13.7334 & 0.645 & 0.03 & -1.376 & - & - & 2.14 & 18.361 & - & - \\
4146604595727511040 & - & 274.5341 & -13.84 & 0.566 & 0.207 & -1.515 & - & - & 1.84 & 18.367 & - & - \\
4146600300760123008 & - & 274.6984 & -13.8177 & 0.558 & 0.292 & -1.502 & - & - & 2.06 & 18.398 & - & - \\
4146600575638066304 & - & 274.6718 & -13.7966 & 0.598 & 0.198 & -1.621 & - & - & 2.02 & 18.464 & - & - \\
4146600571346803712 & - & 274.6771 & -13.7883 & 0.554 & 0.268 & -1.774 & - & - & 2.0 & 18.483 & - & - \\
4146599098169309312 & - & 274.6738 & -13.8017 & 0.538 & 0.404 & -1.457 & - & - & 1.97 & 18.521 & - & - \\
4146598273535563008 & - & 274.6498 & -13.8576 & 0.586 & -0.021 & -1.507 & - & - & - & - & - & - \\
4146599957162705408 & - & 274.7925 & -13.7864 & 0.607 & 0.021 & -1.538 & - & - & - & - & - & - \\
4146600369479617792 & - & 274.6842 & -13.8049 & 0.508 & 0.425 & -1.819 & - & - & - & - & - & - \\
4146612464107585280 & - & 274.6411 & -13.7617 & 0.579 & -0.068 & -1.617 & - & - & - & - & - & - \\
4146405584125976576 & - & 274.852 & -13.9597 & 0.512 & 0.322 & -1.649 & - & - & - & - & - & - \\
4146599132529067520 & O3.5 V((f+)) + O7.5 V & 274.6518 & -13.8007 & 0.952 & 1.972 & -0.771 & 10.0 & 2.0 & - & - & - & 1 \\
4146592707257808384 & O9.5 III + B3-5 V/III & 274.7445 & -13.9912 & 0.517 & 0.523 & -2.084 & 19.2 & 1.0 & - & - & - & 1 \\
4152407474508172160 & O7.5 V(n)z & 275.1421 & -13.9544 & 0.616 & 1.234 & -1.934 & 5.0 & 5.0 & - & - & - & 2 \\
4104201586232296960 & O5.5 V(n)((f)) & 276.4121 & -14.7516 & 0.619 & 5.888 & -4.899 & - & - & - & - & - & 5 \\
4147436097099460224 & O3-4 V (?) & 273.8498 & -13.3266 & 0.571 & -2.361 & -0.11 & - & - & - & - & - & 6 \\
\hline
\end{supertabular}
\tablebib{(1) \citet{Sana2009}; (2) \citet{Evans2005}; (3) \citet{Martayan2008}; (4) \citet{Hillenbrand1993}; (5) \citet{MaizApellaniz2016}; (6) \citet{Gvaramadze2008}}
\end{landscape}
\twocolumn
\end{appendix}

\end{document}